\pdfoutput=1
\documentclass[12pt]{scrartcl}
\usepackage{hyphenat}
\usepackage[T1]{fontenc}         
\usepackage[utf8]{inputenc}
\usepackage[british]{babel}
\areaset{158mm}{238mm}
\usepackage{setspace}
\usepackage[all]{xypic}
\usepackage{scrlayer-scrpage}
\usepackage{tikz}
\usetikzlibrary{cd}

\usepackage{extarrows}
\usetikzlibrary{arrows}

\usepackage{amsmath}
\usepackage{mathtools}
\usepackage{mathrsfs}
\usepackage{physics}

\usepackage{amssymb}
\usepackage[mathscr]{eucal}
\usepackage{atbegshi}
\usepackage[unicode=true,linktoc=all]{hyperref}
\usepackage{cite}
\usepackage{bm}
\usepackage{amsthm}
\usepackage{slashed}
\usepackage{tikz,pgf}
\usetikzlibrary{shapes}
\usetikzlibrary{calc}
\usetikzlibrary{decorations.pathmorphing}
\usetikzlibrary{decorations.pathreplacing,shapes.misc}
\usetikzlibrary{positioning}
\usetikzlibrary{decorations.markings}
\tikzset{->-/.style={decoration={
  markings,
  mark=at position .5 with {\arrow{>}}},postaction={decorate}}}
\tikzset{-<-/.style={decoration={
  markings,
  mark=at position .5 with {\arrow{<}}},postaction={decorate}}}

\usepackage{xcolor}
  \definecolor{rblue}{RGB}{81, 49, 193}
  \definecolor{rorange}{RGB}{255, 147, 40}
  \definecolor{rgreen}{RGB}{176, 233, 0}

\renewcommand{\tilde}{\widetilde}

\usetikzlibrary{decorations.pathmorphing}

\usepackage{lscape}

\newcommand{\bea}{\begin{equation}\begin{aligned}}
\newcommand{\eea}{\end{aligned}\end{equation}}
\newcommand{\beas}{\begin{equation*}\begin{aligned}}
\newcommand{\eeas}{\end{aligned}\end{equation*}}
\newcommand{\be}{\begin{equation}}
\newcommand{\ee}{\end{equation}}

\newcommand{\mathbbm}[1]{\mathds{#1}}
\newcommand{\apv}[3]{#1\begin{bmatrix} #2 \\ #3 \end{bmatrix}}
\newcommand{\smm}[1]{\left(\begin{smallmatrix} #1 \end{smallmatrix}\right)}

\DeclareMathOperator{\Vol}{Vol}

\tikzset{pics/.cd,
handle/.style={code={
\draw[fill=gray!20]  (-2,0) coordinate (-left) 
to [out=260, in=60] (-3,-2) 
to [out=240, in=110] (-3,-4) 
to [out=290,in=180] (0,-6) 
to [out=0,in=250] (3,-4) 
to [out=70,in=300] (3,-2) 
to [out=120,in=280] (2,0)  coordinate (-right);
\pgfgettransformentries{\tmpa}{\tmpb}{\tmp}{\tmp}{\tmp}{\tmp}
\pgfmathsetmacro{\myrot}{-atan2(\tmpb,\tmpa)}
\draw[rotate around={\myrot:(0,-2.5)}] (-1.2,-2.4) to[bend right]  (1.2,-2.4);
\draw[fill=white,rotate around={\myrot:(0,-2.5)}] (-1,-2.5) to[bend right] (1,-2.5) 
to[bend right] (-1,-2.5);
}}}

\numberwithin{equation}{section}

\vspace{0cm}

\title{\vspace{-1cm} Four-manifolds and Symmetry Categories of 2d CFTs}

\author{Vladimir Bashmakov,$^\dagger$ Michele Del Zotto,$^{\dagger,\ddagger}$ \\ and Azeem Hasan,$^{\ddagger}$
\\[1cm]
	\small\slshape$^\dagger$ Department of Physics and Astronomy, Uppsala University,  \\[-0.2cm] 
	\small\slshape Box 516, SE-75120 Uppsala, Sweden\\
	\small\slshape$^{\ddagger}$ Mathematics Institute, Uppsala University,  \\[-0.2cm] 
	\small\slshape Box 480, SE-75106 Uppsala, Sweden\\
	}
\date{}

\def\cC{{\cal C}}

\def\cN{{\cal N}}

\newcommand{\<}{\langle}
\renewcommand{\>}{\rangle}
\newcommand{\beaa}{\begin{eqnarray}}
\newcommand{\eeaa}{\end{eqnarray}}
\usepackage{dsfont}

\definecolor{dgreen}{rgb}{0, 0.55, 0}
\usepackage{pifont}
\usepackage{amssymb}
\definecolor{GreenYellow}{cmyk}{0.6,0,1.,0}%%%PANTONE GREEN
\definecolor{Red}{cmyk}{0,1.,1.,0}%%%PANTONE RED

\usetikzlibrary{shapes.multipart}
\usetikzlibrary{decorations.pathreplacing,calligraphy}
\usetikzlibrary{decorations.pathmorphing}
\tikzset{snake it/.style={decorate, decoration=snake}}
\usepackage{ulem}
\definecolor{darkred}{rgb}{0.8,0.1,0.1}
\hypersetup{colorlinks=true, linkcolor=darkred, citecolor=dgreen, linktoc=page}

\def\sixg{\mathfrak X_{\mathfrak g}}
\def\sixa{\mathfrak X_{N}}
\def\topg{\mathfrak F_{\mathfrak g}}
\def\topa{\mathfrak F_N}
\def\MCG{\mathcal M}

\begin{document}

\maketitle

\paragraph{\hspace{7cm}\large{Abstract}}
\vspace{-1cm}
\begin{abstract}

\noindent In this paper we study the geometric origin of non-invertible symmetries of 2d theories arising from the reduction of 6d $(2,0)$ theories on four-manifolds. This generalizes and extends our previous results in the context of class $\mathcal S$ theories to a wider realm of models. In particular, we find that relative 2d field theories, such as the chiral boson, have a higher dimensional origin in four-manifolds that are not null cobordant. Moreover, we see that for the 2d theories with a 6d origin, the non-invertible symmetries have a geometric origin as a sum over topologies from the perspective of the 7d symmetry TFT. In particular, we show that the Tambara-Yamagami non-invertible symmetries $TY(\mathbb Z_N)$ can be given a geometric origin of this kind. We focus on examples that do not depend on spin structures, but we analyse the simplest of such cases, finding an interesting parallel between the extra choices arising in that context and symmetry fractionalization in Maxwell theories. % Moreover, we give an explicit construction of a symmetry TFT with an $(E_8)_1$ current algebra. 
%We conclude with some comments on how to extend our construction to 3d theories.

\end{abstract}

\vfill{}
--------------------------

May 2023

%PREPRINT UUITP XX/YY
\thispagestyle{empty}

\newpage
\setcounter{tocdepth}{2}
\tableofcontents

\section{Introduction}
Finding a framework to study higher symmetries from a unified perspective is an important open question.\footnote{\ See footnote \ref{footnote:citationdump} for further comments and references.} A recent proposal to characterize the symmetry of a quantum field theory (QFT) in $D$ spacetime dimensions is to exploit a $D+1$ dimensional topological field theory (TFT)\cite{Freed:2022qnc,Apruzzi:2021nmk}, the so-called symmetry TFT. The symmetry TFT encodes the symmetry structure of the QFT in terms of boundary conditions, and it generalizes to higher dimensions the well known correspondence between symmetries of 2d theories and 3d TFTs \cite{Moore:1989yh,Reshetikhin:1991tc,Turaev:1992hq,Fuchs:2002cm,Fuchs:2003id,Fuchs:2004dz,Fuchs:2004xi,Fjelstad:2005ua}. For finite higher group global symmetries (see eg. \cite{Kapustin:2013qsa,Kapustin:2013uxa,Benini:2018reh}) the symmetry TFT are well-defined mathematical objects, the so-called finite homotopy TFTs of \cite{Freed:2009qp}. It is interesting to exhibit concrete examples for symmetries that are more general than finite higher group. The problem of determining the symmetry TFT for intrinsic non-invertible symmetries was recently studied in \cite{Kaidi:2022cpf}, the main outcome of that study is that for a symmetry to be intrinsically non-invertible, the resulting symmetry TFT must differ from a (possibly twisted) Dijkgraaf-Witten (DW) cocycle \cite{Dijkgraaf:1989pz}.\footnote{\, Here we recall that a given QFT has an intrinsic non-invertible symmetry if the latter does not have an origin as an otherwise invertible symmetry upon some topological manipulation on the theory, such as eg. gauging a discrete symmetry \cite{Kaidi:2022uux}.} %\footnote{\ For instance, in the case of internal higher symmetry groups with continuous components there are various explicit constructions in field theory\cite{Cordova:2018cvg,Cordova:2020tij,Brennan:2020ehu} and string/M/F theory \cite{Apruzzi:2021vcu,Apruzzi:2021mlh,DelZotto:2022joo,Cvetic:2022imb} --- see also \cite{Baez:2005sn,Sati:2008eg,Sati:2009ic,Fiorenza:2010mh,Fiorenza:2012tb} --- and yet a nice mathematically rigorous treatment of the corresponding symmetry TFT is not available.} 
An extremely powerful tool to explore and produce examples of intrinsic (as well as non-intrinsic) non-invertible symmetries is provided by higher dimensional SCFTs. In this paper we continue the exploration of the corresponding symmetry TFTs and non-invertible symmetries arising from the study of the dimensional reduction of higher dimensional SCFTs \cite{Bashmakov:2022jtl,Bashmakov:2022uek,Antinucci:2022cdi}, building on the seminal works \cite{Tachikawa:2013hya,Gukov:2020btk}.

While in \cite{Bashmakov:2022jtl,Bashmakov:2022uek,Antinucci:2022vyk} the focus have been four-dimensional theories obtained from compactifications of 6d (2,0) theories on Riemann surfaces without punctures (see eg. \cite{Gaiotto:2009we,Gaiotto:2009hg}), in this project we focus instead on the case of two dimensional theories and four manifolds (see eg. \cite{Gadde:2013sca,Putrov:2015jpa,Gukov:2018iiq,Dedushenko:2017tdw}).  Extending the construction outlined in \cite{Bashmakov:2022uek} we recover several known aspects of symmetry categories of 2d/3d theories from the 6d/7d perspective via dimensional reduction. In particular, the fact that for 2d models many features of symmetry categories are well-known provides a plethora of interesting consistency checks for the technique. 

For the sake of simplicity in this paper we focus on the $\mathfrak{a}_{N-1}$ 6d (2,0) theories $\sixa$. We find that whenever a given 4-manifold is null-bordant, the resulting 2d theory is an absolute CFT. If that is not the case then there are more possibilities, generalizing the well-known example of a chiral boson. The signature of the 4-manifold is related to the chirality of the resulting 2d SCFT in a natural way: four manifolds with non-vanisihing signature can give rise to relative 2d field theories (where the details are encoded in $N$). This gives an interesting parallel to theories with and without conformal blocks, revisiting and extending several well-known related studies \cite{Harvey:1995tg, Verlinde:1995mz, Seiberg:2011dr}. As a first application to non-invertible symmetries in two dimensions \cite{Frohlich:2004ef,Freed:2018cec,Chang:2018iay,Komargodski:2020mxz} we recover,  building on the analysis of \cite{Thorngren:2019iar,Thorngren:2021yso,Kaidi:2022cpf}, the symmetry TFT for the Tambara-Yamagami $\mathbb Z_N$ 2d categorical symmetry from $\sixa$, where we denote $\sixa$ the 6d (2,0) SCFT of $\mathfrak a_{N-1}$-type. The latter arises by looking at one of the simplest 4-manifolds with trivial signature $Y_4 = S^2 \times S^2$. Swapping the two $S^2$'s in this construction is equivalent to an electromagnetic $\mathbb Z_2$ duality, but at a point where the two $S^2$'s have the same volume, this becomes a symmetry of the theory, and we can gauge it: by this procedure one obtains the $TY(\mathbb Z_N)$ symmetry TFT \cite{Kaidi:2022cpf}. Gauging such $\mathbb Z_2$ from the perspective of the 7d CS theory is equivalent to a sum over topologies. This example is interesting because it demonstrates that also in the case of two dimensional compactifications, the non-invertible symmetries which arise as a gauging from the perspective of the symmetry TFT can be interpreted as a sum over topologies, as remarked in the context of the reduction on Riemann surfaces leading to 4d/5d setups  \cite{Bashmakov:2022jtl,Bashmakov:2022uek,Antinucci:2022vyk}. We stress however that the 4-manifold examples is much richer and one can obtain more general situations where 4-manifolds that are not null-bordant nevertheless give rise to gapped boundary conditions. These examples can be easily extended to more 4-manifolds. We present the general theory and we analyze in detail the cases where there is no explicit dependence on the spin structure. We give some more applications in the context of del Pezzo surfaces, 4-manifolds obtained from blow ups of $\mathbb P^2$ at generic points.\footnote{These examples are interesting as the corresponding wrapped M5 branes control the worldvolume theoires of monopole strings of certain well-known 5d SCFTs \cite{Seiberg:1996bd,Douglas:1996xp,Ganor:1996mu}.} We conclude this paper with a discussion of some interesting preliminary results towards extending our construction incorporating the dependence on the Wu structure of the four-manifold in the context of a toy theory with $N=1$, arising from the reduction of a single chiral tensor field on a four-manifold. We find an interesting parallel between parition functions of the chiral boson and the possible parition functions of Maxwell theories which differ by symmetry fractionalization by studying the theory on $\mathbb T^2 \times \mathbb P^2$.

\bigskip

This paper is organized as follows: in section \ref{sec:symm-tft-dimens} we quickly and schematically review the dimensional reduction of the $\sixa$ theory to lower dimensional field theories to set up notations and conventions. In section \ref{sec:7d-view-3d} we discuss in details the case of 4 manifolds. In section \ref{sec:spin-chern-simons} we comment on the dependence of the overall construction on the choice of a Wu structure for the 4-manifold, and we study the case of the dimensional reduction of a single self-dual tensor as an example to illustrate some of the salient features of this type of systems. In section \ref{sec:conclusions} we discuss some future directions and outlook, in particular commenting on how to extend our construction to the 3d cases.

\section{Symmetry TFT from dimensional reduction}
\label{sec:symm-tft-dimens}
We begin with a brief review of how to read off the symmetry TFT of the lower dimensional theories from the symmetry TFT of the $6d\, (2,0)$
theory. This has been covered extensively in e.g. \cite{Tachikawa:2013hya,Gukov:2020btk} (see also \cite{Bashmakov:2022uek}) to which we refer for further details.

\medskip

We will focus on $6d\, (2,0)$ theory $\sixa$. The symmetry TFT of this theory $\topa$ contains a 7d Chern-Simons term which is going to be of utmost importance for our analysis:
\begin{align}
\label{eq:1}
\topa &= \frac{N}{4\pi} \int_{W_{7}} c \wedge \dd c  + \cdots ~,
\end{align}
where $W_{7}$ is a seven manifold\footnote{\, For the sake of simplicity we will require that the fourth Wu class $\nu_{4}$ vanishes for $W_{7}$ but we will relax this condition later --- for a detailed discussion on the dependence of this construction from $\nu_4$ we refer our readers to appendix A of \cite{Gukov:2020btk}.} and we are using a differential form notation $c \in H^{3}(W_{7},U(1)$. The basic observables of this topological quantum field theory are the Wilson surfaces supported on 3-manifolds $\Sigma_{3}$ inside $W_{7}$,
\begin{align}
\label{eq:2}
  \Phi_{q}(\Sigma_{3}) = \exp(iq \oint_{\Sigma_{3}} c)~, && q \in \mathbb{Z}_{N} ~.
\end{align}

If two manifold $\Sigma_{3}$ and $\Sigma_{3}^{\prime}$ form a Hopf link inside $W_{7}$, then the correlation functions involving Wilson surfaces supported on them satisfy:
\begin{align}
\label{eq:3}
  \ev{\Phi_{q}(\Sigma_{3}) \Phi_{q^{\prime}}(\Sigma_{3}^{\prime})\, \dots } &= \exp(\frac{2\pi i}{N} qq^{\prime}\mbox{link}(\Sigma_{3},\Sigma_{3}^{\prime})) \ev{\dots } ~,
\end{align}
where $\dots $ denote any number of insertions which do not link with $\Sigma_{3}$ or $\Sigma_{3}^{\prime}$ and $\mbox{link}$ denotes the linking number. If $W_{7}$ has a $6d$ boundary $X_{6}$ (which we assume to be compact), we can push $\Sigma_{3}$ and $\Sigma_{3}^{\prime}$ to the boundary to obtain equal time commutation relations,
\begin{align}
\label{eq:4}
  \Phi_{q}(\Sigma_{3})\Phi_{q^{\prime}}(\Sigma_{3}^{\prime}) &= \exp(\frac{2\pi i}{N} qq^{\prime}\ev{\Sigma_{3},\Sigma_{3}^{\prime}})\Phi_{q^{\prime}}(\Sigma_{3}^{\prime}) \Phi_{q^{\prime}}(\Sigma_{3}) ~.
\end{align}
The bracket $\ev{\,,\,}$ denotes the intersection pairing inside the boundary $X_{6}$ of $W_{7}$. Since for generic values of $N$ the Wilson surfaces $\Phi(\Sigma_{3})$ realize a non-commutative algebra at the boundary we cannot impose Dirichlet boundary condition for $c$, in what follows we will assume this is the case, but this assumption can be relaxed when $N$ is a perfect square -- see section \ref{sec:topol-bound-cond}. As a topological quantum field theory, the 7d Chern-Simons theory associates to $X_{6}$ a Hilbert space $\topa(X_6)$ on which $\Phi(\Sigma_{3})$ with $\Sigma_{3} \in H_{3}(X_{6},\mathbb{Z}_{n})$ act, thus furnishing a representation of the non-commutative algebra given by \eqref{eq:4}.

Regarding $\Sigma_3$ as an element of $H_{3}(X_{6},\mathbb{Z}_{n})$, $\Phi_{q}(\Sigma_{3}) = \Phi_{1}(q\Sigma_{3})$ so we will define $\Phi(\Sigma_{3}) = \Phi_{1}(\Sigma_{3})$ and drop the subscript. Since the algebra given by \eqref{eq:4} is non-commutative, all of $\Phi(\Sigma_{3})$ can't be simultaneously diagonalized. We can simultaneously diagonalize $\Phi(\Sigma_{3})$ and $\Phi(\Sigma_{3}^{\prime})$ only if $\ev{\Sigma_{3},\Sigma_{3}^{\prime}} = 0 \mod N$. Moreover, if a state $\ket{K}$ is a common eigenvector of $\Phi(\Sigma_{3})$ and $\Phi(\Sigma_{3}^{\prime})$ it is also an eigenvector of $\Phi(q \Sigma_{3} + q^{\prime} \Sigma_{3}^{\prime})$. As a result the cycles $\Sigma_{3}$ that diagonalize $\ket{K}$ define a lattice which is isotropic i.e. no two cycles in it intersect. To define a representation of the Heisenberg algebra, the best we can do is demand that a state is simultaneously fixed by $\Phi(\Sigma_{3})$ with $\Sigma_{3}$ forming a maximal isotropic lattice $\mathcal{L}$ of $H_{3}(X_{6},\mathbb{Z}_{N})$, i.e. this state which we label by $\ket{\mathcal{L}}$ is defined by,
\begin{align}
\label{eq:5}
  \Phi(\Sigma_{3})\ket{\mathcal{L}} = \ket{\mathcal{L}} ~, && \forall \Sigma_{3} \in \mathcal{L} ~.
\end{align}
Starting from this state $\mathcal{L}$ we can obtain other states in the Hilbert space by acting on $\mathcal{L}$ with other $\Phi(\Sigma_{3}^{\prime})$ such that $\Sigma_{3}^{\prime}$ intersects some element of $\mathcal{L}$ i.e. the Hilbert space $\topa(X_6)$ is spanned by
\begin{align}
\label{eq:6}
  \ket{\mathcal{L},\Sigma_{3}^{\prime}} = \Phi(\Sigma_{3}^{\prime})\ket{\mathcal{L}} ~.
\end{align}
Such states are eigenstates of $\Phi(\Sigma_{3})$ with $\Sigma_{3} \in \mathcal{L}$ with eigenvalue being $\exp(\frac{2\pi i}{N} \ev{\Sigma_{3},\Sigma_{3}^{\prime}})$ and two states $\ket{\mathcal{L},\Sigma_{3}^{\prime}}$ and $\ket{\mathcal{L}  , \Sigma_{3}^{\prime\prime}}$ are linearly independent if and only if they belong to different coset in $\mathcal{L}^{\perp} = H_{3}(X_{6},\mathbb{Z}_{N}) / \mathcal{L}$. This procedure provides us with a representation of the algebra in \eqref{eq:4} with dimension $\abs{\mathcal{L}^{\perp}}$.

\subsection{Topological boundary conditions}
\label{sec:topol-bound-cond}
Each vector in the Hilbert space as described above gives us a boundary condition for the $7d$ TQFT. However, they do not determine a (gapped) topological boundary condition in general. The reason is that $X_{6}$ can in general have a non-trivial group of large diffeomorphism i.e diffeomorphism which are not continuously connected to identity, which we refer to as the mapping class group of $X_{6}$, and we denote 
\be
\MCG(X_6) = \pi_0 \text{Diff}(X_6)
\ee
in what follows. Given a non-trivial $F \in \MCG(X_6) $, it generally rotates the sublattice $\mathcal{L}$ into a different sublattice $F(\mathcal{L})$ and the cycle $\Sigma_{3}^{\prime}$ into a different one $F(\Sigma_{3}^{\prime})$. This transformation of states due to large diffeomorphisms is how the symmetry TFT encodes the relative nature of $6d\, (2,0)$ theory. To construct an absolute theory we must assign to each $X_{6}$ a ray in the Hilbert space constructed above which is invariant under all of $\MCG(X_{6})$. If one ends up with a projective representation of $\MCG(X_{6})$, such an invariant state does not exist. 

The requirement of invariance under large diffeomorphisms of all $X_{6}$ is very stringent and for generic $N$ no such boundary condition exists. The exception is when $N = M^{2}$ is a perfect square \cite{Gukov:2020btk,Tachikawa:2013hya}. In this case we choose the isotropic sublattice $\mathcal{L}$ to be $MH_{3}(X_{6},\mathbb{Z}_{M^{2}})$ i.e all cycles which are multiples of $M$. This lattice is invariant under all isomorphism of $H_{3}(X_{6},\mathbb{Z}_{M^{2}})$ and hence all diffeomorphisms of $X_{6}$. However for generic $N$ no such choice exists and the $(2,0)$ theory is relative.

\subsection{Dimensional reduction and boundary conditions}
\label{sec:dimens-reduct-symm}

The relative nature of the $6d$ theory described above can be exploited to construct theories in lower dimensions which differ only in their global structure i.e. these theories have the same spectrum of local operators but they differ in their global properties such as the corresponding spectra of extended operators, or their parition functions on compact manifolds with non-trivial topology. The general procedure is to start with an $X_{6} = Y_{k} \times M_{d}$, where $k=6-d$ and then dimensionally reduce the $6d$ theory along $Y_{k}$ to obtain a $d$ dimensional theory.\footnote{\, With a topological twist to preserve some supersymmetry but the precise details twisting are not important for us here.}  We denote the resulting theory
\be
D_{Y_k} \sixa = \mathcal T_N[Y_k] \ee
This will in general give us another relative theory, defined as an edge mode for the topological field theory $D_{Y_k} \topa$, obtained from dimensionally reducing $\topa$ along $Y_k$, however if we can choose a lattice $\mathcal{L} \subset H_{3}(Y_{k} \times M_{d} , \mathbb{Z}_{N})$ such that it is invariant under large diffeomorphisms of $Y_k$ in a way which allows to factor out the choice of $M_{d}$, then the theory $D_{Y_k} \topa$ can admit a gapped boundary condition which allows us to obtain an absolute theory.\footnote{\, This choice is referred to as a choice of polarization in section 3 of
\cite{Gukov:2020btk}} We  denote the latter interchangeably 
\be
\langle \mathcal L | D_{Y_k} \sixa \rangle  \qquad \text{or} \quad D_{Y_k}^{\mathcal{L}} \sixa \quad \text{or} \quad \mathcal T_N[Y_k]_\mathcal{L}\in \text{(S)QFT}_d\,.
\ee
Since the starting $6d$ theory is relative this means that $H_{3}(Y_{k},\mathbb{Z}_{N})$ is typically not invariant under at least some of the large diffeomorphism of $Y_{k}$. When the dimensionally reduced $d$-dimensional theory is absolute this non-trivial action typically translates to a statement about dualities of the theories thus obtained: 
\be
\mathcal T_N[Y_k]_\mathcal{L} \longleftrightarrow \mathcal T_N[F(Y_k)]_{F(\mathcal{L})}
\ee
where $F \in \MCG(Y_k \times M_d)$, but by construction ends up acting non-trivially only as its induced action on $Y_k$, $\pi_{Y_k}F \in \MCG(Y_k)$. In what follows often we abuse notation, leaving the above remark as understood. This is a well-known fact: we refer our readers to \cite{Gukov:2020btk,Tachikawa:2013hya} for a review and further references. Since in the current work we focus on the $2d$ theories let us briefly remind our readers about this specific case \cite{Gadde:2013sca,Putrov:2015jpa,Gukov:2018iiq}.

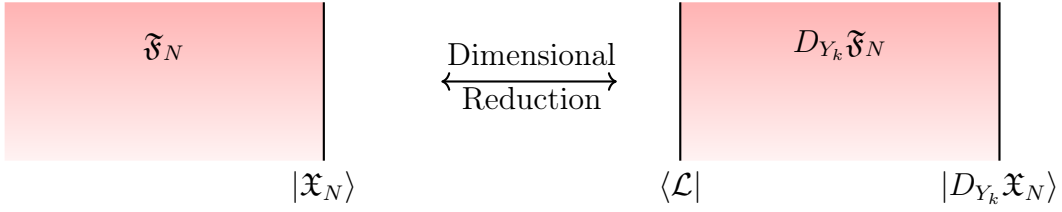
\begin{figure}
\begin{center}
\begin{tikzpicture}[scale=0.7]

	\begin{scope}[xshift=-5.2 in]
	\shade[line width=2pt, top color=red!30, bottom color=red!5] 
	(0,0) to [out=90, in=-90]  (0,3)
	to [out=0,in=180] (6,3)
	to [out = -90, in =90] (6,0)
	to [out=180, in =0]  (0,0);
	
	\node[above] at (3,1.7) {$\topa$};

	\draw[thick] (6,0) -- (6,3);
	\node[below] at (6,0) {$|\sixa\rangle $}; 
	\end{scope}

	\draw[thick, <->] (-1.7,1.5) -- (-5, 1.5);
	\node[above] at (-3.3,1.6) {Dimensional};
	\node[below] at (-3.3,1.6) {Reduction};

	\begin{scope}[xshift=-0.2 in]
	\shade[line width=2pt, top color=red!30, bottom color=red!5] 
	(0,0) to [out=90, in=-90]  (0,3)
	to [out=0,in=180] (6,3)
	to [out = -90, in =90] (6,0)
	to [out=180, in =0]  (0,0);
	
	\node[above] at (3,1.7) {$D_{Y_k}\topa$};
	
	\draw[thick] (0,0) -- (0,3);
	\draw[thick] (6,0) -- (6,3);
	\node[below] at (6,0) {$|D_{Y_k}\sixa\rangle $}; 
	\node[below] at (0,0) {$\langle \mathcal L| $};

	\end{scope}
	
\end{tikzpicture}	
\end{center}
	
	\caption{Dimensional reduction of the 7d/6d relative field theory. Provided a maximally isotropic sublattice $\mathcal L$ exists in a way which is independent of the choice of $M_d$ in the splitting $X_6 = Y_{k} \times M_d$ we obtain a corresponding absolute theory associated to a gapped topological boundary condition for $D_{Y_{k}}\topa$.
	}
	\label{fig:twistdefshrink}
\end{figure}

\subsection{Two dimensional theories}
\label{sec:two-dimens-theor}

Consider the case of $2d$ theories obtained by reducing the $6d$ theory on a four manifold $Y_{4}$ which we assume to be compact and simply connected. This ensures that the $2d$ theory we obtain is conformal and in particular for absolute theories this means that the partition functions on a torus $\mathbb{T}^{2}$ are modular invariant.\footnote{\ One can consider more general four manifolds with non-trivial one cycles. In that case, we expect the corresponding $2d$ theories to contain gauge fields such that their gauge couplings scale with the size of the cycles. As a result in the limit where we send the volume of $Y_{4}$ to zero and obtain a genuine $2d$ theory, the gauge fields decouple. Hence these gauge fields are not important for the low energy limit, but capture some KK modes.} Moreover restricting to simply connected four manifolds ensures that the $2d$ theory only depends on the conformal class of the metric on $Y_{4}$. The mapping class group of diffeomorphism preserving the conformal structure is bigger than that of large diffeomorphism preserving the metric. Although it is not in general a tractable object for internal manifolds $Y_k$ with $k>2$, in some simple examples we can determine some of its relevant subgroups. That is typically enough to constrain the symmetries. The aim of this paper is to exploit this geometric control when it is available to construct interesting topological defects in these 2d theories. Since the defects in two dimensional conformal field theories are much better understood than their higher dimensional counterparts, this allows us to make contact with existing literature. 

\medskip

We begin by constructing the symmetry TFT on a three manifold $W_{3}$ that bounds $M_{2}$. This comes from reducing the 7d abelian Chern-Simons theory on $Y_{4} \times W_{3}$. Choosing a basis $A_{i}$ of $2$-cycles on $Y_{4}$ we can write $c$ as $c|_{W_{3}} = \sum_{i} a_{i}$ with,
\begin{align}
\label{eq:14}
  a_{i} &= \oint_{A_{i}} c ~.
\end{align}
Integrating over the $4$-manifold gives us an abelian Chern-Simons theory in 3-dimensions with the action,
\begin{align}
\label{eq:15}
  S_{3d} = \frac{N}{4\pi}\int_{W_{3}} Q^{ij}\,a_{i} \wedge \dd a_{j}
\end{align}
where $Q$ is the intersection form of $Y_{4}$. $Q$ is always unimodular and symmetric \cite{Milnor}.

For simply connected $Y_{4}$,
\begin{align}
\label{eq:16}
  H_{3}(Y_{4} \times M_{2} , \mathbb{Z}_{N}) \cong H_{2}(Y_{4},\mathbb{Z}_{N}) \otimes H_{1}(M_{2},\mathbb{Z}_{N}) ~.
\end{align}
For any Riemann surface the only subgroups of integral first homology that are invariant under the mapping class group are the trivial group and the whole homology itself. As a result a maximal isotropic sublattice of $H_{3}(Y_{4} \times M_{2},\mathbb{Z}_{N})$ for generic $N$ must take the form $L \otimes H_{1}(M_{2},\mathbb{Z}_{N})$ where $L$ is an sublattice of $H_{2}(Y_{4},\mathbb{Z}_{N})$ isotropic with respect to intersection form $Q$ and with number of elements given by $\sqrt{\abs{H_{2}(Y_{4},\mathbb{Z}_{N})}}$. From \eqref{eq:15} it can be seen that such an isotropic lattice also determines a topological boundary condition for the symmetry TFT. However, since the intersection form $Q$ is symmetric, it is much harder to find such a sublattice.

To illustrate this point, for any bilinear form $Q$ over any field $\mathbb F$, the dimension of maximal isotropic sublattice of $Q$ is an invariant of $Q$ called Witt index. For $\mathbb F$ the field of rational numbers and $Q$ symmetric, the Witt index is given by the minimum of the number of positive eigenvalues and negative eigenvalues. Hence $H_{2}(Y_{4},\mathbb{Q})$ has an isotropic subspace of dimension $\frac{1}{2}\dim(H_{2}(Y_{4},\mathbb{Q}))$ only if signature of $Q$ vanishes. This should be contrasted with antisymmetric bilinear forms for which isotropic subspaces of this dimension always exist. The bound on the dimension of an isotropic subspace of homology with rational coefficients also limits the rank of an integral isotropic sublattice.

\medskip

For our purposes we will mostly consider $N=p$ a prime and hence $\mathbb F = \mathbb{F}_{p}$ is a finite field. For this case isotropic subspaces of dimension $\frac{1}{2} \dim(H_{2}(Y_{4},\mathbb{F}_{p}))$ can exist even when the signature of the intersection form on $H_{2}(Y_{4},\mathbb{Z})$ is not zero. These isotropic subspaces do not lift to the isotropic subspaces of $H_{2}(Y_{4},\mathbb{Z})$ instead such a lift contains cycles with intersections which are non-zero multiples of $p$. This is to be contrasted with the situation familiar in the construction of  class $\mathcal S$ theories: isotropic integrals lifts always exist for the maximal isotropic subspaces of the antisymmetric intersection forms of one-cycles on a Riemann surfaces, and this was a crucial ingredient for the geometrization of gapped topological boundary conditions in that context. The problem of determining gapped topological boundary conditions becomes much more involved in the case of dimensional reduction on four manifolds, a fact which is reflected in the richness of both four manifolds and two dimensional CFT. Exploring this interplay is among the main results of the current work.

\medskip

Lastly we note that if we relax the condition that $Y_{4}$ is simply connected, then the resulting theory has 1-form symmetries which lead to the phenomenon of decomposition. The operators which implement these operations are ``topological points.'' These 1-form symmetries were studied in \cite{Komargodski:2020mxz} to understand deconfinement in 2-dimensional adjoint QCD. It would be interesting to geometrize the results obtained there to understand the low energy behavior of a wider range gauge theories in two dimensions: we leave this study for future work.

\subsection{Exceptional Symmetries and Non Invertible Defects}
\label{sec:except-symm}
In the preceding section we have described how to obtain the symmetry TFT of the 2d theories which arise as the dimensional reduction of $6d\, (2,0)$ theory. However this description is incomplete: while the result is accurate for theories at generic values of the moduli associated to the internal manifold $Y_k$, it can be refined for certain special theories which we now describe. The symmetry TFT we have obtained can have automorphisms: for the case of dimensional reduction on a Riemann surface $\Sigma_{g}$ relevent for class $\mathcal{S}$ theories these automorphism correspond to group of $2g \times 2g$ matrices preserving the symplectic form i.e. the intersection form of one cycles on the Riemann surface. Similarly, for the case of Chern-Simons theory obtained by reduction on a $4$-manifold, this is the group of matrices preserving the intersection form $Q$.

An element $F$ of the mapping class group of the manifold $Y_{k}$ on which we reduce acts on the TFT through such a automorphism which leaves the symmetry TFT invariant. However its action on the boundary state is typically non-trivial, thus changing the theory on boundary to a dual theory. For the theories of class $\mathcal{S}$ this actions changes various gauge couplings to a dual value and at the same time it changes the maximal isotropic sublattice $L$ of $H_{1}(\Sigma_{g},\mathbb{Z}_{N})$ (which determines the isotropic sublattice and hence the global structure of the model) to a different one. For the case of two dimensional theories the action of $F$ on the boundary corresponds eg. to an operations that generalize changing the radii of various compact bosons to a dual value, and just like the 4d case, in order to give the complete statement of duality we must supply this change with a change in the isotropic sublattice of $H_{2}(Y_{4},\mathbb{Z}_{N})$ which determines the global structure.

As a result the mapping class group element $F$ generally gives rise to an operator in the symmetry TFT which when pushed to the boundary becomes a topological interface between two dual theories. However things are different if the continuous conformal moduli\footnote{e.g. the gauge couplings for $4d$ theories and the radii for $2d$ theories} are stabilized by a subgroup $\mathcal{G}$ of the mapping class group. In this case the defects corresponding to a mapping class group elements $G \in \mathcal{G}$ implement a duality symmetry. We can divide these symmetries into two broad classes:
\begin{itemize}
  \item If $\mathcal G$ fixes the global structure then the corresponding interface is a defect in the theory implementing a (possibly anomalous) $\mathcal{G}$ symmetry. Two such defects fuse according to the multiplication in $\mathcal G$;
  \item If $\mathcal G$ changes the global structure, then we can undo this change by half space gauging \cite{Choi:2021kmx} to again obtain a topological defect in the theory. Due to half space gauging the resulting defect becomes non-invertible.
\end{itemize}

Because of these additional symmetries, which are only present at these special points, the symmetry TFT has to be modified. This is done by gauging a suitable subgroup of $\mathcal{G}$, as explained in \cite{Kaidi:2022cpf}. The topological defects corresponding to these mapping class elements are constructed as condensates of specific 3d Wilson lines for $\sixa$, following the same procedure already outlined in \cite{Bashmakov:2022uek}. Since these are crucial for our construction we review them quickly in the following section. 

\subsection{Codensation Surgery Defects}

\begin{figure}
\begin{center}
\begin{tikzpicture}[scale=0.7]

	\begin{scope}[xshift=-5.2 in]
	\shade[line width=2pt, top color=green!30, bottom color=green!5] 
	(0,0) to [out=90, in=-90]  (0,3)
	to [out=0,in=180] (6,3)
	to [out = -90, in =90] (6,0)
	to [out=180, in =0]  (0,0);
	
	\draw[thick] (0,0) -- (0,3);
	\draw[thick] (6,0) -- (6,3);
	\draw[ultra thick,blue] (3,0) -- (3,1.5);
	\node at (3,1.5) [circle,fill,blue, inner sep=1.5pt]{};
	\node[below] at (0,0) {$\langle \mathcal L| $};
	\node[below] at (6,0) {$|D_{Y_k} \sixa\rangle $}; 
	\node[below] at (3,0) {$\cC_F$};
	\end{scope}

	\draw[thick, snake it, <->] (-1.7,1.5) -- (-5, 1.5);
	\node[above] at (-3.3,1.6) {Expand\,/\,Shrink};

	\begin{scope}[xshift=3.3 in]
	\draw[very thick, blue] (-7,0) -- (-7,1.5);
	\draw[very thick, green] (-7,1.5) -- (-7,3);
	\node[above] at (-7,3) {$\langle \mathcal L|D_{Y_k} \sixa\rangle$};
	\node[below] at (-7,0) {$\langle F(\mathcal L)|D_{F(Y_k)} \sixa\rangle$};
	\node at (-7,1.5) [circle,fill,blue, inner sep=1.5pt]{};
	\node[left,blue] at (-7,1.5) {$\cN_F$};
	\end{scope}
	
\end{tikzpicture}	
\end{center}
	
	\caption{The surgery defect $\cC_F$ is given a topological boundary and one obtains a twist defect. Shrinking the slab, this twist defect becomes a duality interface $\cN_F$.
	}
	\label{fig:twistsh2}
\end{figure}
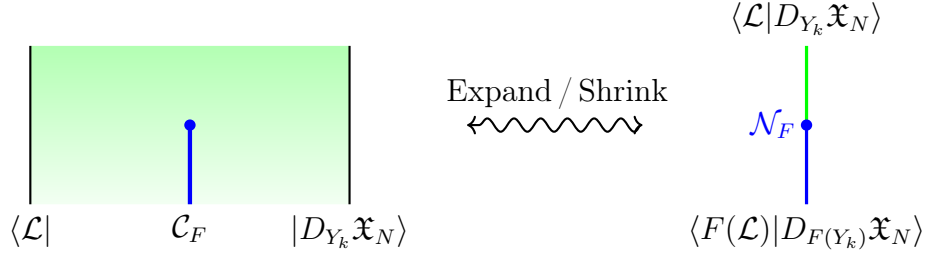
\label{sec:codensation-defect}
Obtaining a non-invertible symmetry along the lines described above requires combining the action of a mapping class with a discrete gauging. This discrete gauging operation acts only on the global structure i.e. the isotropic sublattice $L$ and not on the continuous coupling and can be encoded as right action of a matrix $F$ which preserves the intersection form. The resulting defect $\mathcal{C}_{F}$ can be implemented as a condensation defect \cite{Gaiotto:2019xmp,Roumpedakis:2022aik} following the procedure outlined in \cite{Kaidi:2022cpf,Bashmakov:2022uek}. The construction of $\mathcal{C}_{F}$ can be streamlined using the $6d$ picture. The basic idea is to construct these defect out of the operators $\Phi(\Sigma_{3})$  through the isomorphism\footnote{We assume as before that $Y_{4}$ is simply connected and hence in particular the homology is torsion free.}
\begin{align}
\label{eq:18}
  H_{3}(Y_{4} \times M_{2}, \mathbb{Z}_{N}) \cong H_{1}(M_{2} , H_{2}(Y_{4},\mathbb{Z}_{N})) ~.
\end{align}
This allows to regard $\Phi(\Sigma_{3})$ as flux through a one-cycle of $M_{2}$ with values in $H_{2}(Y_{4},\mathbb{Z}_{N})$. Hence the geometry determines automatically the value of all the allowed fluxes. These include both background fluxes and gauge fluxes and the precise description of which correspond to which is dictated by the choice of the global structure. The intersection pairing on $H_{3}(Y_{4} \times M_{2} , \mathbb{Z}_{N})$ becomes a combination of the antisymmetric pairing between the elements of $H_{1}(M_{2},\mathbb{Z}_{N})$ and the symmetric pairing between the winding and momentum modes of $2d$ compact bosons.\footnote{This pairing is the analogue of Dirac pairing for general dyonic lines in the 4d case.}

The condensation defect $\mathcal{C}_{F}$ that implements the action of mapping class $F$ is defined by,
\begin{align}
\label{eq:19}
  \Phi(\Sigma_{1}) \times \mathcal{C}_{F}(M_{2}) = \mathcal{C}_{F}(M_{2}) \times \Phi(F\Sigma_{1}) ~, && \forall \Sigma_{1} \in H_{1}(M_{2} , H_{2}(Y_{4},\mathbb{Z}_{N})) ~.
\end{align}

Using the fact that $F$ respects the intersection form of $Y_{4}$ i.e. $F^{T}QF = Q$ and assuming for simplicity that $N$ is a prime the $\mathcal C_{F}$ can be written as,
\begin{align}
\label{eq:20}
  \mathcal{C}_{F} &=  \frac{1}{\sqrt{\abs{H_2(Y_4,\mathbb{Z}_N)}}} \sum_{\Sigma_{1} \in H_{1}(M_{2},H_{2}(Y_{4},\mathbb{Z}_{N}))} \exp(\frac{2\pi i}{2N}\ev{F\Sigma_{1},\Sigma_{1}})\Phi((1-F)\Sigma_{1}) ~.
\end{align}
The proof that this is indeed the case and that $\mathcal{C}_{F}$ as written above implements the action in \eqref{eq:19} is a simple modification of the discussion in Appendix D of \cite{Bashmakov:2022uek} where it is carried out with the roles of $Y_{4}$ and $M_{2}$ reversed. It can be checked that the computation rests solely on the fact that $F$ preserves the intersection form of the manifold along which we carry out dimensional reduction.\footnote{\ In the special case of $N=2$ if $F = \left(\begin{smallmatrix} 0 & 1 \\ 1 & 0\end{smallmatrix}\right)$ i.e the case of T-duality for $\mathfrak{a}_{1}$ theory described in appendix \ref{sec:non-invertible-t} the expression for the condensate is different. The treatment for that case can be found in \cite{Kaidi:2022cpf} --- see also section \ref{sec:TY}.}

Now for such elements $F \in \MCG(Y_k)$ corresponding to dualities according to the discussion above, we can construct codimension-1 defects implementing the associated symmetry from the bulk TFT. These condensation defects can now be used to construct various defects in the boundary $d$-dimensional theories by a standard trick \cite{Kaidi:2022cpf}.  Indeed, each of the condensation defects above always admits at least one topological boundary condition and this allows us to end them in the bulk $d+1$ dimensional TFT. The resulting defects are codimension-2 and are better thought of as non-genuine defects attached to a manifold that is one-dimension higher --- see Figure \ref{fig:twistsh2}. Such non-genuine operators were referred to as ``twist defects'' in \cite{Barkeshli:2014cna,Kaidi:2022cpf}. By moving the twist defects to the boundary as shown in Figure \ref{fig:twistsh2}, one obtains an interface between the $d$-dimensional theories that we denote $\cN_F$. The fusion rule of these interfaces can be determined but the result depends on the specifics of the given geometry.

\section{The 7d view of 3d Chern Simons}
\label{sec:7d-view-3d}
As explained in the last section, the $7d$ abelian Chern-Simons theory \eqref{eq:1} when considered on a manifold $Y_{4} \times W_{3}$ and integrated along $Y_{4}$ gives $3d$ Chern-Simons theory with $K$-matrix given by the intersection form of $Y_{4}$. In this section we explain in detail the relationship of the $7d$ Chern-Simons theory and the associated Heisenberg algebra \eqref{eq:4} with various aspects of the $3d$ Chern-Simons theory, especially its topological boundary conditions and the structure of modular tensor category that its observables encode. We illustrate these connections with the help of some simple examples corresponding to $Y_4$ some del Pezzo surface.

\subsection{3d Abelian Chern-Simons theories}
\label{sec:3d-chern-simons}
The abelian Chern-Simons theories were classified in \cite{Belov:2005ze} and we briefly recall the results here\footnote{We normalize the periods of field strength using the common convention is physics literature i.e. $\frac{F}{2\pi}$ has integral periods. This convention was also used in the last section. However, this is different from \cite{Belov:2005ze}, in which the convention is to define $F$ so that it has integral periods}. An abelian Chern-Simons theory in 3d dimensions is characterized by an integral lattice $\Lambda$ of rank $r$ endowed with a symmetric bilinear form $K: \Lambda \times \Lambda \to \mathbb{Z}$. By choosing a basis of $\Lambda$ we can encode $K$ in a symmetric $r \times r$ matrix $K^{ij}$. The action of the this Chern-Simons theory on a 3-manifold is written by introducing $r$ $U(1)$ gauge fields $A_{i}$ and is given by,
\begin{align}
\label{eq:21}
  S_{3d} = \frac{1}{4\pi}\int_{W_{3}} K^{ij} A_{i} \wedge \dd A_{j} ~.
\end{align}

We are interested in how the action written above transforms under the gauge transformations. For closed $W_{3}$, one way to determine this is to choose a bounding $4$-manifold $W_{4}$ and extend the gauge fields to it. Calling these gauge field $\tilde{A}_{i}$ and the corresponding field strengths $\tilde{F}_{i}$ we can write down the four dimensional action,
\begin{align}
\label{eq:22}
  S_{4d} = \frac{1}{4\pi} \int_{W_{4}} K^{ij}\tilde{F}_{i} \wedge \tilde{F}_{j} ~.
\end{align}
This is a $4d$ gauge theory with just the topological $\theta$ terms. The Chern Simons action \eqref{eq:21} is the boundary term for $S_{4d}$, so we need to determine to what extent this action depends on the details of the particular extension we have chosen for the boundary gauge fields. Since, $\int_{W_{4}} K^{ij}\tilde{F}_{i} \wedge \tilde{F}_{j}$ is always an integer times $4\pi^{2}$, the dependence of the exponentiated action $\exp(i S_{4d})$ on the extension gives us two classes of Chern-Simons theories,
\begin{itemize}
  \item A bilinear form $K: \Lambda \times \Lambda \to  \mathbb{Z}$ is called even if $K(x,x) = 0 \mod 2$ for all $x \in \Lambda$. For even $K$, the action is independent of the extension into the bulk and such bilinear forms define ordinary Chern-Simons theories. In this work, we will only deal with these theories.
  \item If $K$ is not even it is called odd. To define Chern-Simons theory for these cases, we can choose a spin structure on $W_{3}$ and demand that $W_{4}$ has a compatible spin structure. This ensures that $\int_{W_{4}} K^{ij}\tilde{F}_{i} \wedge \tilde{F}_{j}$ is $8\pi^{2}$ times an integer and hence the extension is dependent only on the spin structure chosen. Such theories are called spin Chern-Simons theories. In modern language this makes the spin Chern-Simons theory defined by an odd $K$ one of the simpler examples of a relative theory.\footnote{\, We can be more precise about this statement: if we regard a TFT as a functor from oriented bordisms to vector spaces, the theory is relative, but if we instead regard a TFT as functor from spin bordism the theory is absolute. Whether a theory is relative or absolute depends on how much structure we are willing to put on the manifolds.}
\end{itemize}

\subsection{Discriminant group and topological boundary conditions}
\label{sec:discr-subgr-topol}
Like the $7d$ Chern-Simons discussed in Section \ref{sec:symm-tft-dimens}, the observables of abelian $3d$ Chern-Simons theories are Wilson loops. The dual lattice $\Lambda^{*}$ of a lattice $\Lambda$ is the set of linear map $\Lambda \to \mathbb{Z}$. To write down a Wilson loop we pick a closed loop $\Sigma_1$ in the spacetime $W_{3}$ and an element $X \in \Lambda^{*}$. This map has a unique extension to a map $\mathbb{R}^{r} \to \mathbb{R}$ which we also call $X$. The Wilson loop on $\Sigma_1$ with charge $X$ which we will denote $\Phi_{X}(\Sigma_1)$ is defined by,
\begin{align}
\label{eq:23}
  \Phi_{X}(\Sigma_1) = \exp(i \oint_{\Sigma_1}X(A_{i})) ~.
\end{align}
The Wilson loops can be pushed to the boundary $M_2$ of $W_{3}$ and their equal time correlation functions take the form of a Heisenberg algebra similar to \eqref{eq:4}. This setup and the topological boundary conditions for Abelian Chern-Simons theories were studied in \cite{Kapustin:2010hk} and we now review the results of interest to the current work. The partition vector the TFT assigns to boundary furnishes a representation of this non-commutative algebra.

We start by describing the discriminant group $\mathrm{D}$ of a lattice $\Lambda$. By fixing one element we can regard $K(x,\cdot)$ as a map $\Lambda \to \mathbb{Z}$, i.e for a fixed $x$, $K(x,\cdot)$ is an element of the dual lattice. Hence $K$ itself can be regarded as a map $K: \Lambda \to \Lambda^{*}$, and its image is generally a subset of $\Lambda^{*}$. The discriminant group $\mathrm{D}$ is the quotient $\Lambda^{*} / \,\mbox{Im}(K)$. It is a finite Abelian group with the number of elements equal to $\abs{\det(K)}$.

As an example let us consider the simplest case: that of a $U(1)$ Chern-Simons theory at level $N$. The lattice $\Lambda$ in this case is the group of integers $\mathbb{Z}$ with the bilinear form given by $K(x,y) = Nxy$. The dual lattice $\Lambda^{*}$ is also isomorphic to $\mathbb{Z}$ with $n$ representing the map $x \mapsto n x$. The $K$ then corresponds to multiplication of these maps by $N$ and hence the discriminant group is $\mathbb{Z} / N\mathbb{Z}$ i.e. $\mathbb{Z}_{N}$. The Wilson loops for the $7d$ Chern-Simons theory \eqref{eq:2} correspond precisely to the those with charges in discriminant group.

Restricting to the discriminant subgroup greatly simplifies the study of the Wilson loops since their is now only a finite number of them. It is also enough to classify the topological boundary conditions. Indeed one of the main results of \cite{Kapustin:2010hk} is that the boundary conditions of a 3d Chern-Simons theory correspond to the Lagrangian subgroups of $\mathrm{D}$. A lagrangian subgroup $L$ of $\mathrm{D}$ is a subgroup which is isotropic with respect to the intersection form $K$ and has precisely $\sqrt{\abs{\mathrm{D}}}$ elements.

\subsection{Categorical structure of Chern-Simons}
\label{sec:categ-struct-chern}
Three dimensional quantum field theories correspond to algebraic structures called modular tensor categories. Abelian CS theories are prototypical examples of 3d TFT and the corresponding modular tensor categories are well known (see e.g. \cite{Stirling:2008bq, Lee:2018eqa, Kapustin:2010hk}). The data needed to define the modular tensor category $\mathcal{C}$ are,
\begin{enumerate}
  \item\label{item:5} A group of anyons which for abelian Chern-Simons theories is the discriminant group $\mathrm{D}$.
  \item\label{item:6} For each anyon $a$, its topological spin $\theta(a)$. For the CS theories described above, it is,
        \begin{align}
          \label{eq:24}
          \theta(a) = \exp(2\pi i \frac{1}{2}K^{-1}(a,a)) ~,
        \end{align}
        This expressions is well defined on $\mathrm{D}$ if $K$ is even i.e. when the CS theory is an ordinary CS theory and not a spin CS theory.
  \item\label{item:7} The braiding phase for each pair of anyons $a$ and $b$. For CS theories it is given by,
        \begin{align}
          \label{eq:25}
          B(a,b) = \exp(2\pi i K^{-1}(a,b)) ~.
        \end{align}
\end{enumerate}
The spin and braiding phases of anyons are required to satisfy several consistency conditions which we do not describe here but which can be found in the references cited above. In terms of these data we can reconstruct the F-symbols and the hexagon relations which specify the corresponding symmetry category, as discussed in detail in \cite{Lee:2018eqa} to which we refer our readers. Therefore the information which we recover from geometry completely determines the theories we are after provided $N$ is even or $Q$ has even diagonal elements. Whenever that is not the case more care is needed: the resulting theories are spin Chern Simons and the construction crucially depends on the choice of a trivialization for the Wu structure on the 4-manifold of interest. We discuss some details about this in section \ref{sec:spin-chern-simons} for the case of a 6d theory of rank one corresponding to a free chiral self-dual tensor in 6d. % Our primary interest is the symmetries of the $2d$ theory which live on the boundary of abelian CS theories. This categorification of abelian CS theories allows us to construct the symmetry categories for the boundary theory: it is the Drinfeld center of $\mathcal{C}$.

Let us discuss in slightly more detail on the aforementioned symmetries of Chern-Simons theories. The description in terms of $K-$matrix is not unique: the transformations of the form
\begin{equation}
   K\,\rightarrow\,V\,K\,V^T,\qquad   a\,\rightarrow\,V a,
\end{equation}
with $V\in GL(N,\,\mathbb{Z})$ is an invertible matrix, does not change the braiding or fusion structure. We will be especially intereste in those transformations, also called \textit{automorphisms}, which leave the $K-$matrix invariant:
\begin{equation}
    K\,=\,U\,K\,U^T.
\end{equation}
These transformations correspond to \textit{anyonic symmetry transformations}: they permute lines with the same spins and leave invariant the fusion rules \cite{Khan:2014waa, Teo:2015xla}. We will denote the group of automorphisms by $\text{Aut}(K)$. Not all the elements in $\text{Aut}(K)$ correspond to a non-trivial transformation on anyons, though. Indeed, the transformations of the form
\begin{equation}
    a\,\rightarrow\,a\,+\,K\,b,
\end{equation}
which leave an anyon invariant up to a local (with integer spin and trivial braiding) particle. Transformations of this kind are referred to as \textit{inner automorphisms} and form a normal subgroup of the automorphism group, $\text{Inner}(K)\,\unlhd\,\text{Aut}(K)$. Our ultimate interest is in the quotient of the automorphism group with respect to internal automorphisms, which is known as the \textit{outer automorphism} group:
\begin{equation}
    \text{Outer}(K)\,=\,\frac{\text{Aut}(K)}{\text{Inner}(K)}.
\end{equation}
\subsection{Topological boundary conditions from $6d$}
\label{sec:topol-bound-cond-1}
As explained in Section \ref{sec:two-dimens-theor}, the $2d$ theory obtained by reducing $6d\, (2,0)$ theory of type $\mathfrak{a}_{N}$ on a compact simply connected four manifold $Y_{4}$ has as its symmetry TFT an abelian CS theory with lattice $\Lambda = H_{2}(Y_{4},\mathbb{Z})$ and the K-matrix,
\begin{align}
\label{eq:26}
  K = NQ ~.
\end{align}
Where $Q$ is the intersection form on $Y_{4}$. The discriminant subgroup is then homology with coefficients in $\mathbb{Z}_{N}$ i.e $H_{2}(Y_{4},\mathbb{Z}_{N})$. For square-free $N$ for which there are no topological boundary conditions in $7d$, this discriminant group immediately allows us to match the topological boundary conditions we expected from $6d$ with the result for $3d$ Chern-Simons theory obtained above.

For this let us recall from the general discussion in Section \ref{sec:symm-tft-dimens}, that the global structures for the $6d\, (2,0)$ theory of type $\mathfrak{a}_{N-1}$ on $Y_{6-d} \times Z_{d}$ dimensionally reduced along $Y_{6-d}$ correspond to precisely those maximal isotropic sublattices $\mathcal{L}$ of $H_{3}(Y_{6-d} \times Z_{d},\mathbb{Z}_{N})$ which for all $Z_{d}$ are invariant under the mapping class group of $d$. By Kunneth formula,
\begin{align}
\label{eq:27}
  H_{3}(Y_{4} \times M_2 , \mathbb{Z}_{N}) \cong H_{2}(Y_{4},\mathbb{Z}_{N}) \otimes H_{1}(M_2,\mathbb{Z}_{N}) ~.
\end{align}
Since no non-trivial subgroup of $H_{1}(M_2,\mathbb{Z}_{N})$ is invariant under the mapping class group of a two dimensional manifold $M_2$, the maximal isotropic sublattice must take the form,
\begin{align}
\label{eq:28}
  \mathcal{L} = L \otimes H_{1}(M_2,\mathbb{Z}_{N}) ~.
\end{align}
The fact that $\mathcal{L}$ is an maximal isotropic sublattice then means that $L$ must be a Lagrangian subgroup of $H_{2}(Y_{4},\mathbb{Z}_{N})$ i.e. the discriminant group for the symmetry TFT. If the intersection form $Q$ of $Y_{4}$ is even, this arguments works for all square free $N$, however for odd $Q$ we must restrict to even square free $N$ to obtain a CS theory without dependence on spin structure.

\subsection{The case of $S^2 \times S^2$ and $TY(\mathbb Z_N)$}\label{sec:TY}

In this section we discuss the details of the dictionary in one explicit example, namely $Y_4 = S^2 \times S^2$. Let us begin by considering the case $N=2$ which is slightly different from the general one. This example  will also provide an illustration of the condensate formalism. The action of $D_{S^2 \times S^2} \mathfrak{F}_2$ is
\begin{equation}
    \frac{2}{2\pi}\int\,A_1\wedge dA_2,
\end{equation}
and so the corresponding $K-$matrix is given by two times the intersection pairing on $S^2 \times S^2$:
\begin{equation}
    K\,= 2 Q = \,\left(
    \begin{matrix}
       0 && 2\\
       2 && 0
    \end{matrix}
    \right).
\end{equation}
A simple inspection reveals that $\text{Outer}(K)\,\simeq\,\mathbb{Z}_2$, and is generated by the matrix
\begin{equation}
   F\,=\, \left(
    \begin{matrix}
        0 && 1\\
        1 && 0
    \end{matrix}
    \right).
\end{equation}
From the 6d perspective $F$ arises as the element of $\MCG(S^2 \times S^2)$ which swaps the two 2-spheres in the four-manifold. $H_2(S^2 \times S^2,\mathbb Z_2)$ has two generators $C_1$ and $C_2$, and by considering $\Sigma_3^i = \gamma \cup C_i$ as supports for the 7d Wilson surfaces we obtain the following basic lines in this model:
\begin{gather}
    \Phi(C_1 \cup \gamma) \to \mathit{e}\, = \,e^{i\oint_{\gamma} A_1},\qquad \Phi(C_2 \cup \gamma) \to  \mathit{m}\,=\,e^{i\oint_{\gamma} A_2} \nonumber \\
    \qquad \Phi((C_1 + C_2)\cup \gamma) \to \psi\,=\,e^{i\oint A_1+i\oint A_2},
\end{gather}
These satisfy the obvious Abelian fusion rules which can be derived straighforwardly from the Heisenberg algebra:
\begin{gather}
    \mathit{e}\,\times\,\mathit{e}\,=\,\mathit{m}\,\times\,\mathit{m}\,=\,\psi\,\times\,\psi\,=\,1,\nonumber\\
    \mathit{e}\,\times\,\psi\,=\,\mathit{m},\qquad \mathit{m}\,\times\,\psi\,=\,\mathit{e},\\
    \mathit{e}\,\times\,\mathit{m}\,=\,\psi.\nonumber
\end{gather}
The lines $\mathit{e}$ and $\mathit{m}$ are sometimes called the electric line and the magnetic line, respectively, and the anyonic $\mathbb{Z}_2$ symmetry corresponding to $F$ exchanges them, $\mathit{e}\,\leftrightarrow\,\mathit{m}$, therefore being dubbed electric-magnetic symmetry. The fermion line $\psi$ is invariant under the action of electric-magnetic $\mathbb{Z}_2$.

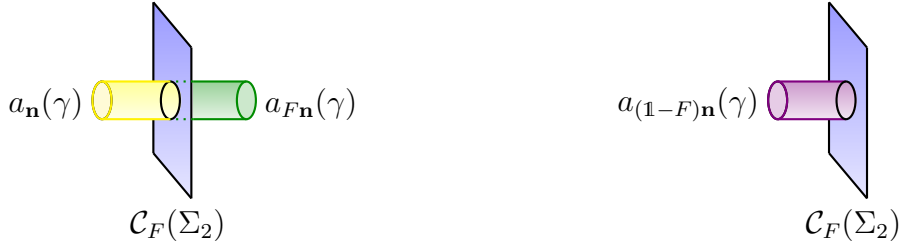
\begin{figure}[t]
    \centering

\begin{tikzpicture}

\begin{scope}[xshift=0 in, yshift=-1.2in]

   \shade[top color=blue!40, bottom color=blue!10,xshift=-0.6in, rotate=90]  (0,-0.7) -- (2,-0.7) -- (2.6,-0.2) -- (0.6,-0.2)-- (0,-0.7);

\draw[thick,xshift=-0.6in,rotate=90] (0,-0.7) -- (2,-0.7);
\draw[thick,xshift=-0.6in,rotate=90] (0,-0.7) -- (0.6,-0.2);
\draw[thick,xshift=-0.6in,rotate=90]  (0.6,-0.2)--(2.6,-0.2);
\draw[thick,xshift=-0.6in,rotate=90]  (2.6,-0.2)-- (2,-0.7);

 \shade[top color=yellow!40, bottom color=yellow!10]  (-2,1.55) -- (-1.1,1.55) 
 to[out=-10,in=10] (-1.1,1.05)  -- (-2,1.05)-- (-2,1.55);
  \shade[top color=dgreen!40, bottom color=dgreen!10] (-0.1,1.05) -- (-0.8,1.05)-- (-0.8,1.55) -- (-0.1,1.55)-- (-0.1,1.05) ;
    \draw[thick] (-1.1,1.3) ellipse (0.125cm and 0.25cm);
     \draw[thick,fill=yellow!20] (-2,1.3) ellipse (0.125cm and 0.25cm);
         \draw[thick,dgreen,fill=dgreen!20] (-0.1,1.3) ellipse (0.125cm and 0.25cm);

\node[below] at (-1, 0){$\cC_{F}(\Sigma_2)$};

 \draw[thick,yellow] (-2,1.3) ellipse (0.125cm and 0.25cm);
 \draw[thick,yellow] (-2,1.55)--(-1.1,1.55);
  \draw[thick,yellow] (-2,1.05)--(-1.1,1.05);

     \draw[thick,dgreen] (-0.1,1.55)--(-0.8,1.55);
  \draw[thick,dgreen] (-0.1,1.05)--(-0.8,1.05);
  \draw[thick,dotted,dgreen] (-0.8,1.55)--(-1.1,1.55);
  \draw[thick,dotted,dgreen] (-0.8,1.05)--(-1.1,1.05);
    
    \node[left] at (-2.1,1.25) {$a_{\mathbf{n}}(\gamma)$};
    
     \node[right] at (0,1.25) {$a_{F\mathbf{n}}(\gamma)$};

\end{scope}

\begin{scope}[xshift=3.5 in, yshift=-1.2in]
  
   \shade[top color=blue!40, bottom color=blue!10,xshift=-0.6in, rotate=90]  (0,-0.7) -- (2,-0.7) -- (2.6,-0.2) -- (0.6,-0.2)-- (0,-0.7);

\draw[thick,xshift=-0.6in,rotate=90] (0,-0.7) -- (2,-0.7);
\draw[thick,xshift=-0.6in,rotate=90] (0,-0.7) -- (0.6,-0.2);
\draw[thick,xshift=-0.6in,rotate=90]  (0.6,-0.2)--(2.6,-0.2);
\draw[thick,xshift=-0.6in,rotate=90]  (2.6,-0.2)-- (2,-0.7);

 \shade[top color=violet!40, bottom color=violet!10]  (-2,1.55) -- (-1.1,1.55) 
 to[out=-10,in=10] (-1.1,1.05)  -- (-2,1.05)-- (-2,1.55);

     \draw[thick,fill=violet!20] (-2,1.3) ellipse (0.125cm and 0.25cm);

\node[below] at (-1, 0){$\cC_{F}(\Sigma_2)$};

 \draw[thick,violet] (-2,1.3) ellipse (0.125cm and 0.25cm);
 \draw[thick,violet] (-2,1.55)--(-1.1,1.55);
  \draw[thick,violet] (-2,1.05)--(-1.1,1.05);
  
   \draw[thick] (-1.1,1.3) ellipse (0.125cm and 0.25cm);
   
       \node[left] at (-2.1,1.25) {$a_{(\mathds{1}-F)\mathbf{n}}(\gamma)$};

\end{scope}

\end{tikzpicture}

    \caption{The automorphism transformation operators $\cC_F(\Sigma_2)$ change the anyon lables in a given way. The folding trick implies that this operators should be able to absorb lines of the form $a_{(\mathds{1}-F)\mathbf{n}}(\gamma)$, and therefore are condensates of them.}
    \label{fig:folding}
\end{figure}

Now, there must exist a topological co-dimension one operator, realizing the electric-magnetic $\mathbb{Z}_2$ action, and, following \cite{Kaidi:2022cpf, Bashmakov:2022uek}, we can represent this operator as a condensate of lines. The idea goes as follows. We can drag an anyon $a_{\mathbf{n}}$ from the left to the right through a surface defect, and along the way it gets transformed into the line $a_{F\mathbf{n}}$. We now exploit the folding trick and reflect the trajectory of the line past the defect, bringing it to the left, but with the form of the charge conjugate line $a_{-F\mathbf{n}}$ instead. Now we do not have anything on the right, and so the line $a_{(\mathds{1}-F)\mathbf{n}}$ gets absorbed (see fig. \ref{fig:folding}).

In the case at hands $\mathbf{n}_e\,=\,\begin{psmallmatrix} 1 \\ 0 \end{psmallmatrix}$, $F\,=\,\begin{psmallmatrix} 0 && 1\\ 1 && 0 \end{psmallmatrix}$, and so the absorbed line has the label $(\mathds{1}-F)\mathbf{n}_e\,=\,\begin{psmallmatrix} 1 \\ -1 \end{psmallmatrix}\,=\,\begin{psmallmatrix} 1 \\ 1 \end{psmallmatrix}\,\text{mod}\,2$, and is recognized as the $\psi$ line. The condensate supported on a subamnifold $\Sigma_2$ is then,
\begin{equation}
    \cC_F(\Sigma_2)\,=\,\frac{1}{|H_0(\Sigma_2, \mathbb{Z}_2)|}\,\sum_{\gamma\in H_1(\Sigma_2,\mathbb{Z}_2)}\,\psi(\gamma).
\end{equation}

The normalization is chosen to provide the following fusion rule for condensates:
\begin{equation}
    \cC_F\,\times\,\cC_F\,=\,\frac{|H^1(\Sigma_2,\mathbb{Z}_2)|}{|H^0(\Sigma_2,\mathbb{Z}_2)|^2}\,=\,\chi(\Sigma_2, \mathbb{Z}_2)^{-1}.
\end{equation}
We recognise here the 2d version of the condensate $\mathcal C_F$ we have constructed in section \ref{sec:codensation-defect} above which we obtain from the $\mathbb Z_2$ action that swaps the two spheres in geometry. It is interesting to compare this definition with the one in equation \eqref{eq:20}. This is the geometrical origin of the EM duality in this model.

Having constructed the transformation operator on manifolds without boundary, we can consider a situation when the operator ends on a curve $\gamma$. This is achieved by imposing Dirichlet boundary conditions on the boundary: in our case $(A_1+A_2)\vert_{\gamma}=0$. This is a non-genuine co-dimension 2 defect, which we will denote $\sigma_0(\gamma)$; it is non-genuine, because it comes with a surface attached. More explicitly, we can define the defect line in the following way:

\begin{equation}
    \sigma_0(\gamma)\,=\,\frac{1}{|H^0(\Sigma_2, \gamma, \mathbb{Z}_2)|}\,\sum_{\gamma\in H_1(\Sigma_2,\mathbb{Z}_2)}\,\psi(\gamma),
\end{equation}
where the normalization is introduced, following \cite{Kaidi:2022cpf}.
We can also locate an electric line $\mathit(e)$ along the boundary $\gamma$: this gives rise to a new defect $\sigma_1(\gamma)$. Can we get to more defects by putting along the boundary the other to lines? $\psi$ trivialises along the boundary due to the boundary conditions. For the magnetic line we have $\int_{\gamma}\,A_2\,=\,\int_{\gamma}\,(A_1\,+\,2A_2)\,=\,\int_{\gamma}\,A_1$. In the first equality we used that $(A_1+A_2)\vert_{\gamma}\,=\,0$, and the second equality is valid mod $2$. Thus, the magnetic line is equivalent to the electric line on the defect boundary, and so again leads to $\sigma_1$. These statements can be summarized by the following fusion rules:

\begin{gather}                              \       \mathit{e}\,\times\,\sigma_{\lambda}\,=\,\mathit{m}\,\times\,\sigma_{\lambda}\,=\sigma_{\lambda+1},\nonumber\\
     \psi\,\times\,\sigma_{\lambda}\,=\,\sigma_{\lambda},
\end{gather}
where $\lambda$ is a mod $2$ index. With slightly more work (see \cite{Kaidi:2022cpf} for the details), one can also obtain the defect fusion rules:
\begin{gather}
    \sigma_0\,\times\,\sigma_0\,=\,\sigma_1\,\times\,\sigma_1\,=\,\chi(M_2, \mathbb{Z}_N)^{-1}\,(1\,+\,\psi),\nonumber\\
    \sigma_0\,\times\,\sigma_1\,=\,\chi(M_2, \mathbb{Z}_N)^{-1}\,(\mathit{e}\,+\mathit{m}).
\end{gather}
Here the overall coefficient, depending on the two-manifold $M_2$, attached to the defect, can be removed by an appropriate counterterm. We are not specifying the curves for line operators in the equations above, assuming that they all have the same support. The fusions of the defects with the same labels gives a higher gauging of $\psi$ line along a curve. The fusion of two diffects with different labels can be obtain by fusing the condensate with the $\mathit{e}$ line (or equivalently with the $\textit{m}$ line).

This discussion generalizes straightforwardly to the $\mathbb{Z}_N$ theories with arbitrary $N > 2$. The action for $D_{S^2 \times S^2} \topa$ that arises from dimensional reduction is

\begin{equation}
    \frac{N}{2\pi}\int\,A_1\wedge dA_2\,.
\end{equation}
Again we see this system has an anyonic symmetry given by $\mathbb{Z}_2$ that exchanges $A_1$ with $A_2$. Sometimes we can also refer to it as the electric-magnetic symmetry. This symmetry is clearly geometrized by swapping the two $S^2$'s. For this example we have $N$ electric, $N$ magnetic and in total $N^2$ dyonic lines, that have a 6d origin from operators $\Phi(\Sigma_3)$ of the following forms:
\begin{equation}
    \label{dyonic_lines}
    \Phi((n C_1 + m C_2)\cup \gamma) \to \mathit{\ell}_{n,m}\,=\,e^{(i n\oint_{\gamma}A_1 + i m \oint_{\gamma} A_2 )}.
\end{equation}
In analogy with the preceding discussion, we consider the surface defects, implementing the electric-magnetic symmetry. The folding trick implies that the line $\mathit{\ell}_{1, -1}$ must be absorbed by this defect (see \cite{Kaidi:2022cpf} for a thorough discussion), which is therefore given by the condensate
\begin{equation}
\label{condensate}
    \cC_F(\Sigma_2)\,=\,\frac{1}{|H_0(\Sigma_2, \mathbb{Z}_N)|}\,\sum_{\gamma\in H_1(\Sigma_2,\mathbb{Z}_H)}\,\mathit{\ell}_{1, -1}(\gamma).
\end{equation}
This is indeed a surgery defect arising from condensation of the type we discussed in section \ref{sec:codensation-defect}. Using the dictionary between the Wilson surfaces of the 7d theory and the 3d Wilson lines in equation \eqref{dyonic_lines}, it can be checked that this indeed reproduces the expression in \eqref{eq:20}. For this we note that in this case the element of $\MCG(S^2 \times S^2)$ that we are representing is 
\be
F = \left(\begin{matrix} 0 & 1 \\ 1 & 0 \end{matrix}\right)\ee
 that acts on homology by exchanging $C_1$ and $C_2$. The image of $1-F$ is generated by $C_1 - C_2$ and hence the lines that participate in the condensate are $\Phi((C_1 - C_2)\cup \gamma)$ with $\gamma \in H_1(\Sigma_2 , \mathbb{Z}_N)$ i.e. $l_{1,-1}(\gamma)$. Since all the phases in \eqref{eq:20} vanish for this example, that expression reduces to \eqref{condensate}. One can check that the fusion of these surface defects is still governed by $\mathbb{Z}_2$ group law.

\begin{figure}
\begin{center}
\begin{tikzpicture}[scale=0.7]

	\begin{scope}[xshift=-5.2 in]
	\shade[line width=2pt, top color=green!30, bottom color=green!5] 
	(0,0) to [out=90, in=-90]  (0,3)
	to [out=0,in=180] (6,3)
	to [out = -90, in =90] (6,0)
	to [out=180, in =0]  (0,0);
	
	\draw[thick] (0,0) -- (0,3);
	\draw[thick] (6,0) -- (6,3);
	\draw[ultra thick,blue] (3,0) -- (3,1.5);
	\node at (3,1.5) [circle,fill,blue, inner sep=1.5pt]{};
	\node[below] at (0,0) {$\langle \mathcal L| $};
	\node[below] at (6,0) {$|D_{S^2\times S^2} \sixa\rangle $}; 
	\node[below] at (3,0) {$\cC_F$};
	\end{scope}

	\draw[thick, snake it, <->] (-1.7,1.5) -- (-5, 1.5);
	\node[above] at (-3.3,1.6) {Expand\,/\,Shrink};

	\begin{scope}[xshift=3.3 in]
	\draw[very thick, blue] (-7,0) -- (-7,1.5);
	\draw[very thick, green] (-7,1.5) -- (-7,3);
	\node[above] at (-7,3) {$\langle \mathcal L|D_{S^2\times S^2} \sixa\rangle$};
	\node[below] at (-7,0) {$\langle F(\mathcal L)|D_{F(S^2\times S^2)} \sixa\rangle$};
	\node at (-7,1.5) [circle,fill,blue, inner sep=1.5pt]{};
	\node[left,blue] at (-7,1.5) {$\cN_F$};
	\end{scope}
	
\end{tikzpicture}	
\end{center}
	
	\caption{The surgery defect $\cC_F$ is given a topological boundary and one obtains a twist defect. Shrinking the slab, this twist defect becomes a duality interface $\cN_F$.
	}
	\label{fig:twistsh}
\end{figure}
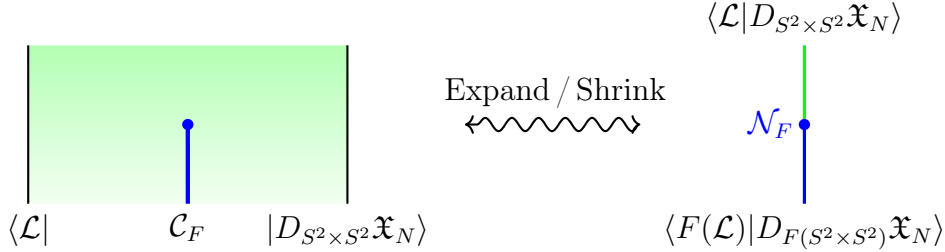

As before, we can put the defect $\cC_F$ on a two-manifold with boundary, imposing Dirichlet boundary conditions (see figure \ref{fig:twistsh}): this will give rise to a non-genuine defect operator, which we dub $\cN_{F}$. Following the derivation in  \cite{Kaidi:2022cpf} we see that also in this case the defect in \eqref{eq:20} gives rise to the EM duality defect, and that from such a condensate insertion with an appropriate twist, we can obtain the non-invertible generator for the non-invertible Tambara-Yamagami symmetry $TY(\mathbb Z_N)$.

%We can fuse this sigma with electric lines (the fusion with magnetic lines has equivalent effect due to the boundary conditions), and get an $N-$tuple of defects:
%\begin{gather}
  %  \mathit{e}\,\times\,\sigma_{\lambda}\,=\,\mathit{m}\,\times\,\sigma_{\lambda}\,=\,\sigma_{\lambda+1},\\
    %\ell_{n,m}\,\times\,\sigma_{\lambda}\,=\,\sigma_{\lambda+n+m},
%\end{gather}
%with $\lambda$ being a mod $N$ index.

%The basic defect fusion rule is
%\begin{equation}
  %  \sigma_0\,\times\,\sigma_0\,=\,\chi(M_2, \mathbb{Z}_N)^{-1}\,\sum_{H_1(M_2,\mathbb{Z}_N)}\,\ell_{(1,-1)}(\gamma),
%\end{equation}
%from which one can obtain
%\begin{equation}
  %  \sigma_{\lambda}\,\times\,\sigma_{\lambda'}\,=\,\chi(M_2, \mathbb{Z}_N)^{-1}\,\sum_{n=0}^{N-1}\,\ell_{(n+\lambda+\lambda', -n)}.%
%\end{equation}

\subsection{Algebra and geometry of topological boundary conditions}
\label{sec:some-algebr-geom}
Finding a Lagrangian subgroup of a finite abelian group is a purely algebraic problem. The same problem for an antisymmetric bilinear form determines the global structures of $4d$ theories of class $\mathcal{S}$. In \cite{Bashmakov:2022uek,Gukov:2020btk} it was realized that for class $\mathcal{S}$ theories, this algebraic problem can be geometrized in an elegant manner: the boundary conditions are engineered by a null bordism of Riemann surface defining the class $\mathcal{S}$ theory. In this section we explore whether such a geometrization is possible for $2d$ theories too. This turns out not to be the case: the bordism of $Y_{4}$ when they exist determine a boundary condition for the $2d$ theory. However such geometric boundary conditions are rather special and (depending on $N$) boundary conditions exist even for those manifolds which are not null bordant. Before expanding on this we recall some basic facts about $4$ manifolds and their intersection forms.

For smooth simply connected manifolds the intersections forms are rather special.\footnote{see e.g. \cite{Stern2001} for an elementary exposition.} They are unimodular i.e. the determinant of $Q$ is $\pm 1$. For topological four manifolds, Freedman showed that every symmetric unimodular bilinear form is realized as the intersection form of a simply connected four manifold. However situation is much more complicated for smooth four manifolds and we illustrate the complexity with the help of some families of four manifold.

\paragraph{$\#^{n} \mathbb{P}^{2}$:}
We start with the example of connected sum of $n$ copies of complex projective plane $\mathbb{P}^{2}$. The second homology has rank $n$ and the intersection form is the $n \times n$ identity matrix. This intersection form is also positive definite i.e. $Q(x,x) > 0$ for all 2 cycles $x$. We can also consider the orientation reversal of these manifolds, which gives a negative definite intersection form i.e $Q(x,x) < 0$. Identity and minus identity are in fact the only definite intersection forms possible. Other definite intersection forms such as $E_{8}$ Cartan matrix matrix are not realized as the intersection form of any smooth manifold.

$\mathbb{P}^{2}$ is not null bordant and in fact generates the oriented bordism group in four dimensions i.e. every $4$ manifold is bordant to one of $\#^{n}\mathbb{P}^{2}$. However as we will see Lagrangian subgroups of $H_{2}(\#^{n}\mathbb{P}^{2} , \mathbb{Z}_{N})$ exist for some values of $N$.

\paragraph{$\#^{n} (S^{2} \times S^{2})$:}
Since intersection form is symmetric its eigenvalues are real. An important invariant of a symmetric bilinear form is the signature defined by $b_{+} - b_{-}$, where $b_{\pm}$ are the number of positive and negative eigenvalues. A four manifold is null bordant iff the signature of its intersection form is $0$. The simplest example of such a manifold is $S^{2} \times S^{2}$ which was considered in Section \ref{sec:two-dimens-theor}. It is also the simplest manifold with an even intersection form which is,
\begin{align}
\label{eq:29}
  H =
  \begin{pmatrix}
    0 & 1 \\
    1 & 0
  \end{pmatrix} ~.
\end{align}
A null bordism of $S^{2} \times S^{2}$ can be constructed by filling one of the two spheres into a ball. Thus we obtain two null bordisms which correspond to the two boundary conditions (which exist for all $N$) described in Section \ref{sec:two-dimens-theor}. For $N=2$ however there is an extra global structure which is not engineered by a bordism.

The connected sum of $n$ copies of $S^{2} \times S^{2}$ have as intersection form $n$ copies of $H$. $2^{n}$ bordisms for it can be constructed by filling a sphere to get a ball into each of $n$ connected components and these correspond to topological boundary conditions which are natural generalization of ones considered for $S^{2} \times S^{2}$. However more topological boundary conditions exist and it would be interesting to figure out whether the extra boundary conditions also correspond to bordisms. We can ask this question in a slightly general setting: let us suppose that there is a Lagrangian subgroup $L$ of $H_{2}(Y_{4},\mathbb{Z}_{N})$ along with a Lagrangian subgroup $\tilde{L}$ of $H_{2}(Y_{4},\mathbb{Z})$\footnote{By Lagrangian subgroup in this context we mean a free abelian of half the rank which is isotropic w.r.t. the intersection form.} such that $L$ is the mod $N$ reduction of $\tilde{L}$. In this case we can hope to engineer a bordism geometrizing the boundary condition specified by $L$ by filling in the collection of two cycles in $\tilde{L}$. This procedure is in direct analogy with how handlebodies are constructed, but there might be obstructions to filling in the all the cycles in $\tilde{L}$ smoothly. The question is whether such obstructions exist and if they do in what circumstances?

\paragraph{$K3$ surfaces:}
Topologically all compact Calabi-Yau two fold are K3 surfaces. Their intersection form contain 2 copies of the Cartan matrix of the $E_{8}$ Lie group and 3 copies of $H$. The two copies of $E_{8}$ contribute $8$ each to the signature. Since $H$ has signature $0$, the signature of $K3$ surfaces is $16$. $16$ is also the minimum positive value of the signature for a spin $4$ manifold, and indeed K3 is the generator of the spin bordism group in four dimensions, i.e. any spin manifold is bordant to the connected sum of some number of K3s.

The lattice $H_{2}(Y_{4},\mathbb{Z})$ with the quadratic form induced by the intersection pairing on $Y_4$ admits Lagrangian sublattices only when the signature of $Y_4$ is $0$. For this reason the topological boundary conditions for the $6d\, (2,0)$ theory of type $\mathfrak{a}_{N-1}$ are not simply classified by a mod $N$ reduction of the Lagrangian subgroups of $H_{2}(K3,\mathbb{Z})$. However, as explained in the following section, there are some Lagrangian sublattices when $N$ is a prime or when $N$ is a product of two primes, and it seems natural to conjecture that they exist for all $N$.\footnote{\ Since an M5 brane wrapping a K3 is dual to a Heterotic string, it is natural to expect to be able to reproduce multiple heterotic strings from multiple M5 branes. However, the theory on a stack of $N$ M5s is, decoupling the center of mass, a 6d (2,0) $A_{N-1}$ SCFT. The choices that we see realized here then might be interesting for studying dualities between M-theory and heterotic strings. It would be interesting to study this in greater detail in the future.}

\subsubsection{Finding Lagrangian subgroups}
\label{sec:find-lagr-subgr}
We now describe when topological boundary conditions exist by exploiting and extending the treatment of Section 4.2 of \cite{GarciaEtxebarria:2019caf}.\footnote{\ We thank Iñaki García Etxebarria for pointing us to \cite{GarciaEtxebarria:2019caf}. The treatment in this section improved considerably thanks to this. The background material for this section can be found in e.g. \cite{ConwayS88,Morrison-Miranda}, here we quickly summarize the results we need.} The salient point is that many bilinear forms which are different over integers become equivalent over $\mathbb{Z}_{N}$. We will restrict ourselves to bilinear forms which are of relevance to four manifolds i.e. we assume that the bilinear form $Q$ is the $\mod N$ reduction of an integral unimodular form.

To start with let us consider a Lagrangian subgroup $L$ of $H_{2}(Y_{4},\mathbb{Z}_{N})$ with $N=N_{1}N_{2}$ with $N_{1}$ and $N_{2}$ coprime. All elements of $L$ whose order is coprime to $N_{2}$ form a subgroup of $L$ which we denote $L_{1}$. Similarly all elements with order coprime to $N_{1}$ form a subgroup of $L$ which we denote $L_{2}$. Any element $g \in L_{1}$ can be written as $N_{2}g_{1}$ and it can be shown that all such $g_{1}$ form a Lagrangian subgroup $\widetilde{L}_{1}$of $H_{2}(Y_{4},\mathbb{Z}_{N_{1}})$. Similarly any element of $g \in L_{2}$ can be written as $N_{1}g_{2}$ and such $g_{2}$ form a Lagrangian subgroup $\widetilde{L}_{2}$ of $H_{2}(Y_{4},\mathbb{Z}_{N_{2}})$. Hence $L$ can be decomposed into a Lagrangian subgroup of $\widetilde{L}_{1}$ of $H_{2}(Y_{4},\mathbb{Z}_{N_{1}})$ and a Lagrangian subgroup $\widetilde{L}_{2}$ of $H_{2}(Y_{4},\mathbb{Z}_{N_{2}})$. The converse is also true, given Lagrangian subgroups $\widetilde{L}_{1}$ and $\widetilde{L}_{2}$ we can assemble them into a Lagrangian subgroup $L$ of $H_{2}(Y_{4},\mathbb{Z}_{N})$. Applying this argument repeatedly to $N = p_{1}^{k_{1}} \dots p_{i}^{k_{i}}$, we obtain that Lagrangian subgroups of $H_{2}(Y_{4},\mathbb{Z}_{N})$ exist if and only if Lagrangian subgroups exist for each of $H_{2}(Y_{4},\mathbb{Z}_{p^{k_{i}}_{i}})$. Hence we can restrict ourselves to finding Lagrangian subgroups for $H_{2}(Y_{4},\mathbb{Z}_{p^{k}})$ with $p$ a prime.

Let us start with the case of $N=p^{k}$ with $p$ an odd prime. In this case there are at most two inequivalent classes of bilinear forms for any given dimension $n$ (which in our case is $b_2(Y_4)$). When $n=2m+1$ is odd, the two classes are represented by,
\begin{align}
\label{eq:58}
  Q^{(n)}_{+} = \mathbbm{1} \oplus \underbrace{H \oplus \cdots \oplus H}_{m \text{ times}}  ~, && Q^{(n)}_{-} =  (-\mathbbm{1}) \oplus \underbrace{H \oplus \cdots \oplus H}_{m \text{ times}}   ~,
\end{align}
here $H$ is as in \eqref{eq:29} and $\mathbbm{1}$ represents the $1 \times 1$ identity matrix.

When $n = 2m$ is even, the two classes are,
\begin{align}
\label{eq:59}
  Q_{+}^{(n)} =\underbrace{H \oplus \cdots \oplus H}_{m \text{ times}} ~, && Q^{(n)}_{-} =  \mathbbm{1}_{2} \oplus \underbrace{H \oplus \cdots \oplus H}_{m-1 \text{ times}} ~.
\end{align}
A bilinear form with determinant $\pm 1$ belongs to the class of $Q^{(n)}_{+}$ if $\det(Q) = (-1)^{\lfloor\frac{n}{2}\rfloor}$ and $Q^{(n)}_{-}$ if $\det(Q) = -(-1)^{\lfloor\frac{n}{2}\rfloor}$. However $Q^{(n)}_{\pm}$ are not always inequivalent. When $p = 1 \mod 4$, $-1$ is a quadratic residue and as a result $Q^{(n)}_{+}$ can be rescaled by a square of $-1$ to get a bilinear form equivalent to $Q^{(n)}_{-}$ and vice versa. Hence for $p = 1 \mod 4$, the intersection form on $H_{2}(Y_{4},\mathbb{Z}_{p^{k}})$ is completely determined the second Betti number $b_{2}(Y_{4})$. While for $p = 3 \mod 4$, $-1$ is not a quadratic residue and $Q^{(n)}_{\pm}$ are genuinely inequivalent.

Next let us discuss the case of $N=2^{k}$ which is slightly more involved. A complete set of invariants for a bilinear form $Q$ in this case is given by the dimension $n$, the evenness,\footnote{\ Unlike the case of odd primes $p^{k}$, evenness in $\mathbb{Z}_{2^{k}}$ is well defined, hence one can distinguish between even and odd bilinear forms.} and the signature modulo $8$. Even forms exist only when $n=2m$ and there is a unique even form, $E^{(m)} = Q_{+}^{(2m)}$, for each $m$.

When $n=2m$, the odd forms have even signatures, while when $n=2m+1$ is odd, the odd forms have odd signatures. For $n=2m$ even, we have odd forms with signatures (0,2,4,6): the odd form of signature $0$ with lowest dimension is $\mbox{diag}(1,-1)$. For signature $2$ it is $\mathbbm{1}_{2}$, for signature $6 = -2 \mod 8$, it is $-\mathbbm{1}_{2}$. Lastly, odd forms with signature $4$ exist only in dimension four and higher with the four dimensional one being $\mathbbm{1}_{4}$. All these forms can be combined with suitable factors of $E^{(m')}$ to give odd forms with the same signature in higher dimensions, the distinct classes of bilinear forms are represented by,
\begin{align}
\label{eq:60}
  Q_{0}^{(2m+2)} &= \mbox{diag}(1,-1) \oplus E^{(m)} ~, \nonumber \\
  Q_{2}^{(2m+2)} &= \mathbbm{1}_{2} \oplus E^{(m)} ~, \nonumber \\
  Q_{4}^{(2m+4)} &= \mathbbm{1}_{4} \oplus E^{(m)} ~, \nonumber \\
  Q_{6}^{(2m+2)} &= (-\mathbbm{1}_{2}) \oplus E^{(m)} ~.
\end{align}

Similarly odd forms with odd signature exist when $n=2m+1$ is odd. A representative for each class is,
\begin{align}
\label{eq:61}
 Q_{1}^{(2m+1)} &= \mathbbm{1} \oplus E^{(m)} ~, && Q_{7}^{(2m+1)} = (-\mathbbm{1}) \oplus E^{(m)} ~, \nonumber \\
 Q_{3}^{(2m+3)} &= \mathbbm{1}_{3} \oplus E^{(m)} ~, && Q_{5}^{(2m+3)} = (-\mathbbm{1}_{3}) \oplus E^{(m)} ~. \nonumber \\
\end{align}
However since $1 = -1$ for $\mathbb{Z}_{2}$, for $k=1$ most of these forms are not distinct. This is because for $k=1$ the signature is only well defined $\mod 2$ and not $\mod 8$ as in the general case. It can be seen that the distinct ones are represented by one of $E^{(m)},Q_{0}^{(2m+2)}$ and $Q_{1}^{(2m+1)}$.

Having described the equivalence classes let us now turn to the problem of finding Lagrangian subspaces. Once again we start with $N=p^{k}$ with $p$ an odd prime. Instead of trying to classify all the Lagrangian subspaces we will restrict to those that have a basis. \footnote{\ More technically we consider only those Lagrangian subgroups which are free submodules of $H_{2}(Y_{4},\mathbb{Z}_{p^{k}})$. If $Y_{4}$ is simply connected $H_{2}(Y_{4},\mathbb{Z}_{p^{k}})$ itself is always a free module. This assumption does have some physical relevance: only in this case the global symmetry is $\mathbb{Z}_{p^{k}}^{\frac{1}{2}b_{2}(Y_{4})}.$} This is a technical assumption but allows us to deal with the problem of finding Lagrangian subgroups using tools of linear algebra. For instance, consider the case $k=1$. Since $\mathbb{Z}_{p}$ is a field all subspaces and hence all Lagrangian subspaces admit a basis. However, this is not necessarily true for $k > 1$, e.g. for $k=2l$ even, $p^{l}H_{2}(Y_{4},\mathbb{Z}_{p^{2l}})$ is a Lagrangian subspace, but it does not have a basis.

Let us start with the case of $n = b_{2}(Y_{4}) = 2m$. Suppose that a Lagrangian subspace $L$ with a basis exists and we extend this basis to a basis of $H_{2}(Y_{4},\mathbb{Z}_{p^{k}})$. In this basis the the matrix representing $Q$ takes the form,
\begin{align}
\label{eq:62}
  Q &=
\begin{pmatrix}
  0 & A^{T} \\
  A & B
\end{pmatrix} ~,
\end{align}
with $A,B$ $m \times m$ matrices. $B$ must be symmetric and since $Q$ is non-degenerate $A$ must be invertible. The Lagrangian subspace $L$ is the row span of the first $m$ rows in this basis. Let us perform a change of basis $Q \to S Q S^{T}$ with $S =
\left(\begin{smallmatrix} A^{-1} & 0 \\ 0 & \mathbbm{1} \end{smallmatrix}\right)$. This basis change preserves $L$ but $Q$ becomes,
\begin{align}
\label{eq:63}
  Q =
\begin{pmatrix}
  0 & \mathbbm{1} \\
  \mathbbm{1} & B
\end{pmatrix} ~.
\end{align}
Hence for any Lagrangian subspace $L$ we can choose a basis of $H_{2}(Y_{4},\mathbb{Z}_{p^{k}})$ such that $L$ is the span of first $m$ rows and the intersection matrix takes the form above. This hold even for for $p=2$. For odd primes, we can do a further change of basis which also preserves $L$, $Q \to S Q S^{T}$ with $S =
\left(\begin{smallmatrix} \mathbbm{1} & 0 \\ -\frac{1}{2}(B + B^{T}) & \mathbbm{1} \end{smallmatrix}\right)$. This changes $Q$ to
\begin{align}
\label{eq:64}
  Q =
  \begin{pmatrix}
    0 & \mathbbm{1} \\
    \mathbbm{1} & 0
  \end{pmatrix} ~,
\end{align}
which is related $Q^{(n)}_{+}$ as defined in \eqref{eq:59} by a permutation of basis elements. Hence if such an $L$ exists $Q$ must be equivalent to $Q^{(m)}_{+}$ and any other Lagrangian subspace with a basis is related to $L$ by some element of $SO(Q^{(n)}_{+},\mathbb{Z}_{p^{k}})$ i.e. the group of matrices preserving $Q^{(n)}_{+}$.

We have shown that when $p = 3 \mod 4$, $Q^{(n)}_{-}$ has no Lagrangian subgroup with a basis, we now show it has no Lagrangian subgroups at all when $k$ is odd. To show this let us suppose that $x = \smm{a \\ b \\ \vec{c}} \in L$ which is an isotropic subspace w.r.t $Q^{(n)}_{-}$. Then $Q^{(n)}_{-}(x,x) = a^{2} + b^{2} = 0$. Since for $p = 3 \mod 4$, a square root of $-1$ doesn't exist, we must have $a^{2} = b^{2} = 0$. For $k$ odd the only solution to $a^{2} = 0$ is $a = 0$. Since all the $\vec{c}$ form a Lagrangian subgroup of $Q^{(n-2)}_{+}$, an isotropic subspace for $Q^{(n)}_{-}$ can have at most the size of Lagrangian subspace of $Q^{(n-2)}_{-}$ but that is smaller than the size of a Lagrangian subgroup of $Q^{(n)}_{-}$. For $k=2l$ even, Lagrangian subgroups exist and it can be seen that any such subgroup takes the form $L \cong (p^{l}\mathbb{Z}_{p^{2l}}^{2} \oplus L^{\prime})$ where $L^{\prime}$ is a Lagrangian subgroup for $Q^{(n-2)}_{+}$.

When $n = 2m+1$ is odd a Lagrangian subspace can exist only when $k=2l$ is even. In that case a Lagrangian subgroup takes the form $L \cong p^{l} \mathbb{Z}_{p^{2l}} \oplus L^{\prime}$ where $L$ is a Lagrangian subgroup for $Q^{(n-1)}_{+}$.

Next we quickly deal with the case of $N=2^{k}$. In this case use methods similar to those employed above it can be checked that for $k$ odd, Lagrangian subspaces exist only for signature $0$ forms i.e. $E^{(m)}$ and $Q^{(2m+2)}_{0}$. For $k$ even the Lagrangian subgroups exist for all $l$. Lastly we draw attention to the case of $k=1$. Over $\mathbb{F}_{2}$ both forms of even dimensions $E^{(m)}$ and $Q^{(2m+2)}_{0}$ admit Lagrangian subgroups while the odd dimensional form $Q^{(2m+1)}_{1}$ does not.

Lastly let us summarize the topological boundary conditions the discussion above allows to construct,
\begin{itemize}
  \item If the intersection form $Q$ of $Y_{4}$ is even, it is equivalent to $E^{(m)}$ over $\mathbb{Z}_{2^{k}}$ and $Q^{(2m)}_{+}$ over $\mathbb{Z}_{p^{k}}$ where $m = \frac{1}{2}b_{2}(Y_{4})$. In this case Lagrangian subgroups exist for all $N$. The intersection form for the Chern-Simons theory given by $NQ$ is even for all $N$ and hence we obtain topological boundary conditions for all $N$. This includes the case of $\#^{n}(S^{2} \times S^{2})$ and K3 surfaces and more generally all spin manifolds.
  \item If the intersection form of $Y_{4}$ is odd we must choose $N$ to be even to make the intersection form $NQ$ of CS theory even. When $b_{2}(Y_{4}) = 2m$ is even there are following case to consider,
        \begin{itemize}
          \item If $N$ is an even perfect square, the topological boundary conditions always exist.
          \item If the signature of $Q$ is $0 \mod 8$ then $\det(Q) = (-1)^{m}$, and as a result the topological boundary conditions exist for all even $N$. The manifolds in this class include the del Pezzo $dP_{1}$ which will be described later and which has the intersection form $\mbox{diag}(-1,1)$. $\#^{8n}\mathbb{P}^{2}$ are further examples.
          \item If the signature of $Q$ is not $0 \mod 8$ but $\det(Q) = (-1)^{m}$, the topological boundary conditions exist for all $N=2M$ when $M$ is odd. Examples of manifolds in this category are $\#^{8n+4}\mathbb{P}^{2}$.
          \item If the signature of $Q$ is not $0 \mod 8$ and $\det(Q) = -(-1)^{m}$, the topological boundary conditions exist when $N=2M$ where $M$ is odd and all primes in the prime decomposition of $M$ are congruent to $1 \mod 4$. $\#^{4n+2}\mathbb{P}^{2}$ belong to this class of manifolds.
        \end{itemize}
  \item If $b_{2}(Y_{4})$ is odd, the topological boundary conditions exist if and only if $N$ is an even perfect square.
\end{itemize}

\subsection{Non-invertible symmetries from del Pezzo surfaces}
\label{sec:non-invert-symm}
After describing the topological boundary conditions and how to determine their existence from both the $6d$ perspective and the perspective of the $3d$ symmetry TFT for generic $Y_{4}$, we discuss the special $Y_{4}$'s which are stabilized by some subgroup of mapping class group $Y_{4}$.

By definition an element $F$ of the mapping class group acts on $H_{2}(Y_{4},\mathbb{Z}_{N})$ as a matrix which we will also denote by $F$ that preserves the intersection form i.e. $F^{T} Q F = Q$. However not every matrix $F$ can be realized as the action of a global diffeomorphism on $H_{2}(Y_{4},\mathbb{Z}_{N})$ and determining which $F$ lift to a mapping class requires great control over geometry. As an easy example of topological obstruction to such a lift let us consider the topology of such a two cycle $C$. Every two cycle $C$ can be realized as a Riemann surface $\Sigma_{g}$ in $Y_{4}$ and the genus $g$ of this Riemann surface can vary with the cycle. A diffeomorphism will always preserve the topology of a submanifold and hence will map a cycle to another cycle realized as a submanifold of the same genus. However this genus is invisible to the abstract homology groups.

This and other difficulties of studying four manifolds and the moduli spaces of conformal structures associated to them, make studying the theories with exceptional symmetries more complicated compared to class $\mathcal{S}$ theories. For class $\mathcal{S}$ theories, the moduli space (i.e. space of gauge couplings) is understood and at least for low genus can be used to give an exhaustive account of theories with exceptional symmetries \cite{Bashmakov:2022uek}. As a result instead of attempting a general treatment we study a class of examples where (a subgroup) of mapping class group and its action on $H_{2}(Y_{4},\mathbb{Z}_{N})$ is known and can be used to construct $2d$ CFT with the corresponding exceptional symmetries which according to the general discussion of Section \ref{sec:symm-tft-dimens} are non-invertible.

\subsubsection{del Pezzo surfaces}
\label{sec:del-pezzo-surfaces}
The del Pezzo surfaces are complex surfaces (thus of real dimension four) with positive first Chern class. Up to homeomorphism there are exactly 9 of them. 8 of these can be obtained from $\mathbb{P}^{2}$ by blowing up points in general position.\footnote{General position here means that no three points lie on a line, no four points on a plane and so on.} We will denote by $dP_{k}$ the manifold obtain by blowing up $k$ points on $\mathbb{P}^{2}$. Topologically it is the  same as $\mathbb{P}^{2} \#^{k} \overline{\mathbb{P}^{2}}$ where $\overline{\mathbb{P}^{2}}$ means $\mathbb{P}^{2}$ with the opposite of standard orientation. The remaining del Pezzo surface is topologically $S^{2} \times S^{2}$ and corresponds to the $\mathbb F_0$ Hirzebruch surface, which we have already described above (see also Appendix \ref{sec:non-invertible-t}). Consequently in this section we will focus on the $dP_{k}$s.

Let us denote the non-trivial cycles of the $dP_{k}$ $E_{i}$ for $1 \le i \le k$ (each $E_i$ is the exceptional cycle corresponding to the $i^{th}$ blown up point) and $H$ (which is the pullback of the hyperplane section of $\mathbb{P}^{2}$ under the projection which shrinks all the cycles $E_{i}$ to points).\footnote{ie. the blown down of all $E_{i}$} The intersection form is,
\begin{align}
\label{eq:33}
  Q(H,H) = 1 ~, && Q(H,E_{i}) = 0 ~, && Q(E_{i},E_{j}) = -\delta_{ij} ~.
\end{align}
This gives us a Lorentzian version of the intersection form on $\#^{n} \mathbb{P}^{2}$. The group preserving it is thus $SO(1,k;\mathbb{Z})$. This intersection form is odd and to simplify our analysis we restrict to the case of $6d\, (2,0)$ of type $\mathfrak{a}_{2N-1}$. From the discussion in Section \ref{sec:find-lagr-subgr}, the topological boundary conditions always exist when $N=2M^{2}$. In addition,
\begin{itemize}
  \item For $dP_{1}$ there are topological boundary conditions when $N=2^{l}p_{1}^{k_{1}} \cdots p_{i}^{k}$ and $p_{i} = 1 \mod 4$ for all $i$.
  \item For $dP_{5}$ there are topological boundary conditions when $N=p_{1}^{k_{1}} \cdots p_{i}^{k}$ and $p_{i} = 1 \mod 4$ for all $i$.
  \item For $dP_{3}$ and $dP_{7}$, there are topological boundary conditions for all $N$.
\end{itemize}
This exhausts the list of $N$ for which we can find topological boundary conditions. In particular, for all the cases not listed above the resulting 2d theories will be relative field theories by construction.

Now in the cases above, we want to identify those cases that lead to non-invertible symmetries. The quantities giving rise to the parameters of interest for the $2d$ CFTs out of the dimensional reduction of the $6d\, (2,0)$ theory on $dP_{k}$ are the relative volumes of the $E_{i}$ cycles, for $1 \le i \le k$,
\begin{align}
\label{eq:34}
  R_{i} &= \Vol(E_{i}) ~, && \sum_{i=1}^{k}R_{i} = 1 ~.
\end{align}
where the constraint on the RHS above comes by requiring that the resulting 2d theory is conformally invariant (and hence depend only on the ratios of the $R_i$ and hence one can always rescale them so that the RHS holds). %One can solve it explicitly, but we won't to avoid complications in what follows.

The case of $dP_{1}$ is somewhat special and we deal with it separately. In this case there are no free parameters. The group $SO(1,1;\mathbb{Z})$ is $\mathbb{Z}_{2}$ generated by $-\mathsf{1}$. This transformation necessarily preserves every lattice and as a result all topological boundary conditions are in separate orbits of the mapping class group.

\subsubsection{Exceptional symmetries from del Pezzo surfaces}
\label{sec:except-symm-1}

From now on we consider the case of $dP_{3},dP_{5}$ and $dP_{7}$. For them there are equivalence classes of maps up to isotopy which arise from the exchange of two blowups points $i$ and $j$. The action of this mapping class on $H_{2}(dP_{k},\mathbb{Z})$ is to exchange $E_{i}$ and $E_{j}$. The action of this diffeomorphism is simply $R_{i} \leftrightarrow R_{j}$ and when $R_{i} = R_{j}$, we obtain a theory with an extra $\mathbb{Z}_{2}$ symmetry. Similarly when more $l$ of $R_{i}$ coincide we obtain an exceptional $S_{l}$ symmetry corresponding to a permutations of the corresponding blowups.

However, following \cite{Iqbal:2001ye} we can construct a bigger group of exceptional symmetries corresponding to the Weyl group of lie algebras $E_{k}$. The first Chern class of $dP_{k}$ is given by,
\begin{align}
\label{eq:35}
  c_{1}(dP_{k}) &= 3H - \sum_{l=1}^{k}E_{l} ~.
\end{align}
Its orthogonal complement $K_{k}$ is the sublattice of $H_{2}(dP_{k},\mathbb{Z})$ consisting of two-cycles $x$ such that $Q(c_{1}(dP_{k}),x) = 0$. It is an integral lattice of rank $k$ with a basis $C_{i}$ given by,
\begin{align}
\label{eq:36}
  C_{i} &= E_{i} - E_{i+1} ~, && 1 \le i \le k-1 ~, \nonumber \\
  C_{k} &= H - E_{1} - E_{2} - E_{3} ~.
\end{align}
The intersection form induced by $Q$ on $K_{k}$ is negative definite and in the basis written above equals to minus the Cartan matrix of the $\mathfrak{e}_{k}$ Lie algebra.\footnote{\ Where in the (by now) standard conventions of \cite{Iqbal:2001ye}: $\mathfrak{e}_{1} = \mathfrak{a}_1$, $\mathfrak{e}_{2} = \mathfrak{a}_1 \times \mathfrak{a}_1$, $\mathfrak{e}_{3} = \mathfrak{a}_1 \times \mathfrak{a}_2$, $\mathfrak{e}_{4} = \mathfrak{a}_4$, $\mathfrak{e}_{5} = \mathfrak{so}_{10}$, $\mathfrak{e}_6$, $\mathfrak{e}_7$, $\mathfrak{e}_{8}$.} Therefore $K_{k}$ as a lattice is isomorphic to the root lattice of the exceptional Lie algebra $\mathfrak{e}_{k}$, but with a negative definite bilinear form instead of positive definite.\footnote{We choose have chosen to define $dP_{k}$ with opposite orientation and this would have made the intersection form on $K_{k}$ positive definite. However we follow the convention of \cite{Iqbal:2001ye} which is also more found commonly in string theory literature.} The group preserving this intersection form on $K_{k}$ is then $\text{Weyl}(\mathfrak{e}_k)$, the Weyl group of $\mathfrak{e}_k$, namely the group generated by the reflections $s_{i}$,
\begin{align}
\label{eq:37}
  s_{i}(x) = x + Q(C_{i},x)C_{i} ~.
\end{align}
Since $Q(C_{i},c_{1}(dP_{k})) = 0$, this action preserves the first Chern class. As a result, the Weyl group of $\mathfrak{e}_{k}$ is the subgroup of $SO(1,k;\mathbb{Z}_{k})$ that in addition to the intersection form also preserves the first Chern class. It can also be realized as a subgroup of diffeomorphism of $dP_{k}$. From \eqref{eq:36} it can be seen that for $i \ne k$, $s_{i}$ corresponds precisely to exchanging $E_{i}$ and $E_{i+1}$. These are the transpositions which generate the $S_{k}$ subgroup of the Weyl group described above. The diffeomorphism that implements $s_{k}$ is less intuitive and we refer to \cite{Iqbal:2001ye} for its details.

In particular, consider the case of the unique maximally symmetric point for $dP_{k}$ (for which all $R_{i}$ are equal, in our normalization $R_{i} = \frac{1}{k}$ for $1 \le i \le k$). Since no $\frac{k+1}{2}$ dimensional sublattice of $H_{2}(dP_{k},\mathbb{F}_{p})$ can be invariant under all the Weyl reflections, $s_{i}$, following the procedure outlined in Section \ref{sec:except-symm} we obtain non-invertible symmetries corresponding to $\text{Weyl}(\mathfrak{e}_k)$ elements realized at this point.

\section{Spin Chern-Simons and abelian theory}
\label{sec:spin-chern-simons}
So far we have avoided dealing with spin Chern-Simons theories and their analogue in $7d$ i.e. Chern-Simons theory with odd level which depends on a Wu$_4$ structure. In this section we illustrate some of the subtleties which arise in this case. Our main focus will be theory of an abelian self dual 2-form gauge field in $6$ dimensions. This theory is a higher dimensional analogue of a chiral boson is two dimensions. It is also relevant for the discussion of $6d\, (2,0)$ theories since the tensor multiplet contains self dual gauge fields and it is the global properties of these self dual gauge fields which are controlled by the Chern-Simons term of symmetry TFT, much in the same way that 3d Chern-Simons controls the global properties of Wess-Zumino-Witten models.\footnote{To be precise the analogy should be with WZW model at level 1, for which the 3d bulk theory is $SU(N)$ CS theory which is related by a level rank duality (see e.g. \cite{Hsin:2016blu}) to the abelian CS theory at level $N$.} We will return on the dependence on the Wu structure for the more general case of the 6d (2,0) $\sixg$ theories in future work.

The connection of the global properties of the theory of self dual gauge fields with Chern-Simons theory was realized in \cite{Verlinde:1995mz} and expanded upon in \cite{Witten:1996md,Witten:1996hc,Witten:1999vg,Belov:2006jd} among other works to elucidate aspects of $S$ duality in four dimensional gauge theories. We start by recapping some of the salient results. The starting point is the fact that for general $7$ manifolds, the periods of field strength $c$ field in \eqref{eq:1} obey a shifted quantization condition, i.e.
\begin{align}
\label{eq:38}
  \oint_{Y_4} \dd c &= \frac{1}{2}\oint_{Y_4} v_{4} \mod 1 ~,
\end{align}
with $v_{4}$ being the fourth (integral) Wu class. If the level $k$ is even then this subtlety goes away since the periods of $v_{4}$ are integral. However for odd $k$, this makes $\frac{1}{2} v_{4}$ an ever present background field that cannot be turned off i.e. unlike the more familiar case where background gauge fields form a group, in this case they have merely the structure of an affine space. This means in particular that to define the Chern-Simons theory on a 7 manifold with boundary $X_{6}$, we must choose a trivialization of $v_{4}$ on the boundary. We will now specialize to level 1 which is the relevant level for the abelian theory. The result of the shifted quantization is that the boundary theory is a relative theory whose partition vector forms a representation of a $\mathbb{Z}_{2}$ Heisenberg algebra i.e. the algebra in \eqref{eq:4} with $N=2$.

Up till now, we have considered the partition vector as an abstract object. The action of Heisenberg algebra on it gives us a handle on many important properties of it. In the case of abelian theory however the different components of the partition vector can be calculated following the recipe of \cite{Witten:1999vg}, which we now recount briefly.

The self dual $2$ form field $B^+$ satisfies the self duality condition, $H = \star\, H$ with $H = \dd B^+$ the field strength and $\star$ represents the Hodge star. We want to compute the Euclidean partition function so that after the Wick rotation, the self duality constraint becomes $H = i \star H$. The difficulty in writing a partition function comes from the fact that the self duality constraint is in tension with the quantization of periods of $H$ i.e. we cannot simultaneously impose $H = i \star H$ and that $\oint_{\Sigma_{3}} H \in 2\pi \mathbb{Z}$ for every $\Sigma_{3} \in H_{3}(X_{6},\mathbb{Z})$. The best that can be done is to impose it on a maximal isotropic sublattice of $H_{3}(X_{6},\mathbb{Z})$. In view of \eqref{eq:38} it is actually enough that this lattice is isotropic modulo $2$ i.e. an element of the partition vector is indexed by a \textit{half rank} sublattice $\mathcal{L}$ of $H_{3}(X_{6}, \mathbb{Z})$ such that its projection to $H_{3}(X_{6},\mathbb{Z}_{2})$ is isotropic. This choice refines the possible choices of global structures for the resulting theories.

In the explicit examples we will always choose $\mathcal{L}$ to be an integral isotropic lattice. The background fields then correspond to the (modulo 2 reduction of) $H_{3}(X_{6},\mathbb{Z}) / \mathcal{L}$ and we can represent them by choosing an isotropic lattice $\mathcal{L}^{\perp}$ such that $H_{3}(X_{6},\mathbb{Z}) = \mathcal{L} \oplus \mathcal{L}^{\perp}$. To write down an element of the partition vector, we choose $\mathcal{A}_{i}$-cycles which form a basis of $\mathcal{L}$ and $\mathcal{B}_{i}$ cycles which form a basis of $\mathcal{L}^{\perp}$ such that the intersection form on $H_{3}(X_{6},\mathbb{Z})$ is,
\begin{align}
\label{eq:40}
  \ev{\mathcal{A}_{i},\mathcal{B}_{j}} = \delta_{ij} ~.
\end{align}
Next we choose to restrict to self dual 3-forms which have integral periods over elements of $\mathcal{L}$. In particular we choose a basis of such forms $\lambda_{j}$ such that,
\begin{align}
\label{eq:41}
  \oint_{\mathcal{A}_{i}} \lambda_{j} &= \delta_{ij} ~.
\end{align}
The periods of the $\lambda_{j}$ over $\mathcal{B}_{i}$ can be arbitrary complex numbers and define the period matrix $\tau_{ij}$ i.e,
\begin{align}
\label{eq:42}
  \tau_{ij} = \oint_{\mathcal{B}_{j}} \lambda_{i} ~.
\end{align}
The period matrix $\tau$ can be shown to be symmetric i.e. $\tau_{ij} = \tau_{ji}$. With period matrix in hand, the elements of partition vector we are after are,
\begin{align}
\label{eq:43}
  \apv{Z}{\theta}{\phi} &= \frac{1}{\Delta} \mathcal{I}_{\mathcal{L}}(\theta,\phi;\tau) ~, \nonumber \\
  \mathcal{I}_{\mathcal{L}}(\theta,\phi;\tau) &= \sum_{n \in \mathcal{L} + \theta} \exp(i\pi \sum_{i,j} \tau_{ij} n^{i}n^{j} + 2\pi i \sum_{i}n^{i} \phi_{i}) ~.
\end{align}
Where $\theta \in (\frac{1}{2}\mathcal{L}) / \mathcal{L}$ and $\phi \in (\frac{1}{2}\mathcal{L}^{\perp}) / \mathcal{L}^{\perp}$. Both $\theta$ and $\phi$ change under the action of the mapping class group of $X_{6}$ and the relative nature of the theory is exhibited in the change of $\mathcal{I}_{L}(\theta,\phi;\tau)$ under this variation. 

The factor $\mathcal{I}_{L}(\theta,\phi;\tau)$ is the analogue of the instantonic contributions to the path integral of $4d$ Maxwell theory. The factor $\Delta$ is the contribution of the non-zero modes to the path integral and as a result if two lattices $\mathcal{L}$ and $\mathcal{L}^{\prime}$ are related by the action of some element of mapping class group, $\Delta$ is same for both of them. As a result we do not attempt to compute it but we emphasize that if $\mathcal{L}$ and $\mathcal{L}^{\prime}$ are in the disconnected orbits under the action of mapping class group there is no reason to expect $\Delta$ to be same for both of them.

We will compute $Z$ above in the examples of the form $\mathbb{T}^{2} \times Y_{4}$ where $Y_{4}$ is a compact simply connected four manifold. In the applications to the $4d$ S-duality, it is more natural (see the discussion in Section \ref{sec:symm-tft-dimens}) to choose $\mathcal{L}$ to be (an integral lift) of the form $L \otimes H_{2}(Y_4,\mathbb{Z}_{2})$ where $L$ is a maximal isotropic sublattice of $H_{1}(\mathbb{T}^{2},\mathbb{Z}_{2})$. The $4d$ partition functions thus obtained also turn out to have well known $2d$ interpretations. However we can also consider $\mathcal{L}$ to be of the form $H_{1}(\mathbb{T}^{2},\mathbb{Z}_{2}) \otimes L$ where $L$ is a Lagrangian subgroup of $H_{2}(Y_4,\mathbb{Z}_{2})$. These partition functions turn out to be different from the first kind but must be recognized as partition functions of the $6d$ theory dimensionally reduced along $Y_4$. 

\medskip

As an explicit example let us consider the case of $6d$ abelian theory on $\mathbb{T}^{2} \times \mathbb{P}^{2}$ and reinterpret the result of \cite{Witten:1999vg} in the language of higher form symmetries. Since $H_{2}(\mathbb{P}^{2},\mathbb{Z}_{2}) = \mathbb{Z}_{2}$ has no Lagrangian subgroup, any maximal isotropic subgroup of $H_{2}(\mathbb{P}^{2},\mathbb{Z}_{2})$ and consequently any maximal isotropic sublattice $\mathcal{L}$ is given by $\mathcal{L} = L \otimes H_{2}(\mathbb{P}^{2},\mathbb{Z}_{2})$ where $L$ is a maximal isotropic sublattice of $H_{1}(\mathbb{T}^{2},\mathbb{Z}_{2})$. Since the intersection form of $\mathbb{P}^{2}$ is odd the $3d$ symmetry TFT is a level 1 U(1) CS theory. Then we expect to obtain a relative $2d$ theory, or, equivalently, a $4d$ theory with multiple global structures. We will take $L$ to be generated by the usual $A$ cycle on $\mathbb T^2$. The lone generator $S$ of $H_{1}(\mathbb{P}^{2},\mathbb{Z})$ has self intersection $1$. A basis of three forms with integral periods on $A \cup S$ is then,
\begin{align}
\label{eq:44}
  \lambda = \dd \overline{z} \wedge \kappa ~,
\end{align}
where $\kappa$ is the generator of integral harmonic two forms on $\mathbb{P}^{2}$ and $\overline{z}$ is the anti-holomorphic coordinate on $\mathbb{T}^{2}$. The period matrix can be seen to have a single entry equal to $\overline{\tau}$. The partition function given by \eqref{eq:43} is then,
\begin{align}
\label{eq:45}
  \mathcal{I}(\theta,\phi;\overline{\tau}) = \sum_{n \in \mathbb{Z} + \theta} \exp(i\pi \overline{\tau} n^{2} + 2\pi i \phi\, n)
\end{align}
The choice of Wu structure is encoded in the values of $\theta$ and $\phi$. Either can be $0$ or $\frac{1}{2}$, and thus there four distinct choices of $\mathcal{I}$. Interestingly these are precisely the four Jacobi theta functions. From the $2d$ perspective, reducing a self dual two form on $\mathbb{P}^{2}$ we obtain a chiral (i.e. self dual) boson in $2d$. This boson lives on the $\mathbb{T}^{2}$ boundary of the spin Chern-Simons theory at level $1$, and hence to properly define the theory we need to choose a spin structure on $\mathbb{T}^{2}$ and a compatible spin structure on $3d$ bulk. The familiar four choices of spin structure correspond to the four possible conformal blocks.  Defining $\overline{q} = e^{i \pi  \overline{\tau}}$ we see that
\begin{itemize}
\item the spin structure given by $0 \in H_1(\mathbb T^2, \mathbb Z_2)$ correspond to $\mathcal{I}(0,0;\overline{\tau}) = \theta_{3}(\overline{q})$;
\item the spin structure given by $A\in H_1(\mathbb T^2, \mathbb Z_2)$ corresponds to $\mathcal{I}(\frac{1}{2},0;\overline{\tau}) = \theta_{2}(\overline{q})$;
\item the spin structure  given by $B\in H_1(\mathbb T^2, \mathbb Z_2)$ corresponds to $\mathcal{I}(0,\frac{1}{2};\overline{\tau}) = \theta_{4}(\overline{q})$;
\item the spin structure given by  $A+B\in H_1(\mathbb T^2, \mathbb Z_2)$ corresponds to $\mathcal{I}(\frac{1}{2},\frac{1}{2};\overline{\tau}) = \theta_{1}(\overline{q})$.
\end{itemize}

Now let us take a look at how the choices of spin structures become refined choices of global structure of the $4d$ theory on $\mathbb{P}^{2}$ --- we will see that this is ultimately related to the fractionalization of the Poincar\'e rotation symmetry, and we indeed produce all the fractionalization classes for Maxwell theory \cite{Brennan:2022tyl,Delmastro:2022pfo}. Let us recall that the Maxwell theory has $U(1)_{e} \times U(1)_{m}$ one-form symmetry and the object charged under it are Wilson and 't Hooft lines respectively. In the $6d$ picture the Wilson lines and the 't Hooft lines supported on a closed loop $\gamma_1$\footnote{\ These curves are not topological so $\gamma_1$ need not be a non-trivial cycle. Indeed for $\mathbb{P}^{2}$ there are no such cycles.} arise from the following integrals of the self-dual 2-forms:
\begin{align} 
\label{eq:39}
  W^{(e)}(\gamma_1)= \text{exp}(i \oint_{A\cup \gamma_1} B^+) && W^{(m)}(\gamma_1) = \text{exp}(i \oint_{B\cup \gamma_1} B^+)~.
\end{align}
From the 7d perspective, we are after the corresponding charge operators. The latter are realised by $\Phi(\Sigma_3)$ such that they have a surface $\Sigma_2$ in $\mathbb P^2$ that links with $\gamma_1$ and wrap a 1-cycle in $\mathbb T^2$, corresponding to whether these are electric or magnetic generators. However here we see that to properly define these surfaces we need to keep track of the dependence of the  $7d$ Chern-Simons on the choice of a trivialization of $v_{4}$ on the boundary $X_{6}$. Different trivializations correspond to choices parametrized by classes in $H^{3}(X_{6},\mathbb{Z}_{2})$. For the space at hand there are four choices, namely $0, \tilde{A} \wedge w_{2}(\mathbb{P}^{2}), \tilde{B} \wedge w_{2}(\mathbb{P}^{2})$ and $(\tilde{A}+\tilde{B}) \wedge w_{2}(\mathbb{P}_{2})$ where $\tilde{A}$ and $\tilde{B}$ are Poincar\'e duals to $A$ and $B$ cycles on $\mathbb T^2$. Now to each of these choices correspond a different representative for the electric and the magnetic fluxes associated to $\Phi(\Sigma_3)$. This is how charge fractionalization is encoded in the choice of a trivialization of $v_4$ on the boundary. If we chose the trivial one, then all the fluxes will be integer and in particular no flux will depend on $w_2(\mathbb P^2)$. Then we can choose all lines to be bosons. Instead in the other three choices, we are forced to give $w_{2}(\mathbb{P}^{2})$ flux for respectively the electric, the magnetic, and both electric and magnetic one-form symmetries in the $4d$ theory. Linking with such a flux changes the statistics of the corresponding line. Since the one form multiplying $w_{2}(\mathbb{P}_{2})$ determines a spin structure on $\mathbb{T}^{2}$, we obtain the following dictionary between the spin structure of $\mathbb{T}^{2}$ and the statistics of Wilson and 't Hooft lines for the Maxwell theory:
\begin{itemize}
  \item If the spin structure is $0$: both Wilson and 't Hooft lines are bosonic;
  \item  If the spin structure is $A$: the 't Hooft line can link with the corresponding one-form symmetry flux and become fermionic, while the Wilson lines stay bosonic;
  \item If the spin structure is $B$: the Wilson lines become fermionic while the 't Hooft lines are bosonic.
  \item If the spin structure is $A+B$: both Wilson and 't Hooft lines are fermionic.
\end{itemize}

The last of these four cases is a version of Maxwell theory called ``all fermion electrodynamics'' and it is known to have an anomaly \cite{Wang:2018qoy}. This fact can be recovered from the $6d$ picture by noticing that the partition function for it is given by $\theta_{1}(\overline{q})$ which is notoriously identically zero, thus signaling an anomaly. %More abstractly this fact can be seen as a consequence of the non-commutative Heisenberg algebra along the same lines as the anomaly of $PSU(2)_{-}$ theory under S-transformation at self dual coupling $\tau = i$.

\bigskip

One can consider more examples of this effect: the 6d partition function, even for the simplest case of the free self-dual tensor field, is richer than just the Maxwell theory, and that is encoded in the further choices that arise from the trivialization of $v_4$ on the boundary. We discuss the examples of $dP_1$ and $S^2 \times S^2$ in detail in appendix \ref{sec:moreH} for the interested reader.

\section{Conclusions and outlook}\label{sec:conclusions}

The study of higher symmetries is undergoing a rapid evolution in recent years.\footnote{\, \label{footnote:citationdump} For a (partial) list of references accounting for the recent progress on this topic we refer our readers to eg. \cite{Gaiotto:2014kfa,Gaiotto:2010be,Kapustin:2013qsa,Kapustin:2013uxa,Aharony:2013hda,DelZotto:2015isa,Sharpe:2015mja,Tachikawa:2017gyf,Cordova:2018cvg,Benini:2018reh,Hsin:2018vcg,Wan:2018bns,Thorngren:2019iar,GarciaEtxebarria:2019caf,Eckhard:2019jgg,Wan:2019soo,Bergman:2020ifi,Morrison:2020ool,Albertini:2020mdx,Hsin:2020nts,Bah:2020uev,DelZotto:2020esg,Hason:2020yqf,Aasen:2020jwb,Bhardwaj:2020phs,Apruzzi:2020zot,Cordova:2020tij,Thorngren:2020aph,DelZotto:2020sop,BenettiGenolini:2020doj,Yu:2020twi,Bhardwaj:2020ymp,DeWolfe:2020uzb,Gukov:2020btk,Iqbal:2020lrt,Hidaka:2020izy,Brennan:2020ehu,Komargodski:2020mxz,Closset:2020afy,Thorngren:2020yht,Closset:2020scj,Bhardwaj:2021pfz,Nguyen:2021naa,Heidenreich:2021xpr,Apruzzi:2021phx,Apruzzi:2021vcu,Hosseini:2021ged,Cvetic:2021sxm,Buican:2021xhs,Bhardwaj:2021zrt,Iqbal:2021rkn,Braun:2021sex,Cvetic:2021maf,Closset:2021lhd,Thorngren:2021yso,Sharpe:2021srf,Bhardwaj:2021wif,Hidaka:2021mml,Lee:2021obi,Lee:2021crt,Hidaka:2021kkf,Koide:2021zxj,Apruzzi:2021mlh,Kaidi:2021xfk,Choi:2021kmx,Bah:2021brs,Gukov:2021swm,Closset:2021lwy,Yu:2021zmu,Apruzzi:2021nmk,Beratto:2021xmn,Bhardwaj:2021mzl,Wang:2021vki,Cvetic:2022uuu,DelZotto:2022fnw,Cvetic:2022imb,DelZotto:2022joo,DelZotto:2022ras,Bhardwaj:2022yxj,Hayashi:2022fkw,Kaidi:2022uux,Roumpedakis:2022aik,Choi:2022jqy, Choi:2022zal,Arias-Tamargo:2022nlf,Cordova:2022ieu,Bhardwaj:2022dyt,Benedetti:2022zbb, DelZotto:2022ohj,Bhardwaj:2022scy,Antinucci:2022eat,Carta:2022spy, Apruzzi:2022dlm, Heckman:2022suy, Choi:2022rfe, Bhardwaj:2022lsg, Lin:2022xod, Bartsch:2022mpm,Apruzzi:2022rei,GarciaEtxebarria:2022vzq,Cherman:2022eml,Heckman:2022muc, Lu:2022ver, Niro:2022ctq, Kaidi:2022cpf,Mekareeya:2022spm,vanBeest:2022fss,Antinucci:2022vyk,Giaccari:2022xgs,Bashmakov:2022uek,Cordova:2022fhg, GarciaEtxebarria:2022jky, Choi:2022fgx, Robbins:2022wlr, Bhardwaj:2022kot, Bhardwaj:2022maz, Bartsch:2022ytj, Gaiotto:2020iye,Robbins:2021ibx, Robbins:2021xce,Huang:2021zvu,Inamura:2021szw,Cherman:2021nox,Sharpe:2022ene,Bashmakov:2022jtl, Lee:2022swr, Inamura:2022lun, Damia:2022bcd, Lin:2022dhv,Burbano:2021loy, Damia:2022rxw, Apte:2022xtu, Nawata:2023rdx, Bhardwaj:2023zix, Kaidi:2023maf, Etheredge:2023ler, Lin:2023uvm, Amariti:2023hev, Bhardwaj:2023wzd, Bartsch:2023pzl, Carta:2023bqn, Zhang:2023wlu, Cao:2023doz, Putrov:2023jqi, Acharya:2023bth,Inamura:2023qzl,Dierigl:2023jdp,Antinucci:2023uzq,Cvetic:2023plv}, as well as to \cite{Cordova:2022ruw} for a recent review.} In this paper we have investigated a technique to study the known topological symmetries of 2d theories from a higher dimensional perspective, thus recovering many known results about these theories from geometry. This analysis confirms the idea outlined in \cite{Bashmakov:2022uek} that indeed null bordisms of the compactification manifold are a tool to induce gapped boundary conditions, and that summing over topologies in that context is an operation equivalent to gauging a discrete symmetry. In particular, we have seen that the Tambara-Yamagami $\mathbb Z_N$ symmetries can be obtained in this way. It would be interesting in the future to have a deeper exploration of the 2d symmetry categories that have a 4d origin. This analysis is based on establishing a dictionary between 4-manifolds and fusion 1-category data along the lines of the lattice approach \cite{Aasen:2020jwb}. In particular, it would be interesting to understand if the Haagerup symmetry category (see eg. \cite{Huang:2021nvb} and references therein) can be given a higher dimensional origin, as it was recently shown that the latter cannot have a simple 3d gauge theory origin as a boundary of a Chern Simons gauge theory \cite{teleman}.

The methods discussed in this paper can be extended straightforwardly to other $\mathfrak g$ and manifolds with different dimensions. More precisely, we exploit the fact that 6d (2,0) theories $\sixg$ for $\mathfrak g$ generic are relative field theories \cite{Witten:2009at,Freed:2012bs}: $\sixg$ exists as a boundary condition for a 7d TFT, $\topg$, and for $\mathfrak g$ generic $\topg$ does not have gapped boundary conditions. For this reason $\sixg$ does not assign to six manifolds $X_6$ a partition function, but rather a partition vector in the Hilbert space $\topg(X_6)$, a collection of conformal blocks. Upon reduction on a $k$ dimensional manifold $Y_k$, we obtain a new system $D_{Y_k}\sixg$ which is also a quantum field theory in $d = 6-k$ dimensions. This theory also comes with a symmetry TFT which is obtained via dimensional reduction of $\topg$, $D_{Y_k}\mathfrak F_{\mathfrak g}$. Interestingly $D_{Y_k}\mathfrak F_{\mathfrak g}$ can have gapped boundary conditions even if $\topg$ does not, and the latter arise from null-bordisms of $Y_k$, $k+1$ dimensional manifolds that have boundary $Y_k$, which we can use to smoothly interpolate between $Y_k$ and a point. In the context of the 4d class $\mathcal S$ theories that we considered in \cite{Bashmakov:2022jtl,Bashmakov:2022uek}, $Y_2$ is a genus $g$ riemann surface without punctures, which always admit (several) such null-bordisms, realized via three-dimensional handlebodies. For two dimensional field theories instead, only those four manifolds that have vanishing signature $\sigma(Y_4) = b_2^+ - b_2^-$ can be null-bordant. If one considers the theory $D_{Y_4} \sixg$ for $\mathfrak g = \mathfrak a_{N-1}$ with $N$ prime and $\sigma(Y_4) = 0$, the resulting 2d theory is always an absolute theory. It would be interesting to study these correspondences more in detail in the future for other values of $N$ and general $\mathfrak g$. Also here we expect then to be able to formulate the symmetry structure of the theories $D_{Y_k}\mathfrak X_{\mathfrak g}$ in terms of the boundary conditions of $D_{Y_k}\mathfrak F_{\mathfrak g}$ and the condensates of defects of $\topg$ to give rise to interesting generalizations of duality defects in various dimensions. In particular, it would be interesting to start establishing dictionaries between $k$-manifolds and $(5-k)$-fusion category data.

\section*{Acknowledgements}

We would like to thank Justin Kaidi for initial collaboration on this project as well as for many illuminating discussions that helped us shaping these ideas and refining these techniques. We especially thank Iñaki García Etxebarria for comments on the previous version of this manuscrip which lead to a substantial improvement of the results in Section 3.6. We also thank Christian Copetti, Luis Diogo, Babak Haghighat, Tzu-Chen Huang, Alice Hedenlund, Kantaro Ohmori, and Jian Qiu for many useful conversations. The work of MDZ and AH is supported by the European Research Council (ERC) under the European Union's Horizon 2020 research and innovation program (grant agreement No. 851931). MDZ also acknowledges support from the Simons Foundation Grant $\#$888984 (Simons Collaboration on Global Categorical Symmetries). The work of VB is supported from the Knut and Alice Wallenberg Foundation under grant KAW 2021.0170, VR grant 2018-04438 and the Olle Engkvists Stiftelse grant n. 2180108.

%Another direction that we think would be interesting to extend our study is to include in the analysis the 6d (1,0) theories, for which, however less is known about the corresponding symmetry TFT \cite{Apruzzi:2022dlm}.
%Heckman:2017uxe,

\appendix

\section{A non-invertible T-duality}
\label{sec:non-invertible-t}
As the simplest example of this general framework let us consider the example of theory obtained by dimensional reduction of $6d\, (2,0)$ theory of type $\mathfrak{a}_{1}$ on $S^{2} \times S^{2}$. In this case the conformal structure of $S^{2} \times S^{2}$ is fixed by the sole dimensionless parameter $R^{2}$ which is the ratio of the areas of two spheres. The two spheres intersect each other so the intersection form is the Pauli matrix $\sigma_{1}$.

The symmetry TFT is a $U(1)^{2}$ Chern-Simons theory with only off diagonal couplings,
\begin{align}
\label{eq:17}
  S_{3d} = \frac{1}{4\pi} \int_{W_{3}} (a_{1}\wedge \,\dd a_{2} + a_{2} \wedge \,\dd a_{1})
\end{align}

The mapping class group is $\mathbb{Z}_{2}$ corresponding to exchange of two spheres which we call $S_{1}$ and $S_{2}$. Hence it changes $R \to \frac{1}{R}$ and corresponds to a T-duality which for abelian case considered from the six dimensional perspective in \cite{Verlinde:1995mz}.

The global structures of this $2d$ theory correspond to isotropic sublattices of $H_{2}(S^{2} \times S^{2} , \mathbb{Z}_{2})$. Notice that modulo $2$, the intersection form of $S^{2} \times S^{2}$ is same as the intersection form of $\mathbb{T}^{2}$. So in precise analogy with $4d\, \mathcal{N}=4$ SYM with gauge algebra $\mathfrak{su}(2)$ there are three global structures, generated by $S_{1},S_{2}$ and $S_{1}+S_{2}$ respectively.

The first two global structures, determined by $S_{1}$ and $S_{2}$ respectively are exchanged by the T-duality and as a result they have non-invertible T-duality at the self dual point $R=1$. While the third one generated by $S_{1} + S_{2}$ is fixed by the T-duality and at $R=1$ T-duality becomes an anomalous symmetry. This is very similar to the treatment of $\mathcal{N}=4\, \mathfrak{su}(2)$ SYM under S-transformation of $SL(2,\mathbb{Z})$ mapping class group at $\tau = i$. However there is a crucial difference, in $4d$ case the non-invertible symmetries can be related to the anomalous symmetry using $T$-transformation, this however is not the case for the 2d theory since mapping class group has two disconnected orbits.

We can also consider the dimensional reduction of $\mathfrak{a}_{p-1}$ theory on $S^{2} \times S^{2}$. For $p > 2$ there are only two global structures generated by $S_{1}$ and $S_{2}$ respectively. The two global structures are interchanged under $T$-duality and at $R=1$ we obtain a non-invertible symmetry.

\section{More on Spin CS and the self-dual tensor}\label{sec:moreH}

\subsection{$dP_{1}$}
\label{sec:dp_1}
As the next case let us consider the abelian theory on $\mathbb{T}^{2} \times dP_{1}$. We again start with a lattice $\mathcal{L}$ which is relevant for $4d$ S-duality and is generated by $A \cup H$ and $A \cup E_{1}$ in the notation of Section \ref{sec:del-pezzo-surfaces}. A basis of self dual form with integral periods over this sublattice is,
\begin{align}
\label{eq:46}
  F_{1} &= \dd \overline{z} \wedge \omega_{+} ~, && F_{2} = \dd z \wedge \omega_{-} ~.
\end{align}
Where $\omega_{+}$ and $\omega_{-}$ are 2-forms Poincar\'e dual to $H$ and $E_{1}$ respectively. Choosing $B \cup H$ and $B \cup E_{1}$ for the basis for $\mathcal{L}^{\perp}$ we obtain, the period matrix,
\begin{align}
\label{eq:47}
  \begin{pmatrix}
    -\overline{\tau} && 0 \\
  0 && \tau
\end{pmatrix} ~.
\end{align}

As a result $\mathcal{I}(\theta,\phi;\tau,\overline{\tau})$ is given by,
\begin{align}
\label{eq:48}
  \mathcal{I}(\theta,\phi;\tau,\overline{\tau}) = \sum_{n_{1} \in \mathbb{Z} + \theta_{1} , n_{2} \in \mathbb{Z} + \theta_{2}} \exp(\pi i (n_{1}^{2} \tau - n_{2}^{2} \overline{\tau}) + 2\pi i (n_{1} \phi_{1} + n_{2} \phi_{2}) ) ~.
\end{align}
As expected \cite{Witten:1995gf}, since the intersection form of $dP_{1}$ is odd, this partition function is not invariant under $\tau \to \tau + 1$ but only under $\tau \to \tau + 2$.

However it is possible to write another partition function for $dP_{1}$ too which is not related to \eqref{eq:48} by any duality. $E+H$ has self intersection $0$ and as a result the lattice generated by $A \cup (H+E_{1})$ and $B \cup (H+E_{1})$ is isotropic. This is a lattice of the form $\mathcal{L} = H_{1}(\mathbb{T}^{2} , \mathbb{Z}) \otimes L$ with $L$ a Lagrangian subgroup of $H_{2}(dP_{1},\mathbb{Z})$. However unlike the cases we have considered above no such basis which is of the form $H_{1}(\mathbb{T}^{2} , \mathbb{Z}) \otimes L^{\perp}$ can be written for $\mathcal{L}^{\perp}$. So we choose a basis for $\mathcal{L}^{\perp}$ which is $B \cup H$ and $A \cup E_{1}$. The resulting period matrix is,
\begin{align}
  \label{eq:49}
  \frac{1}{2\tau_{2}}
  \begin{pmatrix}
    i \tau\overline{\tau} & i\overline{\tau} \\
    i \overline{\tau} & i
  \end{pmatrix} ~,
\end{align}
where $\tau = \tau_{1} + i \tau_{2}$. The resulting $\mathcal{I}(\theta,\phi;\tau,\overline{\tau})$ is,
\begin{align}
\label{eq:50}
  \mathcal{I}(\theta,\phi,\tau,\overline{\tau}) &= \sum_{n_{1} \in \mathbb{Z} + \theta_{1} , n_{2} \in \mathbb{Z} + \theta_{2}}\exp(-\left(\frac{\tau\overline{\tau}}{2\tau_{2}}n_{1}^{2} + \frac{1}{2\tau_{2}} n_{2}^{2}\right) - \frac{\overline{\tau}}{\tau_{2}} n_{1}n_{2}) ~, \nonumber \\
                        &= \sum_{n_{1} \in \mathbb{Z} + \theta_{1} , n_{2} \in \mathbb{Z} + \theta_{2}}\exp(-\pi \frac{\left(\tau_{1}n_{1}-n_{2} \right)^{2} + \tau_{2}^{2}n_{1}^{2}}{2\tau_{2}} + \pi i (n_{1}n_{2} + 2 n_{1} \phi_{1} + 2 n_{2} \phi_{2})) ~.
\end{align}
Due to the term involving $n_{1}n_{2}$, this partition function is not invariant under $\tau \to \tau + 1$ even when $\phi = 0$ i.e when background fields have been switched off. This illustrates the point that for a spin Chern-Simons theory a Lagrangian subgroup of the discriminant group does not necessarily give a topological boundary condition. Just like the expression in \eqref{eq:48} it is invariant under $\tau \to \tau + 2$. However the two have a different behavior for $\tau \to -\frac{1}{\tau}$. The expression above is invariant while that in \eqref{eq:48} transforms as a modular form.

\subsection{$S^{2} \times S^{2}$}
\label{sec:s2-s2}
As a last example let us consider case $6d$ theory on $\mathbb{T}^{2} \times S^{2} \times S^{2}$. The two dimensional theory obtained by dimensionally reducing along $S^{2} \times S^{2}$ has a symmetry TFT which is an ordinary Chern-Simons and hence we should expect an modular invariant partition function. Similarly the Maxwell theory on $S^{2} \times S^{2}$ is invariant under $\tau \to \tau + 1$. However the partition function the $6d$ theory gives on $\mathbb{T}^{2} \times S^{2} \times S^{2}$ is not unique. This is because $H_{3}(\mathbb{T}^{2} \times S^{2} \times S^{2})$ has multiple maximal isotropic sublattices with at least two distinct orbits under $SL(2,\mathbb{Z}) \times \mathbb{Z}_{2}$ mapping class group. As in the previous example the first class consists of $\mathcal{L}$ of the form $L \times H_{2}(S^{2} \times S^{2},\mathbb{Z})$. Such an $\mathcal{L}$ transforms into another lattice of this same class under the action of $SL(2,\mathbb{Z})$ mapping class group of $\mathbb{T}^{2}$ and is invariant under the $\mathbb{Z}_{2}$ mapping class group of $S^{2} \times S^{2}$. The second class consists of $\mathcal{L}$ of the form $H_{1}(\mathbb{T}^{2},\mathbb{Z}) \times L^{\prime}$ and the lattices in this class are invariant under the $SL(2,\mathbb{Z})$ but can transform under the $\mathbb{Z}_{2}$ mapping classes of $S^{2} \times S^{2}$. There are no mapping classes that transform a lattice in one class to the other and as we will see the partition functions obtained from the two classes are different.\footnote{The multitude of the partition functions is a consequence of the fact that the manifolds we have considered are not symmetric enough that all maximally isotropic sublattices of $H_{3}(X_{6},\mathbb{Z})$ lies in the same orbit of mapping class group. For the case of $\mathbb{T}^{6}$ which has a mapping class group $SL(6,\mathbb{Z})$ big enough for this, the partition function for theory of two form is known \cite{Dolan:1998qk} and it is $SL(6,\mathbb{Z})$ invariant and hence unique.}

Let us consider a example of the first class with a basis of $\mathcal{L}$ given by $A \cup C_{1}$ and $A \cup C_{2}$ where $C_{1}$ and $C_{2}$ are the two spheres in $S^{2} \times S^{2}$ regarded as 2-cycles. A basis for $\mathcal{L}^{\perp}$ is given by $B \cup C_{2}$ and $A \cup C_{1}$. The Hodge star on $S^{2} \times S^{2}$ depends on $R^{2}$ the relative volume of two spheres and is given by,
\begin{align}
\label{eq:51}
  \star C_{1} = R^2\, C_{2} ~, && \star C_{2} = \frac{1}{R^2} \, C_{1} ~.
\end{align}
A basis of self dual forms with integral periods on $\mathcal{L}$ is given by,
\begin{align}
\label{eq:52}
  F_{1} = \dd x \wedge \omega_{1} + i R^2\, \dd y \wedge \omega_{2} ~, && F_{2} = \dd x \wedge \omega_{2} + \frac{i}{R^2}\,\dd y \wedge \omega_{1} ~,
\end{align}
with $\omega_{1}$ and $\omega_{2}$ being the Poincare duals of $C_{1}$ and $C_{2}$. The corresponding period matrix is,
\begin{align}
\label{eq:53}
  \begin{pmatrix}
    \frac{i\tau_{2}}{R^{2}} && \tau_{1} \\
    \tau_{1} &&   iR^{2} \tau_{2}
  \end{pmatrix} ~.
\end{align}
Which results in,\footnote{Since $\theta$ and $\phi$ do not influence the modular properties we have set them to zero here in order to emphasize the modular properties.}
\begin{align}
\label{eq:54}
  \mathcal{I}(\tau,\overline{\tau},R) =  \sum_{n_{1},n_{2}} \exp(- \pi \left(\frac{n_{1}^{2}}{R^{2}} + n_{2}^{2}R^{2}\right) \tau_{2} - 2\pi i n_{1}n_{2}\tau_{1}) ~.
\end{align}
This is explicitly invariant under $R \to \frac{1}{R}$ and under $\tau \to \tau + 1$ and transforms as modular form under $\tau \to \frac{1}{\tau}$ as expected for the partition function of $4d$ Maxwell theory.\footnote{Correctly matching the weight requires inclusion of the factor of $\Delta$ which we are ignoring for now.}

Although writing this partition function requires choosing a maximal isotropic sublattice $L$ of $H_{1}(\mathbb{T}^{2},\mathbb{Z})$ the result itself is independent of $L$ and hence gives a well defined $2d$ theory which can recognized as usual compact boson of radius $R^{2}$.

Nevertheless by choosing an $\mathcal{L}$ from the second class we can write a different partition function. For this we choose $\mathcal{L}$ to be generated by $A \cup C_{1}$ and $B \cup C_{2}$ with the basis for $\mathcal{L}^{\perp}$ given by $B \cup C_{2}$ and $A \cup C_{1}$. A basis of self dual three forms with integral periods on $\mathcal{L}$ is given by,
\begin{align}
\label{eq:55}
  F_{1} &= \frac{1}{2 i \tau_{2}} \left(\overline{\tau} \dd z \wedge (\omega_{1} - R^{2} \omega_{2}) - \tau \dd \overline{z} \wedge (\omega_{1} + R^{2} \omega_{2}) \right) ~, \nonumber \\
  F_{2} &= \frac{i}{2 \tau_{2}}\left(\dd z \wedge (\omega_{1} - R^{2} \omega_{2}) -  \dd \overline{z} \wedge (\omega_{1} + R^{2} \omega_{2})\right) ~.
\end{align}
This data results in the period matrix,
\begin{align}
  \label{eq:56}
  \frac{1}{\tau_{2}}
  \begin{pmatrix}
  i \tau\overline{\tau}R^{2} & -i \tau_{1}R^{2} \\
  -i \tau_{1} R^{2} & iR^{2}
\end{pmatrix} ~,
\end{align}
and hence the partition function correspond to,
\begin{align}
\label{eq:57}
  \mathcal{I}(R,\tau,\overline{\tau}) &= \sum_{n_{1},n_{2}}\exp(- \frac{((\tau_{1}n_{1}-n_{2})^{2} + \tau_{2}^{2}n_{1}^{2})R^{2}}{\tau_{2}} ) ~.
\end{align}
This expression can be verified to be invariant under $SL(2,\mathbb{Z})$ mapping class group of $\mathbb{T}^{2}$ and hence gives the partition function of an absolute theory.

\bibliographystyle{ytphys}
\bibliography{dualitydefects2.bib}

\providecommand{\href}[2]{#2}\begingroup\raggedright\begin{thebibliography}{100}

\bibitem{Freed:2022qnc}
D.~S. Freed, G.~W. Moore, and C.~Teleman, ``{Topological symmetry in quantum
  field theory},'' \href{http://arxiv.org/abs/2209.07471}{{\ttfamily
  arXiv:2209.07471 [hep-th]}}.

\bibitem{Apruzzi:2021nmk}
F.~Apruzzi, F.~Bonetti, I.~n.~G. Etxebarria, S.~S. Hosseini, and
  S.~Schafer-Nameki, ``{Symmetry TFTs from String Theory},''
  \href{http://arxiv.org/abs/2112.02092}{{\ttfamily arXiv:2112.02092
  [hep-th]}}.

\bibitem{Moore:1989yh}
G.~W. Moore and N.~Seiberg, ``{Taming the Conformal Zoo},''
  \href{http://dx.doi.org/10.1016/0370-2693(89)90897-6}{{\em Phys. Lett. B}
  {\bfseries 220} (1989) 422--430}.

\bibitem{Reshetikhin:1991tc}
N.~Reshetikhin and V.~G. Turaev, ``{Invariants of three manifolds via link
  polynomials and quantum groups},''
  \href{http://dx.doi.org/10.1007/BF01239527}{{\em Invent. Math.} {\bfseries
  103} (1991) 547--597}.

\bibitem{Turaev:1992hq}
V.~G. Turaev and O.~Y. Viro, ``{State sum invariants of 3 manifolds and quantum
  6j symbols},'' \href{http://dx.doi.org/10.1016/0040-9383(92)90015-A}{{\em
  Topology} {\bfseries 31} (1992) 865--902}.

\bibitem{Fuchs:2002cm}
J.~Fuchs, I.~Runkel, and C.~Schweigert, ``{TFT construction of RCFT correlators
  1. Partition functions},''
  \href{http://dx.doi.org/10.1016/S0550-3213(02)00744-7}{{\em Nucl. Phys. B}
  {\bfseries 646} (2002) 353--497},
  \href{http://arxiv.org/abs/hep-th/0204148}{{\ttfamily arXiv:hep-th/0204148}}.

\bibitem{Fuchs:2003id}
J.~Fuchs, I.~Runkel, and C.~Schweigert, ``{TFT construction of RCFT
  correlators. 2. Unoriented world sheets},''
  \href{http://dx.doi.org/10.1016/j.nuclphysb.2003.11.026}{{\em Nucl. Phys. B}
  {\bfseries 678} (2004) 511--637},
  \href{http://arxiv.org/abs/hep-th/0306164}{{\ttfamily arXiv:hep-th/0306164}}.

\bibitem{Fuchs:2004dz}
J.~Fuchs, I.~Runkel, and C.~Schweigert, ``{TFT construction of RCFT
  correlators. 3. Simple currents},''
  \href{http://dx.doi.org/10.1016/j.nuclphysb.2004.05.014}{{\em Nucl. Phys. B}
  {\bfseries 694} (2004) 277--353},
  \href{http://arxiv.org/abs/hep-th/0403157}{{\ttfamily arXiv:hep-th/0403157}}.

\bibitem{Fuchs:2004xi}
J.~Fuchs, I.~Runkel, and C.~Schweigert, ``{TFT construction of RCFT correlators
  IV: Structure constants and correlation functions},''
  \href{http://dx.doi.org/10.1016/j.nuclphysb.2005.03.018}{{\em Nucl. Phys. B}
  {\bfseries 715} (2005) 539--638},
  \href{http://arxiv.org/abs/hep-th/0412290}{{\ttfamily arXiv:hep-th/0412290}}.

\bibitem{Fjelstad:2005ua}
J.~Fjelstad, J.~Fuchs, I.~Runkel, and C.~Schweigert, ``{TFT construction of
  RCFT correlators. V. Proof of modular invariance and factorisation},'' {\em
  Theor. Appl. Categor.} {\bfseries 16} (2006) 342--433,
  \href{http://arxiv.org/abs/hep-th/0503194}{{\ttfamily arXiv:hep-th/0503194}}.

\bibitem{Kapustin:2013qsa}
A.~Kapustin and R.~Thorngren, ``{Topological Field Theory on a Lattice,
  Discrete Theta-Angles and Confinement},''
  \href{http://dx.doi.org/10.4310/ATMP.2014.v18.n5.a4}{{\em Adv. Theor. Math.
  Phys.} {\bfseries 18} no.~5, (2014) 1233--1247},
  \href{http://arxiv.org/abs/1308.2926}{{\ttfamily arXiv:1308.2926 [hep-th]}}.

\bibitem{Kapustin:2013uxa}
A.~Kapustin and R.~Thorngren, ``{Higher symmetry and gapped phases of gauge
  theories},'' \href{http://arxiv.org/abs/1309.4721}{{\ttfamily arXiv:1309.4721
  [hep-th]}}.

\bibitem{Benini:2018reh}
F.~Benini, C.~C\'ordova, and P.-S. Hsin, ``{On 2-Group Global Symmetries and
  their Anomalies},'' \href{http://dx.doi.org/10.1007/JHEP03(2019)118}{{\em
  JHEP} {\bfseries 03} (2019) 118},
  \href{http://arxiv.org/abs/1803.09336}{{\ttfamily arXiv:1803.09336
  [hep-th]}}.

\bibitem{Freed:2009qp}
D.~S. Freed, M.~J. Hopkins, J.~Lurie, and C.~Teleman, ``{Topological Quantum
  Field Theories from Compact Lie Groups},'' in {\em {A Celebration of Raoul
  Bott's Legacy in Mathematics}}.
\newblock 5, 2009.
\newblock \href{http://arxiv.org/abs/0905.0731}{{\ttfamily arXiv:0905.0731
  [math.AT]}}.

\bibitem{Kaidi:2022cpf}
J.~Kaidi, K.~Ohmori, and Y.~Zheng, ``{Symmetry TFTs for Non-Invertible
  Defects},'' \href{http://arxiv.org/abs/2209.11062}{{\ttfamily
  arXiv:2209.11062 [hep-th]}}.

\bibitem{Dijkgraaf:1989pz}
R.~Dijkgraaf and E.~Witten, ``{Topological Gauge Theories and Group
  Cohomology},'' \href{http://dx.doi.org/10.1007/BF02096988}{{\em Commun. Math.
  Phys.} {\bfseries 129} (1990) 393}.

\bibitem{Kaidi:2022uux}
J.~Kaidi, G.~Zafrir, and Y.~Zheng, ``{Non-Invertible Symmetries of
  $\mathcal{N}=4$ SYM and Twisted Compactification},''
  \href{http://arxiv.org/abs/2205.01104}{{\ttfamily arXiv:2205.01104
  [hep-th]}}.

\bibitem{Bashmakov:2022jtl}
V.~Bashmakov, M.~Del~Zotto, and A.~Hasan, ``{On the 6d Origin of Non-invertible
  Symmetries in 4d},'' \href{http://arxiv.org/abs/2206.07073}{{\ttfamily
  arXiv:2206.07073 [hep-th]}}.

\bibitem{Bashmakov:2022uek}
V.~Bashmakov, M.~Del~Zotto, A.~Hasan, and J.~Kaidi, ``{Non-invertible
  Symmetries of Class $\mathcal{S}$ Theories},''
  \href{http://arxiv.org/abs/2211.05138}{{\ttfamily arXiv:2211.05138
  [hep-th]}}.

\bibitem{Antinucci:2022cdi}
A.~Antinucci, C.~Copetti, G.~Galati, and G.~Rizi, ``{''Zoology'' of
  non-invertible duality defects: the view from class $\mathcal{S}$},''
  \href{http://arxiv.org/abs/2212.09549}{{\ttfamily arXiv:2212.09549
  [hep-th]}}.

\bibitem{Tachikawa:2013hya}
Y.~Tachikawa, ``{On the 6d origin of discrete additional data of 4d gauge
  theories},'' \href{http://dx.doi.org/10.1007/JHEP05(2014)020}{{\em JHEP}
  {\bfseries 05} (2014) 020}, \href{http://arxiv.org/abs/1309.0697}{{\ttfamily
  arXiv:1309.0697 [hep-th]}}.

\bibitem{Gukov:2020btk}
S.~Gukov, P.-S. Hsin, and D.~Pei, ``{Generalized global symmetries of $T[M]$
  theories. Part I},'' \href{http://dx.doi.org/10.1007/JHEP04(2021)232}{{\em
  JHEP} {\bfseries 04} (2021) 232},
  \href{http://arxiv.org/abs/2010.15890}{{\ttfamily arXiv:2010.15890
  [hep-th]}}.

\bibitem{Antinucci:2022vyk}
A.~Antinucci, F.~Benini, C.~Copetti, G.~Galati, and G.~Rizi, ``{The holography
  of non-invertible self-duality symmetries},''
  \href{http://arxiv.org/abs/2210.09146}{{\ttfamily arXiv:2210.09146
  [hep-th]}}.

\bibitem{Gaiotto:2009we}
D.~Gaiotto, ``{N=2 dualities},''
  \href{http://dx.doi.org/10.1007/JHEP08(2012)034}{{\em JHEP} {\bfseries 08}
  (2012) 034}, \href{http://arxiv.org/abs/0904.2715}{{\ttfamily arXiv:0904.2715
  [hep-th]}}.

\bibitem{Gaiotto:2009hg}
D.~Gaiotto, G.~W. Moore, and A.~Neitzke, ``{Wall-crossing, Hitchin Systems, and
  the WKB Approximation},'' \href{http://arxiv.org/abs/0907.3987}{{\ttfamily
  arXiv:0907.3987 [hep-th]}}.

\bibitem{Gadde:2013sca}
A.~Gadde, S.~Gukov, and P.~Putrov, ``{Fivebranes and 4-manifolds},''
  \href{http://dx.doi.org/10.1007/978-3-319-43648-7_7}{{\em Prog. Math.}
  {\bfseries 319} (2016) 155--245},
  \href{http://arxiv.org/abs/1306.4320}{{\ttfamily arXiv:1306.4320 [hep-th]}}.

\bibitem{Putrov:2015jpa}
P.~Putrov, J.~Song, and W.~Yan, ``{(0,4) dualities},''
  \href{http://dx.doi.org/10.1007/JHEP03(2016)185}{{\em JHEP} {\bfseries 03}
  (2016) 185}, \href{http://arxiv.org/abs/1505.07110}{{\ttfamily
  arXiv:1505.07110 [hep-th]}}.

\bibitem{Gukov:2018iiq}
S.~Gukov, D.~Pei, P.~Putrov, and C.~Vafa, ``{4-manifolds and topological
  modular forms},'' \href{http://dx.doi.org/10.1007/JHEP05(2021)084}{{\em JHEP}
  {\bfseries 05} (2021) 084}, \href{http://arxiv.org/abs/1811.07884}{{\ttfamily
  arXiv:1811.07884 [hep-th]}}.

\bibitem{Dedushenko:2017tdw}
M.~Dedushenko, S.~Gukov, and P.~Putrov,
  \href{http://dx.doi.org/10.1093/oso/9780198802013.003.0011}{``{Vertex
  algebras and 4-manifold invariants},''} in {\em {Nigel Hitchin's 70th
  Birthday Conference}}, vol.~1, pp.~249--318.
\newblock 5, 2017.
\newblock \href{http://arxiv.org/abs/1705.01645}{{\ttfamily arXiv:1705.01645
  [hep-th]}}.

\bibitem{Harvey:1995tg}
J.~A. Harvey, G.~W. Moore, and A.~Strominger, ``{Reducing S duality to T
  duality},'' \href{http://dx.doi.org/10.1103/PhysRevD.52.7161}{{\em Phys. Rev.
  D} {\bfseries 52} (1995) 7161--7167},
  \href{http://arxiv.org/abs/hep-th/9501022}{{\ttfamily arXiv:hep-th/9501022}}.

\bibitem{Verlinde:1995mz}
E.~P. Verlinde, ``{Global aspects of electric - magnetic duality},''
  \href{http://dx.doi.org/10.1016/0550-3213(95)00431-Q}{{\em Nucl. Phys. B}
  {\bfseries 455} (1995) 211--228},
  \href{http://arxiv.org/abs/hep-th/9506011}{{\ttfamily arXiv:hep-th/9506011}}.

\bibitem{Seiberg:2011dr}
N.~Seiberg and W.~Taylor, ``{Charge Lattices and Consistency of 6D
  Supergravity},'' \href{http://dx.doi.org/10.1007/JHEP06(2011)001}{{\em JHEP}
  {\bfseries 06} (2011) 001}, \href{http://arxiv.org/abs/1103.0019}{{\ttfamily
  arXiv:1103.0019 [hep-th]}}.

\bibitem{Frohlich:2004ef}
J.~Frohlich, J.~Fuchs, I.~Runkel, and C.~Schweigert, ``{Kramers-Wannier duality
  from conformal defects},''
  \href{http://dx.doi.org/10.1103/PhysRevLett.93.070601}{{\em Phys. Rev. Lett.}
  {\bfseries 93} (2004) 070601},
  \href{http://arxiv.org/abs/cond-mat/0404051}{{\ttfamily
  arXiv:cond-mat/0404051}}.

\bibitem{Freed:2018cec}
D.~S. Freed and C.~Teleman, ``{Topological dualities in the Ising model},''
  \href{http://arxiv.org/abs/1806.00008}{{\ttfamily arXiv:1806.00008
  [math.AT]}}.

\bibitem{Chang:2018iay}
C.-M. Chang, Y.-H. Lin, S.-H. Shao, Y.~Wang, and X.~Yin, ``{Topological Defect
  Lines and Renormalization Group Flows in Two Dimensions},''
  \href{http://dx.doi.org/10.1007/JHEP01(2019)026}{{\em JHEP} {\bfseries 01}
  (2019) 026}, \href{http://arxiv.org/abs/1802.04445}{{\ttfamily
  arXiv:1802.04445 [hep-th]}}.

\bibitem{Komargodski:2020mxz}
Z.~Komargodski, K.~Ohmori, K.~Roumpedakis, and S.~Seifnashri, ``{Symmetries and
  strings of adjoint QCD$_{2}$},''
  \href{http://dx.doi.org/10.1007/JHEP03(2021)103}{{\em JHEP} {\bfseries 03}
  (2021) 103}, \href{http://arxiv.org/abs/2008.07567}{{\ttfamily
  arXiv:2008.07567 [hep-th]}}.

\bibitem{Thorngren:2019iar}
R.~Thorngren and Y.~Wang, ``{Fusion Category Symmetry I: Anomaly In-Flow and
  Gapped Phases},'' \href{http://arxiv.org/abs/1912.02817}{{\ttfamily
  arXiv:1912.02817 [hep-th]}}.

\bibitem{Thorngren:2021yso}
R.~Thorngren and Y.~Wang, ``{Fusion Category Symmetry II: Categoriosities at
  $c$ = 1 and Beyond},'' \href{http://arxiv.org/abs/2106.12577}{{\ttfamily
  arXiv:2106.12577 [hep-th]}}.

\bibitem{Seiberg:1996bd}
N.~Seiberg, ``{Five-dimensional SUSY field theories, nontrivial fixed points
  and string dynamics},''
  \href{http://dx.doi.org/10.1016/S0370-2693(96)01215-4}{{\em Phys. Lett. B}
  {\bfseries 388} (1996) 753--760},
  \href{http://arxiv.org/abs/hep-th/9608111}{{\ttfamily arXiv:hep-th/9608111}}.

\bibitem{Douglas:1996xp}
M.~R. Douglas, S.~H. Katz, and C.~Vafa, ``{Small instantons, Del Pezzo surfaces
  and type I-prime theory},''
  \href{http://dx.doi.org/10.1016/S0550-3213(97)00281-2}{{\em Nucl. Phys. B}
  {\bfseries 497} (1997) 155--172},
  \href{http://arxiv.org/abs/hep-th/9609071}{{\ttfamily arXiv:hep-th/9609071}}.

\bibitem{Ganor:1996mu}
O.~J. Ganor and A.~Hanany, ``{Small E(8) instantons and tensionless noncritical
  strings},'' \href{http://dx.doi.org/10.1016/0550-3213(96)00243-X}{{\em Nucl.
  Phys. B} {\bfseries 474} (1996) 122--140},
  \href{http://arxiv.org/abs/hep-th/9602120}{{\ttfamily arXiv:hep-th/9602120}}.

\bibitem{Milnor}
J.~Milnor, ``{On simply connected 4-manifolds},'' vol.~{Proc. Int. Symp.
  Algebraic. Topology, Mexico}.
\newblock 1958.

\bibitem{Choi:2021kmx}
Y.~Choi, C.~Cordova, P.-S. Hsin, H.~T. Lam, and S.-H. Shao, ``{Noninvertible
  duality defects in 3+1 dimensions},''
  \href{http://dx.doi.org/10.1103/PhysRevD.105.125016}{{\em Phys. Rev. D}
  {\bfseries 105} no.~12, (2022) 125016},
  \href{http://arxiv.org/abs/2111.01139}{{\ttfamily arXiv:2111.01139
  [hep-th]}}.

\bibitem{Gaiotto:2019xmp}
D.~Gaiotto and T.~Johnson-Freyd, ``{Condensations in higher categories},''
  \href{http://arxiv.org/abs/1905.09566}{{\ttfamily arXiv:1905.09566
  [math.CT]}}.

\bibitem{Roumpedakis:2022aik}
K.~Roumpedakis, S.~Seifnashri, and S.-H. Shao, ``{Higher Gauging and
  Non-invertible Condensation Defects},''
  \href{http://arxiv.org/abs/2204.02407}{{\ttfamily arXiv:2204.02407
  [hep-th]}}.

\bibitem{Barkeshli:2014cna}
M.~Barkeshli, P.~Bonderson, M.~Cheng, and Z.~Wang, ``{Symmetry
  Fractionalization, Defects, and Gauging of Topological Phases},''
  \href{http://dx.doi.org/10.1103/PhysRevB.100.115147}{{\em Phys. Rev. B}
  {\bfseries 100} no.~11, (2019) 115147},
  \href{http://arxiv.org/abs/1410.4540}{{\ttfamily arXiv:1410.4540
  [cond-mat.str-el]}}.

\bibitem{Belov:2005ze}
D.~Belov and G.~W. Moore, ``{Classification of Abelian spin Chern-Simons
  theories},'' \href{http://arxiv.org/abs/hep-th/0505235}{{\ttfamily
  arXiv:hep-th/0505235}}.

\bibitem{Kapustin:2010hk}
A.~Kapustin and N.~Saulina, ``{Topological boundary conditions in abelian
  Chern-Simons theory},''
  \href{http://dx.doi.org/10.1016/j.nuclphysb.2010.12.017}{{\em Nucl. Phys. B}
  {\bfseries 845} (2011) 393--435},
  \href{http://arxiv.org/abs/1008.0654}{{\ttfamily arXiv:1008.0654 [hep-th]}}.

\bibitem{Stirling:2008bq}
S.~D. Stirling, {\em Abelian Chern-Simons theory with toral gauge group,
  modular tensor categories, and group categories}.
\newblock PhD thesis, Texas U., Math Dept., 2008.
\newblock \href{http://arxiv.org/abs/0807.2857}{{\ttfamily arXiv:0807.2857
  [hep-th]}}.

\bibitem{Lee:2018eqa}
Y.~Lee and Y.~Tachikawa, ``{A study of time reversal symmetry of abelian
  anyons},'' \href{http://dx.doi.org/10.1007/JHEP07(2018)090}{{\em JHEP}
  {\bfseries 07} (2018) 090}, \href{http://arxiv.org/abs/1805.02738}{{\ttfamily
  arXiv:1805.02738 [hep-th]}}.

\bibitem{Khan:2014waa}
M.~N. Khan, J.~C.~Y. Teo, and T.~L. Hughes, ``{Anyonic Symmetries and
  Topological Defects in Abelian Topological Phases: an application to the
  $ADE$ Classification},''
  \href{http://dx.doi.org/10.1103/PhysRevB.90.235149}{{\em Phys. Rev. B}
  {\bfseries 90} no.~23, (2014) 235149},
  \href{http://arxiv.org/abs/1403.6478}{{\ttfamily arXiv:1403.6478
  [cond-mat.str-el]}}.

\bibitem{Teo:2015xla}
J.~C.~Y. Teo, T.~L. Hughes, and E.~Fradkin, ``{Theory of Twist Liquids: Gauging
  an Anyonic Symmetry},''
  \href{http://dx.doi.org/10.1016/j.aop.2015.05.012}{{\em Annals Phys.}
  {\bfseries 360} (2015) 349--445},
  \href{http://arxiv.org/abs/1503.06812}{{\ttfamily arXiv:1503.06812
  [cond-mat.str-el]}}.

\bibitem{Stern2001}
R.~J. Stern, {\em Instantons and the Topology of 4-Manifolds}, pp.~333--341.
\newblock Springer New York, New York, NY, 2001.

\bibitem{GarciaEtxebarria:2019caf}
I.~n. Garc\'\i{}a~Etxebarria, B.~Heidenreich, and D.~Regalado, ``{IIB flux
  non-commutativity and the global structure of field theories},''
  \href{http://dx.doi.org/10.1007/JHEP10(2019)169}{{\em JHEP} {\bfseries 10}
  (2019) 169}, \href{http://arxiv.org/abs/1908.08027}{{\ttfamily
  arXiv:1908.08027 [hep-th]}}.

\bibitem{ConwayS88}
J.~H. Conway and N.~J.~A. Sloane,
  \href{http://dx.doi.org/10.1007/978-1-4757-2016-7}{{\em Sphere Packings,
  Lattices and Groups}}, vol.~290 of {\em Grundlehren der mathematischen
  Wissenschaften}.
\newblock Springer, 1988.
\newblock \url{https://doi.org/10.1007/978-1-4757-2016-7}.

\bibitem{Morrison-Miranda}
R.~Miranda and D.~Morrison, ``{Embeddings of Integral Quadratic Forms},''.
  \url{https://web.math.ucsb.edu/~drm/manuscripts/eiqf.pdf}.

\bibitem{Iqbal:2001ye}
A.~Iqbal, A.~Neitzke, and C.~Vafa, ``{A Mysterious duality},''
  \href{http://dx.doi.org/10.4310/ATMP.2001.v5.n4.a5}{{\em Adv. Theor. Math.
  Phys.} {\bfseries 5} (2002) 769--808},
  \href{http://arxiv.org/abs/hep-th/0111068}{{\ttfamily arXiv:hep-th/0111068}}.

\bibitem{Hsin:2016blu}
P.-S. Hsin and N.~Seiberg, ``{Level/rank Duality and Chern-Simons-Matter
  Theories},'' \href{http://dx.doi.org/10.1007/JHEP09(2016)095}{{\em JHEP}
  {\bfseries 09} (2016) 095}, \href{http://arxiv.org/abs/1607.07457}{{\ttfamily
  arXiv:1607.07457 [hep-th]}}.

\bibitem{Witten:1996md}
E.~Witten, ``{On flux quantization in M theory and the effective action},''
  \href{http://dx.doi.org/10.1016/S0393-0440(96)00042-3}{{\em J. Geom. Phys.}
  {\bfseries 22} (1997) 1--13},
  \href{http://arxiv.org/abs/hep-th/9609122}{{\ttfamily arXiv:hep-th/9609122}}.

\bibitem{Witten:1996hc}
E.~Witten, ``{Five-brane effective action in M theory},''
  \href{http://dx.doi.org/10.1016/S0393-0440(97)80160-X}{{\em J. Geom. Phys.}
  {\bfseries 22} (1997) 103--133},
  \href{http://arxiv.org/abs/hep-th/9610234}{{\ttfamily arXiv:hep-th/9610234}}.

\bibitem{Witten:1999vg}
E.~Witten, ``{Duality relations among topological effects in string theory},''
  \href{http://dx.doi.org/10.1088/1126-6708/2000/05/031}{{\em JHEP} {\bfseries
  05} (2000) 031}, \href{http://arxiv.org/abs/hep-th/9912086}{{\ttfamily
  arXiv:hep-th/9912086}}.

\bibitem{Belov:2006jd}
D.~Belov and G.~W. Moore, ``{Holographic Action for the Self-Dual Field},''
  \href{http://arxiv.org/abs/hep-th/0605038}{{\ttfamily arXiv:hep-th/0605038}}.

\bibitem{Brennan:2022tyl}
T.~D. Brennan, C.~Cordova, and T.~T. Dumitrescu, ``{Line Defect Quantum Numbers
  \& Anomalies},'' \href{http://arxiv.org/abs/2206.15401}{{\ttfamily
  arXiv:2206.15401 [hep-th]}}.

\bibitem{Delmastro:2022pfo}
D.~Delmastro, J.~Gomis, P.-S. Hsin, and Z.~Komargodski, ``{Anomalies and
  Symmetry Fractionalization},''
  \href{http://arxiv.org/abs/2206.15118}{{\ttfamily arXiv:2206.15118
  [hep-th]}}.

\bibitem{Wang:2018qoy}
J.~Wang, X.-G. Wen, and E.~Witten, ``{A New SU(2) Anomaly},''
  \href{http://dx.doi.org/10.1063/1.5082852}{{\em J. Math. Phys.} {\bfseries
  60} no.~5, (2019) 052301}, \href{http://arxiv.org/abs/1810.00844}{{\ttfamily
  arXiv:1810.00844 [hep-th]}}.

\bibitem{Gaiotto:2014kfa}
D.~Gaiotto, A.~Kapustin, N.~Seiberg, and B.~Willett, ``{Generalized Global
  Symmetries},'' \href{http://dx.doi.org/10.1007/JHEP02(2015)172}{{\em JHEP}
  {\bfseries 02} (2015) 172}, \href{http://arxiv.org/abs/1412.5148}{{\ttfamily
  arXiv:1412.5148 [hep-th]}}.

\bibitem{Gaiotto:2010be}
D.~Gaiotto, G.~W. Moore, and A.~Neitzke, ``{Framed BPS States},''
  \href{http://dx.doi.org/10.4310/ATMP.2013.v17.n2.a1}{{\em Adv. Theor. Math.
  Phys.} {\bfseries 17} no.~2, (2013) 241--397},
  \href{http://arxiv.org/abs/1006.0146}{{\ttfamily arXiv:1006.0146 [hep-th]}}.

\bibitem{Aharony:2013hda}
O.~Aharony, N.~Seiberg, and Y.~Tachikawa, ``{Reading between the lines of
  four-dimensional gauge theories},''
  \href{http://dx.doi.org/10.1007/JHEP08(2013)115}{{\em JHEP} {\bfseries 08}
  (2013) 115}, \href{http://arxiv.org/abs/1305.0318}{{\ttfamily arXiv:1305.0318
  [hep-th]}}.

\bibitem{DelZotto:2015isa}
M.~Del~Zotto, J.~J. Heckman, D.~S. Park, and T.~Rudelius, ``{On the Defect
  Group of a 6D SCFT},''
  \href{http://dx.doi.org/10.1007/s11005-016-0839-5}{{\em Lett. Math. Phys.}
  {\bfseries 106} no.~6, (2016) 765--786},
  \href{http://arxiv.org/abs/1503.04806}{{\ttfamily arXiv:1503.04806
  [hep-th]}}.

\bibitem{Sharpe:2015mja}
E.~Sharpe, ``{Notes on generalized global symmetries in QFT},''
  \href{http://dx.doi.org/10.1002/prop.201500048}{{\em Fortsch. Phys.}
  {\bfseries 63} (2015) 659--682},
  \href{http://arxiv.org/abs/1508.04770}{{\ttfamily arXiv:1508.04770
  [hep-th]}}.

\bibitem{Tachikawa:2017gyf}
Y.~Tachikawa, ``{On gauging finite subgroups},''
  \href{http://dx.doi.org/10.21468/SciPostPhys.8.1.015}{{\em SciPost Phys.}
  {\bfseries 8} no.~1, (2020) 015},
  \href{http://arxiv.org/abs/1712.09542}{{\ttfamily arXiv:1712.09542
  [hep-th]}}.

\bibitem{Cordova:2018cvg}
C.~C\'ordova, T.~T. Dumitrescu, and K.~Intriligator, ``{Exploring 2-Group
  Global Symmetries},'' \href{http://dx.doi.org/10.1007/JHEP02(2019)184}{{\em
  JHEP} {\bfseries 02} (2019) 184},
  \href{http://arxiv.org/abs/1802.04790}{{\ttfamily arXiv:1802.04790
  [hep-th]}}.

\bibitem{Hsin:2018vcg}
P.-S. Hsin, H.~T. Lam, and N.~Seiberg, ``{Comments on One-Form Global
  Symmetries and Their Gauging in 3d and 4d},''
  \href{http://dx.doi.org/10.21468/SciPostPhys.6.3.039}{{\em SciPost Phys.}
  {\bfseries 6} no.~3, (2019) 039},
  \href{http://arxiv.org/abs/1812.04716}{{\ttfamily arXiv:1812.04716
  [hep-th]}}.

\bibitem{Wan:2018bns}
Z.~Wan and J.~Wang, ``{Higher anomalies, higher symmetries, and cobordisms I:
  classification of higher-symmetry-protected topological states and their
  boundary fermionic/bosonic anomalies via a generalized cobordism theory},''
  \href{http://dx.doi.org/10.4310/AMSA.2019.v4.n2.a2}{{\em Ann. Math. Sci.
  Appl.} {\bfseries 4} no.~2, (2019) 107--311},
  \href{http://arxiv.org/abs/1812.11967}{{\ttfamily arXiv:1812.11967
  [hep-th]}}.

\bibitem{Eckhard:2019jgg}
J.~Eckhard, H.~Kim, S.~Schafer-Nameki, and B.~Willett, ``{Higher-Form
  Symmetries, Bethe Vacua, and the 3d-3d Correspondence},''
  \href{http://dx.doi.org/10.1007/JHEP01(2020)101}{{\em JHEP} {\bfseries 01}
  (2020) 101}, \href{http://arxiv.org/abs/1910.14086}{{\ttfamily
  arXiv:1910.14086 [hep-th]}}.

\bibitem{Wan:2019soo}
Z.~Wan, J.~Wang, and Y.~Zheng, ``{Higher anomalies, higher symmetries, and
  cobordisms II: Lorentz symmetry extension and enriched bosonic / fermionic
  quantum gauge theory},''
  \href{http://dx.doi.org/10.4310/AMSA.2020.v5.n2.a2}{{\em Ann. Math. Sci.
  Appl.} {\bfseries 05} no.~2, (2020) 171--257},
  \href{http://arxiv.org/abs/1912.13504}{{\ttfamily arXiv:1912.13504
  [hep-th]}}.

\bibitem{Bergman:2020ifi}
O.~Bergman, Y.~Tachikawa, and G.~Zafrir, ``{Generalized symmetries and
  holography in ABJM-type theories},''
  \href{http://dx.doi.org/10.1007/JHEP07(2020)077}{{\em JHEP} {\bfseries 07}
  (2020) 077}, \href{http://arxiv.org/abs/2004.05350}{{\ttfamily
  arXiv:2004.05350 [hep-th]}}.

\bibitem{Morrison:2020ool}
D.~R. Morrison, S.~Schafer-Nameki, and B.~Willett, ``{Higher-Form Symmetries in
  5d},'' \href{http://dx.doi.org/10.1007/JHEP09(2020)024}{{\em JHEP} {\bfseries
  09} (2020) 024}, \href{http://arxiv.org/abs/2005.12296}{{\ttfamily
  arXiv:2005.12296 [hep-th]}}.

\bibitem{Albertini:2020mdx}
F.~Albertini, M.~Del~Zotto, I.~n. Garc\'\i{}a~Etxebarria, and S.~S. Hosseini,
  ``{Higher Form Symmetries and M-theory},''
  \href{http://dx.doi.org/10.1007/JHEP12(2020)203}{{\em JHEP} {\bfseries 12}
  (2020) 203}, \href{http://arxiv.org/abs/2005.12831}{{\ttfamily
  arXiv:2005.12831 [hep-th]}}.

\bibitem{Hsin:2020nts}
P.-S. Hsin and H.~T. Lam, ``{Discrete theta angles, symmetries and
  anomalies},'' \href{http://dx.doi.org/10.21468/SciPostPhys.10.2.032}{{\em
  SciPost Phys.} {\bfseries 10} no.~2, (2021) 032},
  \href{http://arxiv.org/abs/2007.05915}{{\ttfamily arXiv:2007.05915
  [hep-th]}}.

\bibitem{Bah:2020uev}
I.~Bah, F.~Bonetti, and R.~Minasian, ``{Discrete and higher-form symmetries in
  SCFTs from wrapped M5-branes},''
  \href{http://dx.doi.org/10.1007/JHEP03(2021)196}{{\em JHEP} {\bfseries 03}
  (2021) 196}, \href{http://arxiv.org/abs/2007.15003}{{\ttfamily
  arXiv:2007.15003 [hep-th]}}.

\bibitem{DelZotto:2020esg}
M.~Del~Zotto, I.~n. Garc\'\i{}a~Etxebarria, and S.~S. Hosseini, ``{Higher form
  symmetries of Argyres-Douglas theories},''
  \href{http://dx.doi.org/10.1007/JHEP10(2020)056}{{\em JHEP} {\bfseries 10}
  (2020) 056}, \href{http://arxiv.org/abs/2007.15603}{{\ttfamily
  arXiv:2007.15603 [hep-th]}}.

\bibitem{Hason:2020yqf}
I.~Hason, Z.~Komargodski, and R.~Thorngren, ``{Anomaly Matching in the Symmetry
  Broken Phase: Domain Walls, CPT, and the Smith Isomorphism},''
  \href{http://dx.doi.org/10.21468/SciPostPhys.8.4.062}{{\em SciPost Phys.}
  {\bfseries 8} no.~4, (2020) 062},
  \href{http://arxiv.org/abs/1910.14039}{{\ttfamily arXiv:1910.14039
  [hep-th]}}.

\bibitem{Aasen:2020jwb}
D.~Aasen, P.~Fendley, and R.~S.~K. Mong, ``{Topological Defects on the Lattice:
  Dualities and Degeneracies},''
  \href{http://arxiv.org/abs/2008.08598}{{\ttfamily arXiv:2008.08598
  [cond-mat.stat-mech]}}.

\bibitem{Bhardwaj:2020phs}
L.~Bhardwaj and S.~Sch\"afer-Nameki, ``{Higher-form symmetries of 6d and 5d
  theories},'' \href{http://dx.doi.org/10.1007/JHEP02(2021)159}{{\em JHEP}
  {\bfseries 02} (2021) 159}, \href{http://arxiv.org/abs/2008.09600}{{\ttfamily
  arXiv:2008.09600 [hep-th]}}.

\bibitem{Apruzzi:2020zot}
F.~Apruzzi, M.~Dierigl, and L.~Lin, ``{The Fate of Discrete 1-Form Symmetries
  in 6d},'' \href{http://dx.doi.org/10.21468/SciPostPhys.12.2.047}{{\em SciPost
  Phys.} {\bfseries 12} (2022) 047},
  \href{http://arxiv.org/abs/2008.09117}{{\ttfamily arXiv:2008.09117
  [hep-th]}}.

\bibitem{Cordova:2020tij}
C.~Cordova, T.~T. Dumitrescu, and K.~Intriligator, ``{2-Group Global Symmetries
  and Anomalies in Six-Dimensional Quantum Field Theories},''
  \href{http://dx.doi.org/10.1007/JHEP04(2021)252}{{\em JHEP} {\bfseries 04}
  (2021) 252}, \href{http://arxiv.org/abs/2009.00138}{{\ttfamily
  arXiv:2009.00138 [hep-th]}}.

\bibitem{Thorngren:2020aph}
R.~Thorngren, ``{Topological quantum field theory, symmetry breaking, and
  finite gauge theory in 3+1D},''
  \href{http://dx.doi.org/10.1103/PhysRevB.101.245160}{{\em Phys. Rev. B}
  {\bfseries 101} no.~24, (2020) 245160},
  \href{http://arxiv.org/abs/2001.11938}{{\ttfamily arXiv:2001.11938
  [cond-mat.str-el]}}.

\bibitem{DelZotto:2020sop}
M.~Del~Zotto and K.~Ohmori, ``{2-Group Symmetries of 6D Little String Theories
  and T-Duality},'' \href{http://dx.doi.org/10.1007/s00023-021-01018-3}{{\em
  Annales Henri Poincare} {\bfseries 22} no.~7, (2021) 2451--2474},
  \href{http://arxiv.org/abs/2009.03489}{{\ttfamily arXiv:2009.03489
  [hep-th]}}.

\bibitem{BenettiGenolini:2020doj}
P.~Benetti~Genolini and L.~Tizzano, ``{Instantons, symmetries and anomalies in
  five dimensions},'' \href{http://dx.doi.org/10.1007/JHEP04(2021)188}{{\em
  JHEP} {\bfseries 04} (2021) 188},
  \href{http://arxiv.org/abs/2009.07873}{{\ttfamily arXiv:2009.07873
  [hep-th]}}.

\bibitem{Yu:2020twi}
M.~Yu, ``{Symmetries and anomalies of (1+1)d theories: 2-groups and symmetry
  fractionalization},'' \href{http://dx.doi.org/10.1007/JHEP08(2021)061}{{\em
  JHEP} {\bfseries 08} (2021) 061},
  \href{http://arxiv.org/abs/2010.01136}{{\ttfamily arXiv:2010.01136
  [hep-th]}}.

\bibitem{Bhardwaj:2020ymp}
L.~Bhardwaj, Y.~Lee, and Y.~Tachikawa, ``{$SL(2,\mathbb{Z})$ action on QFTs
  with $\mathbb{Z}_2$ symmetry and the Brown-Kervaire invariants},''
  \href{http://dx.doi.org/10.1007/JHEP11(2020)141}{{\em JHEP} {\bfseries 11}
  (2020) 141}, \href{http://arxiv.org/abs/2009.10099}{{\ttfamily
  arXiv:2009.10099 [hep-th]}}.

\bibitem{DeWolfe:2020uzb}
O.~DeWolfe and K.~Higginbotham, ``{Generalized symmetries and 2-groups via
  electromagnetic duality in $AdS/CFT$},''
  \href{http://dx.doi.org/10.1103/PhysRevD.103.026011}{{\em Phys. Rev. D}
  {\bfseries 103} no.~2, (2021) 026011},
  \href{http://arxiv.org/abs/2010.06594}{{\ttfamily arXiv:2010.06594
  [hep-th]}}.

\bibitem{Iqbal:2020lrt}
N.~Iqbal and N.~Poovuttikul, ``{2-group global symmetries, hydrodynamics and
  holography},'' \href{http://arxiv.org/abs/2010.00320}{{\ttfamily
  arXiv:2010.00320 [hep-th]}}.

\bibitem{Hidaka:2020izy}
Y.~Hidaka, M.~Nitta, and R.~Yokokura, ``{Global 3-group symmetry and 't Hooft
  anomalies in axion electrodynamics},''
  \href{http://dx.doi.org/10.1007/JHEP01(2021)173}{{\em JHEP} {\bfseries 01}
  (2021) 173}, \href{http://arxiv.org/abs/2009.14368}{{\ttfamily
  arXiv:2009.14368 [hep-th]}}.

\bibitem{Brennan:2020ehu}
T.~D. Brennan and C.~Cordova, ``{Axions, higher-groups, and emergent
  symmetry},'' \href{http://dx.doi.org/10.1007/JHEP02(2022)145}{{\em JHEP}
  {\bfseries 02} (2022) 145}, \href{http://arxiv.org/abs/2011.09600}{{\ttfamily
  arXiv:2011.09600 [hep-th]}}.

\bibitem{Closset:2020afy}
C.~Closset, S.~Giacomelli, S.~Schafer-Nameki, and Y.-N. Wang, ``{5d and 4d
  SCFTs: Canonical Singularities, Trinions and S-Dualities},''
  \href{http://dx.doi.org/10.1007/JHEP05(2021)274}{{\em JHEP} {\bfseries 05}
  (2021) 274}, \href{http://arxiv.org/abs/2012.12827}{{\ttfamily
  arXiv:2012.12827 [hep-th]}}.

\bibitem{Thorngren:2020yht}
R.~Thorngren and Y.~Wang, ``{Anomalous symmetries end at the boundary},''
  \href{http://dx.doi.org/10.1007/JHEP09(2021)017}{{\em JHEP} {\bfseries 09}
  (2021) 017}, \href{http://arxiv.org/abs/2012.15861}{{\ttfamily
  arXiv:2012.15861 [hep-th]}}.

\bibitem{Closset:2020scj}
C.~Closset, S.~Schafer-Nameki, and Y.-N. Wang, ``{Coulomb and Higgs Branches
  from Canonical Singularities: Part 0},''
  \href{http://dx.doi.org/10.1007/JHEP02(2021)003}{{\em JHEP} {\bfseries 02}
  (2021) 003}, \href{http://arxiv.org/abs/2007.15600}{{\ttfamily
  arXiv:2007.15600 [hep-th]}}.

\bibitem{Bhardwaj:2021pfz}
L.~Bhardwaj, M.~Hubner, and S.~Schafer-Nameki, ``{1-form Symmetries of 4d N=2
  Class S Theories},''
  \href{http://dx.doi.org/10.21468/SciPostPhys.11.5.096}{{\em SciPost Phys.}
  {\bfseries 11} (2021) 096}, \href{http://arxiv.org/abs/2102.01693}{{\ttfamily
  arXiv:2102.01693 [hep-th]}}.

\bibitem{Nguyen:2021naa}
M.~Nguyen, Y.~Tanizaki, and M.~\"Unsal, ``{Noninvertible 1-form symmetry and
  Casimir scaling in 2D Yang-Mills theory},''
  \href{http://dx.doi.org/10.1103/PhysRevD.104.065003}{{\em Phys. Rev. D}
  {\bfseries 104} no.~6, (2021) 065003},
  \href{http://arxiv.org/abs/2104.01824}{{\ttfamily arXiv:2104.01824
  [hep-th]}}.

\bibitem{Heidenreich:2021xpr}
B.~Heidenreich, J.~McNamara, M.~Montero, M.~Reece, T.~Rudelius, and
  I.~Valenzuela, ``{Non-invertible global symmetries and completeness of the
  spectrum},'' \href{http://dx.doi.org/10.1007/JHEP09(2021)203}{{\em JHEP}
  {\bfseries 09} (2021) 203}, \href{http://arxiv.org/abs/2104.07036}{{\ttfamily
  arXiv:2104.07036 [hep-th]}}.

\bibitem{Apruzzi:2021phx}
F.~Apruzzi, M.~van Beest, D.~S.~W. Gould, and S.~Sch\"afer-Nameki,
  ``{Holography, 1-form symmetries, and confinement},''
  \href{http://dx.doi.org/10.1103/PhysRevD.104.066005}{{\em Phys. Rev. D}
  {\bfseries 104} no.~6, (2021) 066005},
  \href{http://arxiv.org/abs/2104.12764}{{\ttfamily arXiv:2104.12764
  [hep-th]}}.

\bibitem{Apruzzi:2021vcu}
F.~Apruzzi, L.~Bhardwaj, J.~Oh, and S.~Schafer-Nameki, ``{The Global Form of
  Flavor Symmetries and 2-Group Symmetries in 5d SCFTs},''
  \href{http://arxiv.org/abs/2105.08724}{{\ttfamily arXiv:2105.08724
  [hep-th]}}.

\bibitem{Hosseini:2021ged}
S.~S. Hosseini and R.~Moscrop, ``{Maruyoshi-Song flows and defect groups of $
  {\mathrm{D}}_{\mathrm{p}}^{\mathrm{b}} $(G) theories},''
  \href{http://dx.doi.org/10.1007/JHEP10(2021)119}{{\em JHEP} {\bfseries 10}
  (2021) 119}, \href{http://arxiv.org/abs/2106.03878}{{\ttfamily
  arXiv:2106.03878 [hep-th]}}.

\bibitem{Cvetic:2021sxm}
M.~Cvetic, M.~Dierigl, L.~Lin, and H.~Y. Zhang, ``{Higher-form symmetries and
  their anomalies in M-/F-theory duality},''
  \href{http://dx.doi.org/10.1103/PhysRevD.104.126019}{{\em Phys. Rev. D}
  {\bfseries 104} no.~12, (2021) 126019},
  \href{http://arxiv.org/abs/2106.07654}{{\ttfamily arXiv:2106.07654
  [hep-th]}}.

\bibitem{Buican:2021xhs}
M.~Buican and H.~Jiang, ``{1-form symmetry, isolated $ \mathcal{N} $ = 2 SCFTs,
  and Calabi-Yau threefolds},''
  \href{http://dx.doi.org/10.1007/JHEP12(2021)024}{{\em JHEP} {\bfseries 12}
  (2021) 024}, \href{http://arxiv.org/abs/2106.09807}{{\ttfamily
  arXiv:2106.09807 [hep-th]}}.

\bibitem{Bhardwaj:2021zrt}
L.~Bhardwaj, M.~Hubner, and S.~Schafer-Nameki, ``{Liberating Confinement from
  Lagrangians: 1-form Symmetries and Lines in 4d N=1 from 6d N=(2,0)},''
  \href{http://dx.doi.org/10.21468/SciPostPhys.12.1.040}{{\em SciPost Phys.}
  {\bfseries 12} (2022) 040}, \href{http://arxiv.org/abs/2106.10265}{{\ttfamily
  arXiv:2106.10265 [hep-th]}}.

\bibitem{Iqbal:2021rkn}
N.~Iqbal and J.~McGreevy, ``{Mean string field theory: Landau-Ginzburg theory
  for 1-form symmetries},'' \href{http://arxiv.org/abs/2106.12610}{{\ttfamily
  arXiv:2106.12610 [hep-th]}}.

\bibitem{Braun:2021sex}
A.~P. Braun, M.~Larfors, and P.-K. Oehlmann, ``{Gauged 2-form symmetries in 6D
  SCFTs coupled to gravity},''
  \href{http://dx.doi.org/10.1007/JHEP12(2021)132}{{\em JHEP} {\bfseries 12}
  (2021) 132}, \href{http://arxiv.org/abs/2106.13198}{{\ttfamily
  arXiv:2106.13198 [hep-th]}}.

\bibitem{Cvetic:2021maf}
M.~Cvetic, J.~J. Heckman, E.~Torres, and G.~Zoccarato, ``{Reflections on the
  matter of 3D N=1 vacua and local Spin(7) compactifications},''
  \href{http://dx.doi.org/10.1103/PhysRevD.105.026008}{{\em Phys. Rev. D}
  {\bfseries 105} no.~2, (2022) 026008},
  \href{http://arxiv.org/abs/2107.00025}{{\ttfamily arXiv:2107.00025
  [hep-th]}}.

\bibitem{Closset:2021lhd}
C.~Closset and H.~Magureanu, ``{The $U$-plane of rank-one 4d $\mathcal{N}=2$ KK
  theories},'' \href{http://dx.doi.org/10.21468/SciPostPhys.12.2.065}{{\em
  SciPost Phys.} {\bfseries 12} (2022) 065},
  \href{http://arxiv.org/abs/2107.03509}{{\ttfamily arXiv:2107.03509
  [hep-th]}}.

\bibitem{Sharpe:2021srf}
E.~Sharpe, ``{Topological operators, noninvertible symmetries and
  decomposition},'' \href{http://arxiv.org/abs/2108.13423}{{\ttfamily
  arXiv:2108.13423 [hep-th]}}.

\bibitem{Bhardwaj:2021wif}
L.~Bhardwaj, ``{2-Group Symmetries in Class S},''
  \href{http://arxiv.org/abs/2107.06816}{{\ttfamily arXiv:2107.06816
  [hep-th]}}.

\bibitem{Hidaka:2021mml}
Y.~Hidaka, M.~Nitta, and R.~Yokokura, ``{Topological axion electrodynamics and
  4-group symmetry},''
  \href{http://dx.doi.org/10.1016/j.physletb.2021.136762}{{\em Phys. Lett. B}
  {\bfseries 823} (2021) 136762},
  \href{http://arxiv.org/abs/2107.08753}{{\ttfamily arXiv:2107.08753
  [hep-th]}}.

\bibitem{Lee:2021obi}
Y.~Lee and Y.~Zheng, ``{Remarks on compatibility between conformal symmetry and
  continuous higher-form symmetries},''
  \href{http://dx.doi.org/10.1103/PhysRevD.104.085005}{{\em Phys. Rev. D}
  {\bfseries 104} no.~8, (2021) 085005},
  \href{http://arxiv.org/abs/2108.00732}{{\ttfamily arXiv:2108.00732
  [hep-th]}}.

\bibitem{Lee:2021crt}
Y.~Lee, K.~Ohmori, and Y.~Tachikawa, ``{Matching higher symmetries across
  Intriligator-Seiberg duality},''
  \href{http://dx.doi.org/10.1007/JHEP10(2021)114}{{\em JHEP} {\bfseries 10}
  (2021) 114}, \href{http://arxiv.org/abs/2108.05369}{{\ttfamily
  arXiv:2108.05369 [hep-th]}}.

\bibitem{Hidaka:2021kkf}
Y.~Hidaka, M.~Nitta, and R.~Yokokura, ``{Global 4-group symmetry and 't Hooft
  anomalies in topological axion electrodynamics},''
  \href{http://arxiv.org/abs/2108.12564}{{\ttfamily arXiv:2108.12564
  [hep-th]}}.

\bibitem{Koide:2021zxj}
M.~Koide, Y.~Nagoya, and S.~Yamaguchi, ``{Non-invertible topological defects in
  4-dimensional $\mathbb {Z}_2$ pure lattice gauge theory},''
  \href{http://dx.doi.org/10.1093/ptep/ptab145}{{\em PTEP} {\bfseries 2022}
  no.~1, (2022) 013B03}, \href{http://arxiv.org/abs/2109.05992}{{\ttfamily
  arXiv:2109.05992 [hep-th]}}.

\bibitem{Apruzzi:2021mlh}
F.~Apruzzi, L.~Bhardwaj, D.~S.~W. Gould, and S.~Schafer-Nameki, ``{2-Group
  Symmetries and their Classification in 6d},''
  \href{http://arxiv.org/abs/2110.14647}{{\ttfamily arXiv:2110.14647
  [hep-th]}}.

\bibitem{Kaidi:2021xfk}
J.~Kaidi, K.~Ohmori, and Y.~Zheng, ``{Kramers-Wannier-like duality defects in
  (3+1)d gauge theories},'' \href{http://arxiv.org/abs/2111.01141}{{\ttfamily
  arXiv:2111.01141 [hep-th]}}.

\bibitem{Bah:2021brs}
I.~Bah, F.~Bonetti, E.~Leung, and P.~Weck, ``{M5-branes Probing Flux
  Backgrounds},'' \href{http://arxiv.org/abs/2111.01790}{{\ttfamily
  arXiv:2111.01790 [hep-th]}}.

\bibitem{Gukov:2021swm}
S.~Gukov, D.~Pei, C.~Reid, and A.~Shehper, ``{Symmetries of 2d TQFTs and
  Equivariant Verlinde Formulae for General Groups},''
  \href{http://arxiv.org/abs/2111.08032}{{\ttfamily arXiv:2111.08032
  [hep-th]}}.

\bibitem{Closset:2021lwy}
C.~Closset, S.~Sch\"afer-Nameki, and Y.-N. Wang, ``{Coulomb and Higgs Branches
  from Canonical Singularities, Part 1: Hypersurfaces with Smooth Calabi-Yau
  Resolutions},'' \href{http://arxiv.org/abs/2111.13564}{{\ttfamily
  arXiv:2111.13564 [hep-th]}}.

\bibitem{Yu:2021zmu}
M.~Yu, ``{Gauging Categorical Symmetries in 3d Topological Orders and Bulk
  Reconstruction},'' \href{http://arxiv.org/abs/2111.13697}{{\ttfamily
  arXiv:2111.13697 [hep-th]}}.

\bibitem{Beratto:2021xmn}
E.~Beratto, N.~Mekareeya, and M.~Sacchi, ``{Zero-form and one-form symmetries
  of the ABJ and related theories},''
  \href{http://arxiv.org/abs/2112.09531}{{\ttfamily arXiv:2112.09531
  [hep-th]}}.

\bibitem{Bhardwaj:2021mzl}
L.~Bhardwaj, S.~Giacomelli, M.~H\"ubner, and S.~Sch\"afer-Nameki, ``{Relative
  Defects in Relative Theories: Trapped Higher-Form Symmetries and Irregular
  Punctures in Class S},'' \href{http://arxiv.org/abs/2201.00018}{{\ttfamily
  arXiv:2201.00018 [hep-th]}}.

\bibitem{Wang:2021vki}
J.~Wang and Y.-Z. You, ``{Gauge Enhanced Quantum Criticality Between Grand
  Unifications: Categorical Higher Symmetry Retraction},''
  \href{http://arxiv.org/abs/2111.10369}{{\ttfamily arXiv:2111.10369
  [hep-th]}}.

\bibitem{Cvetic:2022uuu}
M.~Cvetic, M.~Dierigl, L.~Lin, and H.~Y. Zhang, ``{One Loop to Rule Them All:
  Eight and Nine Dimensional String Vacua from Junctions},''
  \href{http://arxiv.org/abs/2203.03644}{{\ttfamily arXiv:2203.03644
  [hep-th]}}.

\bibitem{DelZotto:2022fnw}
M.~Del~Zotto, J.~J. Heckman, S.~N. Meynet, R.~Moscrop, and H.~Y. Zhang,
  ``{Higher Symmetries of 5d Orbifold SCFTs},''
  \href{http://arxiv.org/abs/2201.08372}{{\ttfamily arXiv:2201.08372
  [hep-th]}}.

\bibitem{Cvetic:2022imb}
M.~Cvetic, J.~J. Heckman, M.~H\"ubner, and E.~Torres, ``{0-form, 1-form, and
  2-group symmetries via cutting and gluing of orbifolds},''
  \href{http://dx.doi.org/10.1103/PhysRevD.106.106003}{{\em Phys. Rev. D}
  {\bfseries 106} no.~10, (2022) 106003},
  \href{http://arxiv.org/abs/2203.10102}{{\ttfamily arXiv:2203.10102
  [hep-th]}}.

\bibitem{DelZotto:2022joo}
M.~Del~Zotto, I.~n.~G. Etxebarria, and S.~Schafer-Nameki, ``{2-Group Symmetries
  and M-Theory},'' \href{http://arxiv.org/abs/2203.10097}{{\ttfamily
  arXiv:2203.10097 [hep-th]}}.

\bibitem{DelZotto:2022ras}
M.~Del~Zotto and I.~n. Etxebarria~Garc\'\i{}a, ``{Global Structures from the
  Infrared},'' \href{http://arxiv.org/abs/2204.06495}{{\ttfamily
  arXiv:2204.06495 [hep-th]}}.

\bibitem{Bhardwaj:2022yxj}
L.~Bhardwaj, L.~Bottini, S.~Schafer-Nameki, and A.~Tiwari, ``{Non-Invertible
  Higher-Categorical Symmetries},''
  \href{http://arxiv.org/abs/2204.06564}{{\ttfamily arXiv:2204.06564
  [hep-th]}}.

\bibitem{Hayashi:2022fkw}
Y.~Hayashi and Y.~Tanizaki, ``{Non-invertible self-duality defects of
  Cardy-Rabinovici model and mixed gravitational anomaly},''
  \href{http://arxiv.org/abs/2204.07440}{{\ttfamily arXiv:2204.07440
  [hep-th]}}.

\bibitem{Choi:2022jqy}
Y.~Choi, H.~T. Lam, and S.-H. Shao, ``{Non-invertible Global Symmetries in the
  Standard Model},'' \href{http://arxiv.org/abs/2205.05086}{{\ttfamily
  arXiv:2205.05086 [hep-th]}}.

\bibitem{Choi:2022zal}
Y.~Choi, C.~Cordova, P.-S. Hsin, H.~T. Lam, and S.-H. Shao, ``{Non-invertible
  Condensation, Duality, and Triality Defects in 3+1 Dimensions},''
  \href{http://arxiv.org/abs/2204.09025}{{\ttfamily arXiv:2204.09025
  [hep-th]}}.

\bibitem{Arias-Tamargo:2022nlf}
G.~Arias-Tamargo and D.~Rodriguez-Gomez, ``{Non-Invertible Symmetries from
  Discrete Gauging and Completeness of the Spectrum},''
  \href{http://arxiv.org/abs/2204.07523}{{\ttfamily arXiv:2204.07523
  [hep-th]}}.

\bibitem{Cordova:2022ieu}
C.~Cordova and K.~Ohmori, ``{Non-Invertible Chiral Symmetry and Exponential
  Hierarchies},'' \href{http://arxiv.org/abs/2205.06243}{{\ttfamily
  arXiv:2205.06243 [hep-th]}}.

\bibitem{Bhardwaj:2022dyt}
L.~Bhardwaj, M.~Bullimore, A.~E.~V. Ferrari, and S.~Schafer-Nameki,
  ``{Anomalies of Generalized Symmetries from Solitonic Defects},''
  \href{http://arxiv.org/abs/2205.15330}{{\ttfamily arXiv:2205.15330
  [hep-th]}}.

\bibitem{Benedetti:2022zbb}
V.~Benedetti, H.~Casini, and J.~M. Magan, ``{Generalized symmetries and
  Noether\textquoteright{}s theorem in QFT},''
  \href{http://dx.doi.org/10.1007/JHEP08(2022)304}{{\em JHEP} {\bfseries 08}
  (2022) 304}, \href{http://arxiv.org/abs/2205.03412}{{\ttfamily
  arXiv:2205.03412 [hep-th]}}.

\bibitem{DelZotto:2022ohj}
M.~Del~Zotto, M.~Liu, and P.-K. Oehlmann, ``{Back to heterotic strings on ALE
  spaces. Part I. Instantons, 2-groups and T-duality},''
  \href{http://dx.doi.org/10.1007/JHEP01(2023)176}{{\em JHEP} {\bfseries 01}
  (2023) 176}, \href{http://arxiv.org/abs/2209.10551}{{\ttfamily
  arXiv:2209.10551 [hep-th]}}.

\bibitem{Bhardwaj:2022scy}
L.~Bhardwaj and D.~S.~W. Gould, ``{Disconnected 0-Form and 2-Group
  Symmetries},'' \href{http://arxiv.org/abs/2206.01287}{{\ttfamily
  arXiv:2206.01287 [hep-th]}}.

\bibitem{Antinucci:2022eat}
A.~Antinucci, G.~Galati, and G.~Rizi, ``{On Continuous 2-Category Symmetries
  and Yang-Mills Theory},'' \href{http://arxiv.org/abs/2206.05646}{{\ttfamily
  arXiv:2206.05646 [hep-th]}}.

\bibitem{Carta:2022spy}
F.~Carta, S.~Giacomelli, N.~Mekareeya, and A.~Mininno, ``{Dynamical
  consequences of 1-form symmetries and the exceptional Argyres-Douglas
  theories},'' \href{http://dx.doi.org/10.1007/JHEP06(2022)059}{{\em JHEP}
  {\bfseries 06} (2022) 059}, \href{http://arxiv.org/abs/2203.16550}{{\ttfamily
  arXiv:2203.16550 [hep-th]}}.

\bibitem{Apruzzi:2022dlm}
F.~Apruzzi, ``{Higher Form Symmetries TFT in 6d},''
  \href{http://arxiv.org/abs/2203.10063}{{\ttfamily arXiv:2203.10063
  [hep-th]}}.

\bibitem{Heckman:2022suy}
J.~J. Heckman, C.~Lawrie, L.~Lin, H.~Y. Zhang, and G.~Zoccarato, ``{6d SCFTs,
  Center-Flavor Symmetries, and Stiefel--Whitney Compactifications},''
  \href{http://arxiv.org/abs/2205.03411}{{\ttfamily arXiv:2205.03411
  [hep-th]}}.

\bibitem{Choi:2022rfe}
Y.~Choi, H.~T. Lam, and S.-H. Shao, ``{Non-invertible Time-reversal
  Symmetry},'' \href{http://arxiv.org/abs/2208.04331}{{\ttfamily
  arXiv:2208.04331 [hep-th]}}.

\bibitem{Bhardwaj:2022lsg}
L.~Bhardwaj, S.~Schafer-Nameki, and J.~Wu, ``{Universal Non-Invertible
  Symmetries},'' \href{http://arxiv.org/abs/2208.05973}{{\ttfamily
  arXiv:2208.05973 [hep-th]}}.

\bibitem{Lin:2022xod}
L.~Lin, D.~G. Robbins, and E.~Sharpe, ``{Decomposition, condensation defects,
  and fusion},'' \href{http://arxiv.org/abs/2208.05982}{{\ttfamily
  arXiv:2208.05982 [hep-th]}}.

\bibitem{Bartsch:2022mpm}
T.~Bartsch, M.~Bullimore, A.~E.~V. Ferrari, and J.~Pearson, ``{Non-invertible
  Symmetries and Higher Representation Theory I},''
  \href{http://arxiv.org/abs/2208.05993}{{\ttfamily arXiv:2208.05993
  [hep-th]}}.

\bibitem{Apruzzi:2022rei}
F.~Apruzzi, I.~Bah, F.~Bonetti, and S.~Schafer-Nameki, ``{Noninvertible
  Symmetries from Holography and Branes},''
  \href{http://dx.doi.org/10.1103/PhysRevLett.130.121601}{{\em Phys. Rev.
  Lett.} {\bfseries 130} no.~12, (2023) 121601},
  \href{http://arxiv.org/abs/2208.07373}{{\ttfamily arXiv:2208.07373
  [hep-th]}}.

\bibitem{GarciaEtxebarria:2022vzq}
I.~n. Garc\'\i{}a~Etxebarria, ``{Branes and Non-Invertible Symmetries},''
  \href{http://dx.doi.org/10.1002/prop.202200154}{{\em Fortsch. Phys.}
  {\bfseries 70} no.~11, (2022) 2200154},
  \href{http://arxiv.org/abs/2208.07508}{{\ttfamily arXiv:2208.07508
  [hep-th]}}.

\bibitem{Cherman:2022eml}
A.~Cherman, T.~Jacobson, and M.~Neuzil, ``{1-form symmetry versus large N
  QCD},'' \href{http://dx.doi.org/10.1007/JHEP02(2023)192}{{\em JHEP}
  {\bfseries 02} (2023) 192}, \href{http://arxiv.org/abs/2209.00027}{{\ttfamily
  arXiv:2209.00027 [hep-th]}}.

\bibitem{Heckman:2022muc}
J.~J. Heckman, M.~H\"ubner, E.~Torres, and H.~Y. Zhang, ``{The Branes Behind
  Generalized Symmetry Operators},''
  \href{http://arxiv.org/abs/2209.03343}{{\ttfamily arXiv:2209.03343
  [hep-th]}}.

\bibitem{Lu:2022ver}
D.-C. Lu and Z.~Sun, ``{On Triality Defects in 2d CFT},''
  \href{http://arxiv.org/abs/2208.06077}{{\ttfamily arXiv:2208.06077
  [hep-th]}}.

\bibitem{Niro:2022ctq}
P.~Niro, K.~Roumpedakis, and O.~Sela, ``{Exploring Non-Invertible Symmetries in
  Free Theories},'' \href{http://arxiv.org/abs/2209.11166}{{\ttfamily
  arXiv:2209.11166 [hep-th]}}.

\bibitem{Mekareeya:2022spm}
N.~Mekareeya and M.~Sacchi, ``{Mixed Anomalies, Two-groups, Non-Invertible
  Symmetries, and 3d Superconformal Indices},''
  \href{http://arxiv.org/abs/2210.02466}{{\ttfamily arXiv:2210.02466
  [hep-th]}}.

\bibitem{vanBeest:2022fss}
M.~van Beest, D.~S.~W. Gould, S.~Schafer-Nameki, and Y.-N. Wang, ``{Symmetry
  TFTs for 3d QFTs from M-theory},''
  \href{http://dx.doi.org/10.1007/JHEP02(2023)226}{{\em JHEP} {\bfseries 02}
  (2023) 226}, \href{http://arxiv.org/abs/2210.03703}{{\ttfamily
  arXiv:2210.03703 [hep-th]}}.

\bibitem{Giaccari:2022xgs}
S.~Giaccari and R.~Volpato, ``{A fresh view on string orbifolds},''
  \href{http://dx.doi.org/10.1007/JHEP01(2023)173}{{\em JHEP} {\bfseries 01}
  (2023) 173}, \href{http://arxiv.org/abs/2210.10034}{{\ttfamily
  arXiv:2210.10034 [hep-th]}}.

\bibitem{Cordova:2022fhg}
C.~Cordova, S.~Hong, S.~Koren, and K.~Ohmori, ``{Neutrino Masses from
  Generalized Symmetry Breaking},''
  \href{http://arxiv.org/abs/2211.07639}{{\ttfamily arXiv:2211.07639
  [hep-ph]}}.

\bibitem{GarciaEtxebarria:2022jky}
I.~n. Garc\'\i{}a~Etxebarria and N.~Iqbal, ``{A Goldstone theorem for
  continuous non-invertible symmetries},''
  \href{http://arxiv.org/abs/2211.09570}{{\ttfamily arXiv:2211.09570
  [hep-th]}}.

\bibitem{Choi:2022fgx}
Y.~Choi, H.~T. Lam, and S.-H. Shao, ``{Non-invertible Gauss Law and Axions},''
  \href{http://arxiv.org/abs/2212.04499}{{\ttfamily arXiv:2212.04499
  [hep-th]}}.

\bibitem{Robbins:2022wlr}
D.~G. Robbins, E.~Sharpe, and T.~Vandermeulen, ``{Decomposition,
  trivially-acting symmetries, and topological operators},''
  \href{http://dx.doi.org/10.1103/PhysRevD.107.085017}{{\em Phys. Rev. D}
  {\bfseries 107} no.~8, (2023) 085017},
  \href{http://arxiv.org/abs/2211.14332}{{\ttfamily arXiv:2211.14332
  [hep-th]}}.

\bibitem{Bhardwaj:2022kot}
L.~Bhardwaj, S.~Schafer-Nameki, and A.~Tiwari, ``{Unifying Constructions of
  Non-Invertible Symmetries},''
  \href{http://arxiv.org/abs/2212.06159}{{\ttfamily arXiv:2212.06159
  [hep-th]}}.

\bibitem{Bhardwaj:2022maz}
L.~Bhardwaj, L.~E. Bottini, S.~Schafer-Nameki, and A.~Tiwari, ``{Non-Invertible
  Symmetry Webs},'' \href{http://arxiv.org/abs/2212.06842}{{\ttfamily
  arXiv:2212.06842 [hep-th]}}.

\bibitem{Bartsch:2022ytj}
T.~Bartsch, M.~Bullimore, A.~E.~V. Ferrari, and J.~Pearson, ``{Non-invertible
  Symmetries and Higher Representation Theory II},''
  \href{http://arxiv.org/abs/2212.07393}{{\ttfamily arXiv:2212.07393
  [hep-th]}}.

\bibitem{Gaiotto:2020iye}
D.~Gaiotto and J.~Kulp, ``{Orbifold groupoids},''
  \href{http://dx.doi.org/10.1007/JHEP02(2021)132}{{\em JHEP} {\bfseries 02}
  (2021) 132}, \href{http://arxiv.org/abs/2008.05960}{{\ttfamily
  arXiv:2008.05960 [hep-th]}}.

\bibitem{Robbins:2021ibx}
D.~G. Robbins, E.~Sharpe, and T.~Vandermeulen, ``{Quantum symmetries in
  orbifolds and decomposition},''
  \href{http://dx.doi.org/10.1007/JHEP02(2022)108}{{\em JHEP} {\bfseries 02}
  (2022) 108}, \href{http://arxiv.org/abs/2107.12386}{{\ttfamily
  arXiv:2107.12386 [hep-th]}}.

\bibitem{Robbins:2021xce}
D.~G. Robbins, E.~Sharpe, and T.~Vandermeulen, ``{Anomaly resolution via
  decomposition},'' \href{http://dx.doi.org/10.1142/S0217751X21502201}{{\em
  Int. J. Mod. Phys. A} {\bfseries 36} no.~29, (2021) 2150220},
  \href{http://arxiv.org/abs/2107.13552}{{\ttfamily arXiv:2107.13552
  [hep-th]}}.

\bibitem{Huang:2021zvu}
T.-C. Huang, Y.-H. Lin, and S.~Seifnashri, ``{Construction of two-dimensional
  topological field theories with non-invertible symmetries},''
  \href{http://dx.doi.org/10.1007/JHEP12(2021)028}{{\em JHEP} {\bfseries 12}
  (2021) 028}, \href{http://arxiv.org/abs/2110.02958}{{\ttfamily
  arXiv:2110.02958 [hep-th]}}.

\bibitem{Inamura:2021szw}
K.~Inamura, ``{On lattice models of gapped phases with fusion category
  symmetries},'' \href{http://dx.doi.org/10.1007/JHEP03(2022)036}{{\em JHEP}
  {\bfseries 03} (2022) 036}, \href{http://arxiv.org/abs/2110.12882}{{\ttfamily
  arXiv:2110.12882 [cond-mat.str-el]}}.

\bibitem{Cherman:2021nox}
A.~Cherman, T.~Jacobson, and M.~Neuzil, ``{Universal Deformations},''
  \href{http://dx.doi.org/10.21468/SciPostPhys.12.4.116}{{\em SciPost Phys.}
  {\bfseries 12} no.~4, (2022) 116},
  \href{http://arxiv.org/abs/2111.00078}{{\ttfamily arXiv:2111.00078
  [hep-th]}}.

\bibitem{Sharpe:2022ene}
E.~Sharpe, ``{An introduction to decomposition},''
  \href{http://arxiv.org/abs/2204.09117}{{\ttfamily arXiv:2204.09117
  [hep-th]}}.

\bibitem{Lee:2022swr}
S.-J. Lee and P.-K. Oehlmann, ``{Geometric Bounds on the 1-Form Gauge
  Sector},'' \href{http://arxiv.org/abs/2212.11915}{{\ttfamily arXiv:2212.11915
  [hep-th]}}.

\bibitem{Inamura:2022lun}
K.~Inamura, ``{Fermionization of fusion category symmetries in 1+1
  dimensions},'' \href{http://arxiv.org/abs/2206.13159}{{\ttfamily
  arXiv:2206.13159 [cond-mat.str-el]}}.

\bibitem{Damia:2022bcd}
J.~A. Damia, R.~Argurio, and E.~Garcia-Valdecasas, ``{Non-Invertible Defects in
  5d, Boundaries and Holography},''
  \href{http://arxiv.org/abs/2207.02831}{{\ttfamily arXiv:2207.02831
  [hep-th]}}.

\bibitem{Lin:2022dhv}
Y.-H. Lin, M.~Okada, S.~Seifnashri, and Y.~Tachikawa, ``{Asymptotic density of
  states in 2d CFTs with non-invertible symmetries},''
  \href{http://arxiv.org/abs/2208.05495}{{\ttfamily arXiv:2208.05495
  [hep-th]}}.

\bibitem{Burbano:2021loy}
I.~M. Burbano, J.~Kulp, and J.~Neuser, ``{Duality Defects in $E_8$},''
  \href{http://arxiv.org/abs/2112.14323}{{\ttfamily arXiv:2112.14323
  [hep-th]}}.

\bibitem{Damia:2022rxw}
J.~A. Damia, R.~Argurio, and L.~Tizzano, ``{Continuous Generalized Symmetries
  in Three Dimensions},'' \href{http://arxiv.org/abs/2206.14093}{{\ttfamily
  arXiv:2206.14093 [hep-th]}}.

\bibitem{Apte:2022xtu}
A.~Apte, C.~Cordova, and H.~T. Lam, ``{Obstructions to Gapped Phases from
  Non-Invertible Symmetries},''
  \href{http://arxiv.org/abs/2212.14605}{{\ttfamily arXiv:2212.14605
  [hep-th]}}.

\bibitem{Nawata:2023rdx}
S.~Nawata, M.~Sperling, H.~E. Wang, and Z.~Zhong, ``{3d $\mathcal{N}=4$ mirror
  symmetry with 1-form symmetry},''
  \href{http://arxiv.org/abs/2301.02409}{{\ttfamily arXiv:2301.02409
  [hep-th]}}.

\bibitem{Bhardwaj:2023zix}
L.~Bhardwaj, M.~Bullimore, A.~E.~V. Ferrari, and S.~Schafer-Nameki,
  ``{Generalized Symmetries and Anomalies of 3d N=4 SCFTs},''
  \href{http://arxiv.org/abs/2301.02249}{{\ttfamily arXiv:2301.02249
  [hep-th]}}.

\bibitem{Kaidi:2023maf}
J.~Kaidi, E.~Nardoni, G.~Zafrir, and Y.~Zheng, ``{Symmetry TFTs and Anomalies
  of Non-Invertible Symmetries},''
  \href{http://arxiv.org/abs/2301.07112}{{\ttfamily arXiv:2301.07112
  [hep-th]}}.

\bibitem{Etheredge:2023ler}
M.~Etheredge, I.~Garcia~Etxebarria, B.~Heidenreich, and S.~Rauch, ``{Branes and
  symmetries for $\mathcal N=3$ S-folds},''
  \href{http://arxiv.org/abs/2302.14068}{{\ttfamily arXiv:2302.14068
  [hep-th]}}.

\bibitem{Lin:2023uvm}
Y.-H. Lin and S.-H. Shao, ``{Bootstrapping Non-invertible Symmetries},''
  \href{http://arxiv.org/abs/2302.13900}{{\ttfamily arXiv:2302.13900
  [hep-th]}}.

\bibitem{Amariti:2023hev}
A.~Amariti, D.~Morgante, A.~Pasternak, S.~Rota, and V.~Tatitscheff, ``{One-form
  symmetries in $\mathcal{N} = 3$$S$-folds},''
  \href{http://arxiv.org/abs/2303.07299}{{\ttfamily arXiv:2303.07299
  [hep-th]}}.

\bibitem{Bhardwaj:2023wzd}
L.~Bhardwaj and S.~Schafer-Nameki, ``{Generalized Charges, Part I: Invertible
  Symmetries and Higher Representations},''
  \href{http://arxiv.org/abs/2304.02660}{{\ttfamily arXiv:2304.02660
  [hep-th]}}.

\bibitem{Bartsch:2023pzl}
T.~Bartsch, M.~Bullimore, and A.~Grigoletto, ``{Higher representations for
  extended operators},'' \href{http://arxiv.org/abs/2304.03789}{{\ttfamily
  arXiv:2304.03789 [hep-th]}}.

\bibitem{Carta:2023bqn}
F.~Carta, S.~Giacomelli, N.~Mekareeya, and A.~Mininno, ``{Comments on
  Non-invertible Symmetries in Argyres-Douglas Theories},''
  \href{http://arxiv.org/abs/2303.16216}{{\ttfamily arXiv:2303.16216
  [hep-th]}}.

\bibitem{Zhang:2023wlu}
C.~Zhang and C.~C\'ordova, ``{Anomalies of $(1+1)D$ categorical symmetries},''
  \href{http://arxiv.org/abs/2304.01262}{{\ttfamily arXiv:2304.01262
  [cond-mat.str-el]}}.

\bibitem{Cao:2023doz}
W.~Cao, L.~Li, M.~Yamazaki, and Y.~Zheng, ``{Subsystem Non-Invertible Symmetry
  Operators and Defects},'' \href{http://arxiv.org/abs/2304.09886}{{\ttfamily
  arXiv:2304.09886 [cond-mat.str-el]}}.

\bibitem{Putrov:2023jqi}
P.~Putrov and J.~Wang, ``{Categorical Symmetry of the Standard Model from
  Gravitational Anomaly},'' \href{http://arxiv.org/abs/2302.14862}{{\ttfamily
  arXiv:2302.14862 [hep-th]}}.

\bibitem{Acharya:2023bth}
B.~S. Acharya, M.~Del~Zotto, J.~J. Heckman, M.~Hubner, and E.~Torres,
  ``{Junctions, Edge Modes, and $G_2$-Holonomy Orbifolds},''
  \href{http://arxiv.org/abs/2304.03300}{{\ttfamily arXiv:2304.03300
  [hep-th]}}.

\bibitem{Inamura:2023qzl}
K.~Inamura and K.~Ohmori, ``{Fusion Surface Models: 2+1d Lattice Models from
  Fusion 2-Categories},'' \href{http://arxiv.org/abs/2305.05774}{{\ttfamily
  arXiv:2305.05774 [cond-mat.str-el]}}.

\bibitem{Dierigl:2023jdp}
M.~Dierigl, J.~J. Heckman, M.~Montero, and E.~Torres, ``{R7-Branes as Charge
  Conjugation Operators},'' \href{http://arxiv.org/abs/2305.05689}{{\ttfamily
  arXiv:2305.05689 [hep-th]}}.

\bibitem{Antinucci:2023uzq}
A.~Antinucci, G.~Galati, G.~Rizi, and M.~Serone, ``{Symmetries and topological
  operators, on average},'' \href{http://arxiv.org/abs/2305.08911}{{\ttfamily
  arXiv:2305.08911 [hep-th]}}.

\bibitem{Cvetic:2023plv}
M.~"Cvetic, J.~J. Heckman, M.~H\"ubner, and E.~Torres, ``{Fluxbranes,
  Generalized Symmetries, and Verlinde's Metastable Monopole},''
  \href{http://arxiv.org/abs/2305.09665}{{\ttfamily arXiv:2305.09665
  [hep-th]}}.

\bibitem{Cordova:2022ruw}
C.~Cordova, T.~T. Dumitrescu, K.~Intriligator, and S.-H. Shao, ``{Snowmass
  White Paper: Generalized Symmetries in Quantum Field Theory and Beyond},'' in
  {\em {Snowmass 2021}}.
\newblock 5, 2022.
\newblock \href{http://arxiv.org/abs/2205.09545}{{\ttfamily arXiv:2205.09545
  [hep-th]}}.

\bibitem{Huang:2021nvb}
T.-C. Huang, Y.-H. Lin, K.~Ohmori, Y.~Tachikawa, and M.~Tezuka, ``{Numerical
  Evidence for a Haagerup Conformal Field Theory},''
  \href{http://dx.doi.org/10.1103/PhysRevLett.128.231603}{{\em Phys. Rev.
  Lett.} {\bfseries 128} no.~23, (2022) 231603},
  \href{http://arxiv.org/abs/2110.03008}{{\ttfamily arXiv:2110.03008
  [cond-mat.stat-mech]}}.

\bibitem{teleman}
C.~Teleman, ``{The Haagerup TQFT is not a gauge theory --- unpublished note,
  available from the
  \href{https://math.berkeley.edu/~teleman/math/Haagerup.pdf}{author's
  website}}.''.

\bibitem{Witten:2009at}
E.~Witten, ``{Geometric Langlands From Six Dimensions},''
  \href{http://arxiv.org/abs/0905.2720}{{\ttfamily arXiv:0905.2720 [hep-th]}}.

\bibitem{Freed:2012bs}
D.~S. Freed and C.~Teleman, ``{Relative quantum field theory},''
  \href{http://dx.doi.org/10.1007/s00220-013-1880-1}{{\em Commun. Math. Phys.}
  {\bfseries 326} (2014) 459--476},
  \href{http://arxiv.org/abs/1212.1692}{{\ttfamily arXiv:1212.1692 [hep-th]}}.

\bibitem{Witten:1995gf}
E.~Witten, ``{On S duality in Abelian gauge theory},''
  \href{http://dx.doi.org/10.1007/BF01671570}{{\em Selecta Math.} {\bfseries 1}
  (1995) 383}, \href{http://arxiv.org/abs/hep-th/9505186}{{\ttfamily
  arXiv:hep-th/9505186}}.

\bibitem{Dolan:1998qk}
L.~Dolan and C.~R. Nappi, ``{A Modular invariant partition function for the
  five-brane},'' \href{http://dx.doi.org/10.1016/S0550-3213(98)00537-9}{{\em
  Nucl. Phys. B} {\bfseries 530} (1998) 683--700},
  \href{http://arxiv.org/abs/hep-th/9806016}{{\ttfamily arXiv:hep-th/9806016}}.

\end{thebibliography}\endgroup

\end{document}